\documentclass[11pt,a4paper,twoside=false,numbers=noenddot,headings=small,bibliography=totoc,chapterprefix=false]{scrbook}
\usepackage[utf8]{inputenc}
\usepackage[T1]{fontenc}
\usepackage{lmodern}
\usepackage{microtype}
\usepackage[a4paper,textwidth=135mm,textheight=215mm,marginratio=1:1]{geometry}
\usepackage{amsmath,amssymb}
\usepackage{booktabs}
\usepackage{array}
\usepackage{graphicx}
\usepackage{xcolor}
\usepackage{xurl}
\usepackage[most]{tcolorbox}
\newtcolorbox{takeaways}[1][Key takeaways]{enhanced,colback=black!3,colframe=black!55,boxrule=0.5pt,arc=1pt,left=6pt,right=6pt,top=4pt,bottom=4pt,fonttitle=\bfseries\small,title=#1,before skip=8pt,after skip=10pt}
\newtcolorbox{guidelines}[1]{breakable,enhanced,colback=white,colframe=black!70,boxrule=0.6pt,arc=0pt,left=6pt,right=6pt,top=4pt,bottom=4pt,fonttitle=\bfseries\small,title=#1,before skip=8pt,after skip=10pt}
\newcommand{\warning}[1]{\paragraph{Warning: #1.}}
\usepackage[style=authoryear-comp,backend=biber,maxcitenames=2,maxbibnames=10,giveninits=true,uniquename=false,uniquelist=false,sorting=nyt,dashed=false]{biblatex}
\usepackage[colorlinks=true,linkcolor=black!70!blue,citecolor=black!60!green,urlcolor=black!60!blue]{hyperref}
\hypersetup{pdftitle={A Tutorial on Generative Models for Music Audio: From Text-to-Music to Live Performance},pdfauthor={Mateusz Modrzejewski, Joanna Kurkowska, Dorien Herremans}}
\newcommand{\windowend}{September 2026}
\setkomafont{disposition}{\normalfont\bfseries}
\setkomafont{descriptionlabel}{\normalfont\itshape}
\title{A Tutorial on Generative Models for Music Audio}
\subtitle{From Text-to-Music to Live Performance}
\author{Mateusz Modrzejewski\thanks{Warsaw University of Technology, Institute of Computer Science, and ZAiKS Lab; mateusz.modrzejewski@pw.edu.pl} \and Joanna Kurkowska\thanks{ZAiKS Lab; joanna.kurkowska@zaiks.org.pl} \and Dorien Herremans\thanks{Singapore University of Technology and Design; dorien\_herremans@sutd.edu.sg}}
\date{26 October 2026}

\usepackage{placeins}
\usepackage{float}
\begin{document}
\frontmatter
\maketitle

\chapter*{Abstract}
\markboth{Abstract}{Abstract}
Music audio generation spans instrumental clips, full songs with vocals, edits of existing recordings, and systems that play alongside a musician in real time. This monograph explains the methods behind these systems for readers familiar with neural-network training who are new to music generation. It introduces neural audio codecs and autoencoders, autoregressive and masked token models, diffusion and flow matching, sampling acceleration, and preference optimization in a shared notation. Worked examples connect training objectives and sampling procedures to published systems. The application sections relate the musical requirements of instrumental generation, songs, editing, and live performance to model design. Comparison tables record representations, methods, inputs, tasks, and release information documented in papers and official sources. Symbolic composition, speech, and sound effects are included where they connect to music audio. We explain how to interpret automatic metrics and learned judges, design and report listening tests, and assess evidence about data provenance, memorization, attribution, and output detection. Selected legal decisions and disputes illustrate the implications for training and release. The reviewed systems combine established generative paradigms with new representations, planning stages, and post-training procedures. Differences in data, scale, and evaluation limit what cross-system comparisons can establish about the sources of reported gains.

\chapter*{Preface}
\markboth{Preface}{Preface}

We came to this subject first as players and listeners, and later as researchers in music information retrieval and music technology. The field has changed quickly, and its literature has become difficult to follow as a whole. We wrote this monograph for graduate students, researchers, and engineers entering music audio generation, and for musicians curious about the methods behind the tools. It brings the main objectives and sampling procedures into a shared notation and connects model design to musical aims, practical constraints, and the evidence behind system comparisons. The mathematical sections assume familiarity with neural-network training, transformers, and variational autoencoders. Musical terms are explained where they first matter. Readers who come to the subject through musical practice may find the sections on control, neural instruments, and live interaction a good place to begin. We hope the monograph makes the field easier to enter than we found it, and that it keeps the music in view.

\tableofcontents

\mainmatter

\chapter{Introduction}
\label{sec:intro}

A useful music generator must do more than produce plausible sound. A lyrics-to-song system must coordinate the words, the singing, and any accompaniment across the song's sections. An edit must change the requested feature and preserve the rest of the recording. A system used in performance must respond soon enough for a musician to act on what they hear. These requirements place different demands on representation, generation, and control, and the methods that current systems use to meet them are scattered across papers that often assume familiarity with earlier work. Neural audio generation predates the systems treated here by nearly a decade. Autoregressive waveform models demonstrated that neural networks can synthesize audio sample by sample~\parencite{oord2016wavenet,mehri2017samplernn}, two-stage text-to-speech pipelines established the pattern of generating an intermediate representation and rendering it with a neural vocoder~\parencite{shen2018tacotron2}, and transformers over symbolic tokens captured long-range musical structure~\parencite{huang2019musictransformer}. Jukebox produced full songs with singing directly in the audio domain~\parencite{dhariwal2020jukebox}, but at low fidelity and with sampling times measured in hours. For music, these systems remained proofs of concept rather than usable tools.

The change came in early 2023, when three results arrived within months of each other. MusicLM framed text-to-music generation as hierarchical language modeling over discrete audio tokens~\parencite{agostinelli2023musiclm}, building on the SoundStream codec with residual vector quantization~\parencite{zeghidour2022soundstream}, the semantic-plus-acoustic token decomposition of AudioLM~\parencite{borsos2023audiolm}, and joint music-text embeddings~\parencite{huang2022mulan}. Other widely used open codecs include EnCodec and the Descript Audio Codec~\parencite{defossez2023encodec,kumar2023dac}. In the same year, a parallel line of work showed that latent diffusion reaches comparable quality from continuous representations~\parencite{huang2023noise2music,liu2023audioldm,schneider2024mousai}, and MusicGen demonstrated that a single-stage token language model with a simple codebook interleaving scheme suffices as a strong, open baseline~\parencite{copet2023musicgen}. Together these systems established the two architectural lineages, codec-token language models and latent diffusion, that organize most of the systems tabulated in this monograph (Figure~\ref{fig:paradigm-tier} gives the counts).

The work that followed these systems spread in several directions at once. It expanded from clip-level instrumental generation toward full songs with vocals, lyrics, and verse-chorus structure, first in closed commercial systems~\parencite{suno2026,udio2026,bai2024seedmusic} and then in open models~\parencite{yuan2025yue,ning2025diffrhythm,gong2025acestep,lei2025levo,yang2026heartmula}. The two lineages converged into planner-renderer hybrids, in which a language model plans a coarse structure and a diffusion or flow model renders audio~\parencite{lam2023efficient,yang2025songbloom,lam2025musicot,qwen2026music}. Flow matching, distillation, and adversarial post-training reduced sampling cost to the point of faster-than-real-time and streaming generation~\parencite{prajwal2024musicflow,novack2025presto,evans2026stableaudio3,caillon2025live,novack2026lmdm}. Preference alignment, introduced by large-scale RLHF for music~\parencite{cideron2024musicrl} and by direct preference optimization for text-to-audio~\parencite{majumder2024tango2}, became an increasingly common post-training stage~\parencite{lei2025levo,chen2025diffrhythmplus}. A parallel thread developed fine-grained and time-varying control~\parencite{wu2024musiccontrolnet,tal2024jasco,novack2024ditto} and instruction-based editing~\parencite{han2023instructme,zhang2024musicmagus,zhang2024instructmusicgen}, moving systems from one-shot generators toward production tools.

Evaluation methods have improved more slowly than generation methods. The Fr\'echet Audio Distance~\parencite{kilgour2019fad}, the default automatic metric for most of this period, is sensitive to the choice of embedding and reference set~\parencite{gui2024adapting,tailleur2024correlation,jeong2026empirical} and correlates weakly with human preferences for music~\parencite{huang2025aligning}. The community is responding with metrics validated against human preferences~\parencite{huang2025aligning}, expert-annotated benchmarks~\parencite{liu2025musiceval,yao2025songeval}, live human-preference arenas~\parencite{kim2025musicarena}, and learned aesthetics predictors~\parencite{tjandra2025aesthetics}, but no standard protocol has emerged, particularly for full songs with lyrics.

The conditions under which music generators are trained and released are contested. Record-label litigation over training data~\parencite{riaa2024lawsuits} has continued alongside selected settlements and licensing agreements~\parencite{umg2025udio,wmg2025udio,wmg2025suno,malik2025warner}, while the Munich Regional Court has issued the two judgments discussed in Section~\ref{sec:data}~\parencite{lgmuenchen2025gema,lgmuenchen2026suno}. Commercial agreements change particular relationships between companies. They do not by themselves establish complete training-data provenance, settle other parties' claims, or demonstrate that musicians have consented or been fairly compensated. We therefore examine data documentation, memorization, attribution, and output identification as distinct questions, with their own methods: memorization measurement~\parencite{yuan2025yue,bralios2023generation}, watermarking~\parencite{sanroman2024audioseal}, and synthetic-music detection~\parencite{rahman2025sonics,comanducci2025fakemusiccaps,afchar2024detecting}. Reporting remains limited, although some systems document their training sources~\parencite{evans2025stableaudioopen,nistal2024diffariff}. We also distinguish company announcements, legal allegations, and court findings according to the evidence each provides.

Surveys of parts of this literature exist, and this text leans on them and points to them where they go deeper. The earlier deep-learning surveys organize generation by objective, representation, and architecture~\parencite{briot2017deep,ji2020comprehensive} or by the musical dimension a system generates~\parencite{herremans2018functional}, and were written before audio-domain systems reached usable quality. Among recent surveys, the broadest treatment of music foundation models spans understanding as well as generation~\parencite{ma2024foundation}. Others cover single branches such as symbolic generation with language-model methods~\parencite{le2024natural}, video-to-music generation~\parencite{ji2025comprehensive,wang2025vision}, music understanding models~\parencite{li2024survey}, large audio models across speech, sound, and music~\parencite{latif2023sparks}, generation in live interaction~\parencite{agchar2024survey}, tools and products~\parencite{zhu2023survey}, or general audio generation methods~\parencite{bozic2024survey,chen2024applications}. Evaluation has surveys of its own~\parencite{xiong2023comprehensive,kader2025survey,lerch2025survey}, alignment has one~\parencite{herremans2025aligning}, ethics has one~\parencite{barnett2023ethical}, and a position paper maps the open research areas of music AI in the foundation-model era~\parencite{wei2024prevailing}. The survey closest in scope organizes music generation by guidance modality across symbolic and audio output and tabulates methods by task, year, and architecture~\parencite{li2025survey}. Its coverage of symbolic generation and of multimodal fusion designs is broader than ours, and its newest systems date from early 2025. These surveys provide complementary coverage, and we refer readers to them for fuller treatments of individual branches. The purpose here is to teach the foundations that the current systems share, to show how the design choices behind them appear in published systems, and to leave the reader able to read the next system's report, place it, and judge it.

The monograph has four learning goals. After Sections~\ref{sec:background} to~\ref{sec:taxonomy}, the reader should be able to identify, in a system description, the representation, the generation method, the conditions, and the evidence behind its claims, write down the training objective and sampling procedure of its paradigm, and compute from its codec the token rate (tokens per second of audio), the sequence length, and the step count that set its cost. The worked examples of Section~\ref{sec:notation} do this for MusicGen, and a reading guide does the same for the standard metrics. Sections~\ref{sec:instrumental} to~\ref{sec:realtime} explain the evidence relevant to three practical choices: between the token and the latent lineage for a given deployment, between an adapter (a small trained module inside a frozen model), a control branch, inference-time optimization, and an explicit plan for a given control, and between a large distilled model and a small purpose-built one for a given latency budget. The evidence and the open trade-offs are stated at the end of each of those sections. After Section~\ref{sec:evaluation}, the reader can tell which numbers in a system report's evaluation have been checked against human judgments, and can design and report a listening test. After Section~\ref{sec:data}, the reader can read a model card for what it says about data, license, and memorization. The same section says which of these statements the dated legal record bears on.

That aim shapes how the text is built. The shared objectives and sampling procedures are stated in one notation and shown at work in representative systems, and the figures and tables are built so that they can be read without the surrounding prose. The comparison tables record the design and release information documented in papers, model cards, repositories, and official product sources. Section~\ref{sec:method} explains the selection process, and Appendix~\ref{app:search} records the coverage and observation dates. Missing information remains explicit, so readers can distinguish a reported property from an inference.

Six questions recur across the systems, and we ask the reader to carry them through the text, because each is a question of machine learning before it is a question of music and each has a different answer in the application classes of Sections~\ref{sec:instrumental} to~\ref{sec:realtime}.
\begin{description}
  \item[Q1, representation.] What information does the generated representation keep and discard, and what does that decide about fidelity, sequence length, and control? Codec tokens, continuous latents, and sketches answer differently (Sections~\ref{sec:building-blocks}, \ref{sec:d1}, and~\ref{sec:instrumental}).
  \item[Q2, cost.] How do sequence length, token rate, and the generative paradigm set training and inference cost, and what bounds latency? The objectives of Section~\ref{sec:notation} make the costs explicit. Sections~\ref{sec:instrumental} and~\ref{sec:realtime} report them.
  \item[Q3, conditioning.] How do global conditions, time-varying conditions, and reference audio enter a model, and what does each route cost in training and in fidelity? Sections~\ref{sec:d3} and~\ref{sec:control} answer.
  \item[Q4, planning.] When should a plan be generated before the audio, what should the plan contain, and what does the separation cost? Section~\ref{sec:notation} states the planner-renderer objective. Sections~\ref{sec:song} and~\ref{sec:control} record where the separation helped.
  \item[Q5, measurement.] What can distribution-level metrics establish about generated music, where do they fail, and how well do the alternative instruments agree with human judgments? Section~\ref{sec:evaluation} answers.
  \item[Q6, artifacts.] What changes when a model is a released artifact rather than a paper: openness, provenance, memorization, and the legal record? Sections~\ref{sec:d5} and~\ref{sec:data} answer.
\end{description}
These questions guide the comparisons in the application sections. The comparison framework of Section~\ref{sec:taxonomy} records the design choices behind Q1 to Q4, the application class in which each choice was made, and the evidence about data, access, and release conditions that Q6 asks for. Section~\ref{sec:conclusion} returns to the six questions with the answers the collected systems support. The answers converge on one observation, stated there with its limits: every generative paradigm in the tables was known by 2023, and the systems since differ in what they generate over, in what follows their training, and in whether a planner precedes the renderer. The tables cannot establish which of these changes account for the reported gains, because the systems differ in data, scale, and evaluation.

\begin{figure}[H]
  \centering
  \includegraphics[width=\linewidth,alt={Block diagram of the monograph's five taxonomy dimensions and the sections that cover each.}]{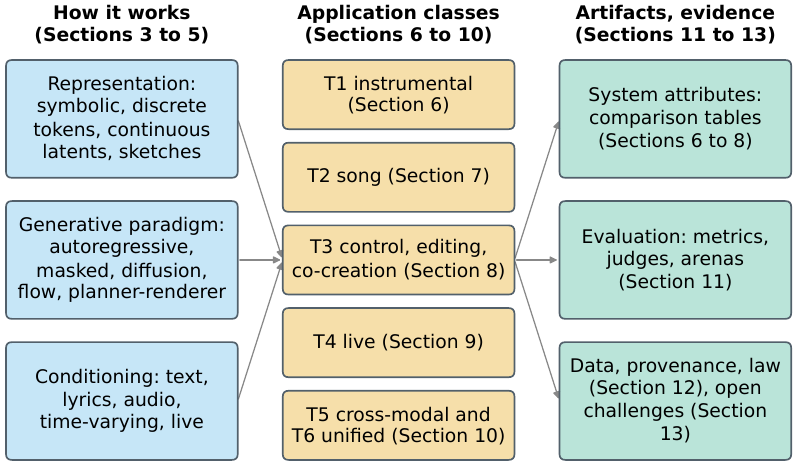}
  \caption{Structure of the monograph. Dimensions D1 to D3 (representation, generative paradigm, conditioning) describe how a model works (Sections~\ref{sec:building-blocks} and~\ref{sec:taxonomy}), and the application classes D4 organize the review (Sections~\ref{sec:instrumental} to~\ref{sec:crossmodal-symbolic}). The release attributes (D5) are recorded in the comparison tables of Sections~\ref{sec:instrumental} to~\ref{sec:realtime}, evaluation is treated in Section~\ref{sec:evaluation}, and data, provenance, and open challenges are treated in Sections~\ref{sec:data} and~\ref{sec:discussion}. Questions Q1 to Q4 of Section~\ref{sec:intro} are taken up in Sections~\ref{sec:building-blocks} to~\ref{sec:crossmodal-symbolic}, Q5 in Section~\ref{sec:evaluation}, and Q6 in Section~\ref{sec:data}.}
  \label{fig:overview}
\end{figure}

Figure~\ref{fig:overview} shows the structure of the monograph. Sections~\ref{sec:background} to~\ref{sec:taxonomy} teach the foundations: the scope, the building blocks, and the taxonomy with its notation and worked examples. Sections~\ref{sec:instrumental} to~\ref{sec:crossmodal-symbolic} review the systems by application class. Sections~\ref{sec:evaluation} and~\ref{sec:data} cover evaluation and data. Sections~\ref{sec:discussion} and~\ref{sec:conclusion} discuss and conclude. Appendix~\ref{app:search} documents the search. The mathematical sections use vectors, conditional probabilities, expectations, and gradients, and they take a neural network to be a parameterized function fitted to training examples. Audio signal processing is not assumed. Where transformers, autoencoders, and other architectures appear, the text explains their role in the pipeline and does not derive them. Four reading paths cover the readers we have in mind. A reader familiar with machine learning but new to music generation should take Sections~\ref{sec:background} to~\ref{sec:notation} in order, then Section~\ref{sec:instrumental} and one application section of interest, and consult Section~\ref{sec:taxonomy} and the tables as needed. A reader building or choosing a system should read Section~\ref{sec:notation}, the section of the intended application class among Sections~\ref{sec:instrumental} to~\ref{sec:realtime} with its closing subsection of lessons and trade-offs, and the comparison tables. Section~\ref{sec:map} explains how to navigate them. A musician or producer who wants the ideas before the formulas can read the openings, running examples, and closing lessons of Sections~\ref{sec:instrumental} to~\ref{sec:realtime}, with Section~\ref{sec:instruments} on neural instruments and Section~\ref{sec:streaming} on live interaction as entry points, and then the evaluation and provenance discussion. The formulas and the statistical protocol assume the stated background. A reader concerned with evaluation, data, or policy can go to Sections~\ref{sec:evaluation} and~\ref{sec:data} directly. They depend on the taxonomy only through the class names.

\paragraph{Scope.} The monograph covers the generation of music audio conditioned on text, lyrics, audio, symbolic signals, or video, including instrumental music, songs with vocals, stems (the separate instrument or voice tracks of a mix) and accompaniments, and edits of existing recordings. It treats symbolic composition only where it meets audio pipelines or co-creation, because its methods and its evaluation differ from those of audio generation, and Section~\ref{sec:symbolic} says where the two meet. Singing-voice synthesis from a score, speech synthesis, and sound-effect generation are excluded unless a system also generates music. We report training data, evaluation, and release status from the original sources. Many industrial systems disclose only part of this information. Coverage and source dates are documented in Section~\ref{sec:method} and Appendix~\ref{app:search}.

\chapter{Background and Scope}
\label{sec:background}

The review focuses on music audio generation from MusicLM~\parencite{agostinelli2023musiclm} onward, with earlier work included where it establishes the methods used by these systems. We take MusicLM as the starting point because it is the point at which text-to-music generation became simultaneously high-fidelity, minute-scale, and specified in enough detail to establish a reusable design. General text-to-audio models are included when they are evaluated on music or serve as the basis for music systems. The other limits of coverage are those of the Scope paragraph in Section~\ref{sec:intro}.

Sections~\ref{sec:building-blocks} and~\ref{sec:notation} develop the representations and objectives, the application sections connect them to musical tasks, and Section~\ref{sec:evaluation} explains the evaluation evidence. Section~\ref{sec:data} separates the technical questions of data and memorization from the dated account of cases and settlements. Section~\ref{sec:method} and Appendix~\ref{app:search} document the collection process and its limits.

\section{Earlier Foundations}
\label{sec:genealogy}

The 2023 systems combined four older lines, each cited here for the design it contributed. Autoregressive waveform models showed that a neural network can synthesize audio sample by sample from a learned distribution, at a cost of one sequential prediction per sample~\parencite{oord2016wavenet,mehri2017samplernn}. The token language models of Section~\ref{sec:instrumental} keep the autoregressive factorization and move it from samples to codec tokens at a rate lower by two orders of magnitude. Two-stage pipelines, in which one model generates an intermediate representation and a second renders it, were established in text-to-speech~\parencite{shen2018tacotron2} and recur in every planner-renderer system of Section~\ref{sec:song}. Transformers over symbolic tokens captured long-range musical structure with relative attention~\parencite{huang2019musictransformer}, which is the model class the token systems scaled, and Jukebox applied the same class to a hierarchy of VQ-VAE audio codes, producing full songs with singing at low fidelity and with sampling times measured in hours~\parencite{dhariwal2020jukebox}. Neural codecs with residual vector quantization~\parencite{zeghidour2022soundstream,defossez2023encodec} and the semantic-plus-acoustic decomposition of AudioLM~\parencite{borsos2023audiolm} are what made the hierarchy fast enough to use. Adversarial synthesis demonstrated fast spectral-domain generation at small scale~\parencite{engel2019gansynth}, and its discriminators return as the adversarial post-training of the current diffusion models. Differentiable signal processing~\parencite{engel2020ddsp} and the RAVE autoencoder~\parencite{caillon2021rave} started the instrument line of Section~\ref{sec:instruments} in parallel with, and independently of, the text-to-music systems. Each of these techniques reappears in modified form in the literature reviewed here.

The infrastructure behind the 2023 systems is covered here as building blocks. First, neural audio codecs with residual vector quantization compress waveforms into discrete token sequences that language models can predict~\parencite{zeghidour2022soundstream,defossez2023encodec,kumar2023dac}, with more recent tokenizers reducing token cost further~\parencite{ji2025wavtokenizer,cheng2026bandtok}. Second, contrastive audio-text embeddings provide the conditioning signal that connects free-form language to audio~\parencite{huang2022mulan,wu2023clap}, and self-supervised music encoders supply features for understanding and evaluation~\parencite{li2024mert,zhu2025muq}. Evaluation is treated in Section~\ref{sec:evaluation}, after the systems, and that section gives the metric and its known limits for the claims about quality in Sections~\ref{sec:instrumental} to~\ref{sec:crossmodal-symbolic}.

\section{Literature Collection}
\label{sec:method}

The bibliography was built by three complementary methods, applied throughout the collection period and closed when the classification was fixed. Citation chasing followed references forward and backward from the landmark systems and covered the proceedings of ISMIR, ICLR, NeurIPS, ICML, ICASSP, Interspeech, AAAI, and IJCAI and the journals TASLP, TISMIR, and TMLR. Structured searches queried the arXiv API with fixed search strings in nineteen topic groups, from surveys and codecs to provenance, preference alignment, neural synthesis instruments, multi-agent composition, motion and games, and a group for the most recent months. Over the collection period they returned 1,115 unique candidates, of which 83 were already cited through citation chasing at the time of the search. Every candidate was screened on title and abstract. During the collection period 253 were added, 163 with a dedicated sentence in the text and 90 in grouped citations. A second reading of the abstracts of every candidate not then added, completed in October 2026 after the classification was closed, found 13 further candidates that meet the inclusion criteria and 137 that merit a grouped mention. These additions are not included in the present classification or its counts. Five of them predate the eligibility cutoff but were identified after the classification had been finalized, and Appendix~\ref{app:screening} names them.

\begin{figure}[H]
  \centering
  \includegraphics[width=0.8\linewidth,alt={Bar chart of the number of cited works per year and per topic in the monograph's bibliography.}]{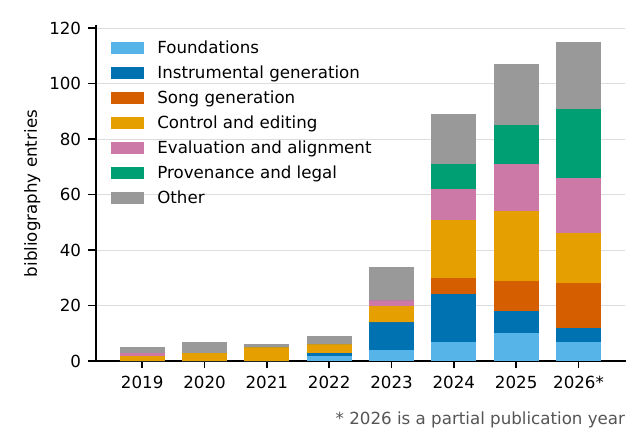}
  \caption{Composition of the reviewed literature by year and category. Counts describe the monograph's bibliography only. Entries before 2023 are the precursors cited for context, and the nine entries published before 2019 are not shown. The year 2026 is a partial publication year, and Section~\ref{sec:method} distinguishes the system cutoff from later contextual additions.}
  \label{fig:corpus}
\end{figure}

Venue monitoring used the accepted-paper list of ISMIR 2026 (140 papers), classified 36 as generative music work and 38 as adjacent, and located public versions for 62. Accepted papers without a public version are not cited. A paper is included if it generates music audio or directly supports it (codecs, embeddings, evaluation, provenance) and if it introduces an architecture, control mechanism, training procedure, evaluation method, or resource that subsequent work builds on. Work available only as a preprint is included when it meets these criteria and is marked as a preprint in the bibliography. Figure~\ref{fig:corpus} shows the composition of the resulting bibliography by year and category.

Candidates were de-duplicated against the bibliography and across groups by arXiv identifier and by normalized title. A candidate was excluded at screening if it generated no music audio and supported no such system, if it was a duplicate or a later version of an included work, or if it contributed no architecture, control mechanism, training procedure, evaluation method, or resource of the kinds listed above. The decision and its reason were recorded for each candidate. Where sources disagree, the monograph follows a fixed order: a peer-reviewed version over a preprint, a preprint over a model card, a model card over product documentation, and product documentation over press or legal reports, except that the openness, license, and withdrawal status of an artifact are taken from its card or repository because they change after the paper. The source is chosen for the claim: a methods paper for an objective, a versioned model card for a checkpoint's license, product documentation for current product behavior, and the court record for a judgment. Appendix~\ref{app:search} gives the search strings, the start date of each group, the closing date of the searches, the per-group counts, and the flow from candidates to included works (Figure~\ref{fig:flow}).

The system comparison includes selected releases with an eligibility cutoff at the end of \windowend{}. Appendix~\ref{app:search} distinguishes that cutoff from the closing date of the structured searches and the dates on which release information was observed. The bibliography also cites some later work for context, and that work does not enter the tables or the counts. The tables describe the documented versions included in this review. They should not be read as continuously updated product specifications.

\chapter{Building Blocks}
\label{sec:building-blocks}

\section{Codecs, Tokenizers, and Autoencoders}
\label{sec:codecs}

A waveform holds tens of thousands of samples for every second of sound, so most systems describe it with a much shorter sequence. An encoder produces that description, a decoder reconstructs audio from it, and the generative model learns to produce new descriptions of the same kind. The representation constrains how much information survives and how much work the generator must do. The exceptions are the waveform-domain cascades and the symbolic systems marked as such in the tables. On the discrete side, SoundStream introduced neural codecs with residual vector quantization (RVQ), turning waveforms into stacks of token sequences~\parencite{zeghidour2022soundstream}. EnCodec provided an open, high-fidelity implementation and became the default tokenizer for music language models~\parencite{defossez2023encodec}, and the Descript Audio Codec improved fidelity per bit substantially~\parencite{kumar2023dac}. AudioLM added the now-standard split between semantic tokens, which carry structure, and acoustic tokens, which carry detail~\parencite{borsos2023audiolm}. A diffusion decoder from tokens to waveform raised the fidelity ceiling of a fixed token stream~\parencite{roman2023discrete}.

Three properties of a codec matter to a generator, and each later tokenizer can be read as an attempt to improve one of them. The first is the token rate, the number of tokens a language model must predict per second of audio, which sets the sequence length and therefore the cost and the reachable duration. The worked example of Section~\ref{sec:notation} computes it for MusicGen. The second is semantic content: whether the tokens carry information about what is played and sung, which a language model can learn to predict, or only about how it sounds. The third is graceful degradation: whether dropping the finer codebooks of an RVQ stack removes detail in a controlled way or breaks the signal. The cost of RVQ stacks, several parallel token streams per frame, has motivated work along all three.

On the token rate, single-codebook codecs reach very low rates~\parencite{ji2025wavtokenizer}, music-specific codecs reach extreme compression~\parencite{xu2025mucodec}, spectral-domain codecs cover full-band stereo~\parencite{li2025spectrostream}, and two-dimensional mel-spectrogram tokenizers share one codebook across frequency bands, and their authors argue that RVQ stacks are poorly matched to autoregressive prediction~\parencite{cheng2026bandtok,li2025meltok}. On semantic content, a codec trained for reconstruction alone may encode information in a form that a language model cannot predict efficiently, which X-Codec addresses by supervising the codec with a semantic encoder~\parencite{ye2024codec}. Source-aware dual-track tokenizers give vocal and accompaniment language models tokens that reconstruct well and remain predictable~\parencite{lin2025duotok}. The most recent step is a semantically supervised low-rate tokenizer: Qwen-Music trains a single 25\,Hz codebook with transcription, chroma, and mel-reconstruction losses so that the tokens keep lyric and pitch information, and leaves the remaining acoustic detail to a downstream generative renderer~\parencite{qwen2026music}. On degradation, multi-scale quantizers place codebooks at different temporal resolutions~\parencite{siuzdak2024snac}, and frequency-ordered training makes truncating an RVQ stack remove predictable bands~\parencite{yang2026harp}. Codec benchmarks~\parencite{wu2024codec} and a systematic review of discrete audio tokenizers~\parencite{mousavi2025discrete} compare these designs. Universal low-frame-rate codecs, semantic codecs for general sound, streaming dialogue codecs, and transformer-only tokenizers trained at scale continue the development~\parencite{hu2026omnicodec,liu2024semanticodec,defossez2024moshi,gong2026moss,he2026tqcodec}, and fast stereo vocoders decode compressed representations into audio~\parencite{zhu2024musichifi}.

A discrete representation chooses entries from a finite vocabulary, and a continuous representation uses vectors of real numbers. In both cases the generator works in the compressed space and a decoder turns its output into audio. The decoder is trained on reconstructions of real recordings, and a separate generator learns to produce plausible new sequences in the same space. A variational autoencoder is one way to learn a continuous space whose distribution is regularized in training, and what follows concerns the resulting representation and decoder rather than the training objective. On the continuous side, diffusion and flow models operate over VAE latents or spectrograms~\parencite{liu2023audioldm,evans2024fast}. The design trend is toward autoencoders that are compact, accurate in reconstruction, and semantically structured at once. Highly compressed latents enable full-length songs in one context~\parencite{evans2024longform}, and frame rates as low as 5\,Hz make block-wise generation tractable~\parencite{jiang2025diffrhythm2}. Semantic-acoustic objectives structure the latent for downstream generation and editing~\parencite{parker2026same,evans2026stableaudio3}, autoencoders trained with a consistency objective (Section~\ref{sec:acceleration}) reconstruct from the latent in a single step~\parencite{pasini2024musiclatent}, and complex-spectral autoencoders address the high-frequency, phase, and stereo failures of high-compression decoders~\parencite{wang2026fourier}.

\section{Audio-Text Embeddings and Music Encoders}
\label{sec:embeddings}

Conditioning is the information supplied to guide generation: a description, a melody, lyrics, or an existing recording. An embedding turns such information into numerical features the generator can use, and when the same embedding also scores the output, the score inherits the distinctions the embedding learned to make. A text prompt reaches a generator through a text encoder or through a joint audio-text embedding trained to place a recording and its description near each other, and the joint embeddings also serve as prompt-adherence scores. MuLan, trained contrastively on 44 million recordings, is the conditioning backbone of MusicLM~\parencite{huang2022mulan}. CLAP is its open counterpart and is also used as an automatic prompt-adherence score~\parencite{wu2023clap}. Self-supervised music encoders provide features for understanding, control, and evaluation: MERT is the common choice~\parencite{li2024mert}, and MuQ uses Mel-RVQ token prediction, reports improvements over MERT, and releases a paired music-text embedding~\parencite{zhu2025muq}. CLaMP~3 extends contrastive music-text pretraining across symbolic, audio, and many languages~\parencite{wu2025clamp}. Benchmarks for these representations~\parencite{yuan2023marble} matter to generation research because metrics, reward models, and planning schemes~\parencite{lam2025musicot} inherit the biases of the encoder they build on (Section~\ref{sec:evaluation}).

\section{Data}
\label{sec:data-building-block}

Training corpora are mostly undisclosed. We defer provenance and licensing to Section~\ref{sec:data}. Two data problems recur on the technical level. First, captions: paired music-text data is scarce, so systems generate pseudo-captions with language models~\parencite{huang2023noise2music} or parse musical metadata out of rich captions~\parencite{melechovsky2024mustango}. Evaluation depends on small human-written caption sets~\parencite{agostinelli2023musiclm,manco2023songdescriber}. Open caption corpora are built by captioning freely licensed music at scale~\parencite{roy2025jamendomaxcaps} or by rewriting expert album reviews~\parencite{alonsojimenez2026allmusiccaps}. Second, quality: open corpora are noisy, and systems increasingly account for recording quality during training, either by conditioning on per-clip quality so that inference can be steered toward high-quality output~\parencite{li2025qamdt} or by filtering and rewarding with learned aesthetics predictors~\parencite{tjandra2025aesthetics}.

\begin{takeaways}
\begin{itemize}
  \item The codec constrains the generator's cost. Its token rate determines sequence length for a given audio duration, while its bitrate and reconstruction quality affect the detail the system can preserve. Semantic content and graceful degradation are the two other properties that later tokenizers compete on.
  \item Continuous latents trade compression against decoder losses in high frequencies, phase, and stereo image, and the newest autoencoders are trained to be compact, accurate in reconstruction, and semantically structured at once.
  \item Text reaches a generator through a text encoder or a contrastive audio-text embedding. When the same embedding also serves as the prompt-adherence score, the system and its metric can share a blind spot.
  \item Training corpora are mostly undisclosed, captions are scarce and often synthetic, and the datasets that are documented under-represent most of the world's music.
\end{itemize}
\end{takeaways}

\chapter{Objectives, Sampling, and Metrics}
\label{sec:notation}

This section gives a common description of music audio generation and fixes the notation. It defines the training objectives of the main generative and preference methods, explains sampling and acceleration, and introduces the standard evaluation metrics. Section~\ref{sec:taxonomy} then uses these definitions to compare systems. Throughout, sampling means producing an output from a learned probability model. An audio sample is one numerical value of a waveform, while musical sampling means reusing a recorded excerpt. The statements are the standard ones from the cited method papers, presented in a shared notation and connected to worked examples so that readers can interpret the system comparisons of Sections~\ref{sec:instrumental} to~\ref{sec:realtime}. A reader who prefers the example to the formula can start from the worked examples, which come before the formal statement wherever the example can stand on its own.

\section{The Generic Problem}
\label{sec:generic}

Every system in this monograph generates music audio by filling four slots.
\begin{enumerate}
  \item \emph{A representation.} A map from a waveform $x$ to the object the model generates, which is a sequence of codec tokens, a continuous latent, a symbolic score, or a sketch, together with a decoder that maps the object back to audio.
  \item \emph{A factorization} of the distribution of that object given the condition $c$: one token at a time, many tokens in parallel, a chain of denoising steps, a vector field integrated from noise to data, or a plan followed by a rendering.
  \item \emph{A route for the condition.} The condition enters as a global embedding attended to at every step, as a sequence aligned in time with the representation, as reference audio that the model continues or inverts, or through the plan.
  \item \emph{A sampler}, with its number of network evaluations, and whatever reduces that number: fewer steps, a distilled student, a cache over blocks.
\end{enumerate}
Section~\ref{sec:taxonomy} uses the first three of these decisions as its comparison questions, records the sampler's consequences in the latency cells of the tables, and adds the application and the release conditions as the context in which a system's results are read. To read a new system report, identify the choices made in each of these four slots. Most of the trade-offs in this monograph are trade-offs between slots. A lower token rate in slot 1 shortens the sequence, which reduces decoding cost in slot 4, usually at a cost in reconstruction quality. A plan in slot 2 can make structure explicit. It also introduces a second model and a possible mismatch between the plans seen in training and the predicted plans seen at inference. A time-aligned condition in slot 3 gives more direct control, but supporting a new control often requires additional training. Figure~\ref{fig:schematics} draws slot 2 for the five paradigms, and Table~\ref{tab:notation} lists the symbols used below.

\begin{table}[H]
  \caption{Notation. The same symbols are used in every section. Where a cited paper uses other symbols, a notation remark in the text gives the correspondence.}
  \label{tab:notation}
  \footnotesize
  \begin{tabular}{@{}p{2.2cm}p{8.8cm}@{}}
    \toprule
    Symbol & Meaning \\
    \midrule
    $x$, $T$, $f_{\mathrm{s}}$ & waveform, its number of samples, the sample rate \\
    $c$ & the condition: text, lyrics, audio, symbolic signals, or video \\
    $E$, $D$ & encoder and decoder of a codec or autoencoder \\
    $e_n$, $N$, $f_{\mathrm{fr}}$ & encoder frames, their number, the frame rate \\
    $Q$, $K$, $\mathcal{C}^{(q)}$ & number of codebooks, codebook size, the $q$-th codebook \\
    $z^{(q)}_n$, $u_i$ & token of stream $q$ at frame $n$; the flattened token sequence \\
    $s_{1:M}$ & a semantic token stream \\
    $h$, $L$ & a continuous latent and its length \\
    $\pi$ & a plan: semantic tokens, a sketch, a melody, or a score \\
    $t$, $\alpha_t$, $\sigma_t$, $\varepsilon$ & time, noise schedule, Gaussian noise \\
    $\varepsilon_\theta$, $v_\theta$, $f_\theta$ & noise predictor, velocity field, consistency function \\
    $w$ & classifier-free guidance weight \\
    $S$ & number of sampling or unmasking steps \\
    $r(x, c)$, $\beta$, $p_{\mathrm{ref}}$ & reward, its trade-off against the reference model, the reference model \\
    $\mu$, $\Sigma$, $k$ & embedding mean and covariance, a kernel \\
    $\phi$, $\tau$ & future visibility and chunk duration of a streaming model \\
    \bottomrule
  \end{tabular}
\end{table}

\section{Representations}

\paragraph{Example first.} MusicGen's codec is an EnCodec model for 32\,kHz monophonic audio~\parencite{copet2023musicgen}. It cuts the waveform into frames with a stride of 640 samples, so it produces $32{,}000 / 640 = 50$ frames per second, and it describes each frame with four tokens, each chosen from a table of 2,048 entries. The language model behind MusicGen must therefore produce $50 \cdot 4 = 200$ tokens for every second of music, or $6{,}000$ tokens for a 30-second clip, and the codec's bitrate is $50 \cdot 4 \cdot \log_2 2048 = 2{,}200$ bits per second. These numbers help explain the duration and runtime trade-offs in the tables. Token rate determines sequence length, while bitrate and codec design constrain reconstruction quality. For a fixed context limit, a lower token rate permits longer outputs and can reduce generation cost, although it may reduce reconstruction quality. The tokenizer research of Section~\ref{sec:building-blocks} works on this trade-off. The rest of this subsection states the general form of what the example computed.

We write $x \in \mathbb{R}^{T}$ for a waveform of $T$ samples and $c$ for the conditioning input (text, lyrics, audio, symbolic signals, or video). Most systems generate over a learned compression of $x$ produced by an encoder $E$ and inverted by a decoder $D$. The waveform-domain cascades and the symbolic systems are the exceptions marked as such in the tables.

A \emph{neural codec} with residual vector quantization (RVQ)~\parencite{zeghidour2022soundstream,defossez2023encodec} maps $x$ to $Q$ parallel token streams. The encoder produces frames $e_1, \dots, e_N \in \mathbb{R}^{d}$ at a frame rate $f_{\mathrm{fr}}$, so that $N = f_{\mathrm{fr}} \cdot T / f_{\mathrm{s}}$ for a sample rate $f_{\mathrm{s}}$, and each frame is quantized in $Q$ stages against codebooks $\mathcal{C}^{(1)}, \dots, \mathcal{C}^{(Q)}$ of size $K$:
\begin{equation}
r^{(0)}_n = e_n, \qquad z^{(q)}_n = \arg\min_{k} \big\lVert r^{(q-1)}_n - \mathcal{C}^{(q)}_k \big\rVert_2, \qquad r^{(q)}_n = r^{(q-1)}_n - \mathcal{C}^{(q)}_{z^{(q)}_n}.
\end{equation}
The codec is trained with a reconstruction loss on $D(z)$ against $x$ (spectral and adversarial terms in the cited codecs), a commitment loss that pulls each encoder residual toward the codebook entry chosen for it, and a codebook update that moves the entries toward the residuals assigned to them. The quantizer itself is not differentiable, and the gradient is passed straight through it to the encoder. The first stream carries most of the energy of the frame and the later streams refine the residual, which is why a truncated stack still decodes and why a language model can predict the streams in order of importance. In musical terms, the stack is a sketch followed by finer and finer corrections, and dropping the last corrections leaves a coarser but recognizable rendering. The bitrate is $f_{\mathrm{fr}} \cdot Q \cdot \log_2 K$ bits per second, and the token rate seen by a generator is $f_{\mathrm{fr}} \cdot Q$ tokens per second. Both figures appear in the codec discussion of Section~\ref{sec:building-blocks}. A \emph{semantic tokenizer} is trained to produce a single stream $s_{1:M}$ that keeps structure (pitch, lyrics, timing) and leaves the remaining acoustic detail to a renderer. What it keeps is a property of its supervision. The split between semantic and acoustic tokens is the one introduced by AudioLM~\parencite{borsos2023audiolm}. A \emph{continuous autoencoder} maps $x$ to a latent $h \in \mathbb{R}^{L \times d}$ with $L \ll T$, trained with a reconstruction loss and, in the variational case, a Kullback--Leibler penalty that keeps the latent close to a Gaussian prior. These three outputs, with symbolic tokens as a fourth, are the classes of the representation dimension D1. In the example, $f_{\mathrm{fr}} = 50$, $Q = 4$, and $K = 2048$.

\paragraph{Notation remark.} The codec papers write the number of codebooks as $N_q$ (SoundStream) or $K$ (EnCodec and MusicGen), and MusicGen's paper calls the parallel streams ``codebook streams''. Our $Q$ is their number of codebooks and our $K$ their codebook size. Our $N$ counts frames, which EnCodec writes as $T / \mathrm{stride}$.

\warning{semantic and acoustic tokens are not two codecs} Both are discrete streams at a frame rate, and a paper may call either one ``tokens''. A semantic stream is built so that a language model can predict it (what it keeps depends on its training; the semantic tokenizers reviewed here keep pitch, lyrics, and timing and drop timbre and room), and it cannot be decoded to audio without a renderer. An acoustic stream is built so that a decoder can reconstruct the waveform, and it is harder to predict. A system that ``generates tokens'' may be doing either, and the tables mark the difference in the D1 column (semantic streams fall under the hybrid class when a renderer follows).

\section{Generative Paradigms}

\begin{figure}[H]
  \centering
  \includegraphics[width=\linewidth,alt={Schematic diagrams of the five generative paradigms: autoregressive token models; masked decoding; diffusion; flow matching; and planner-renderer hybrids.}]{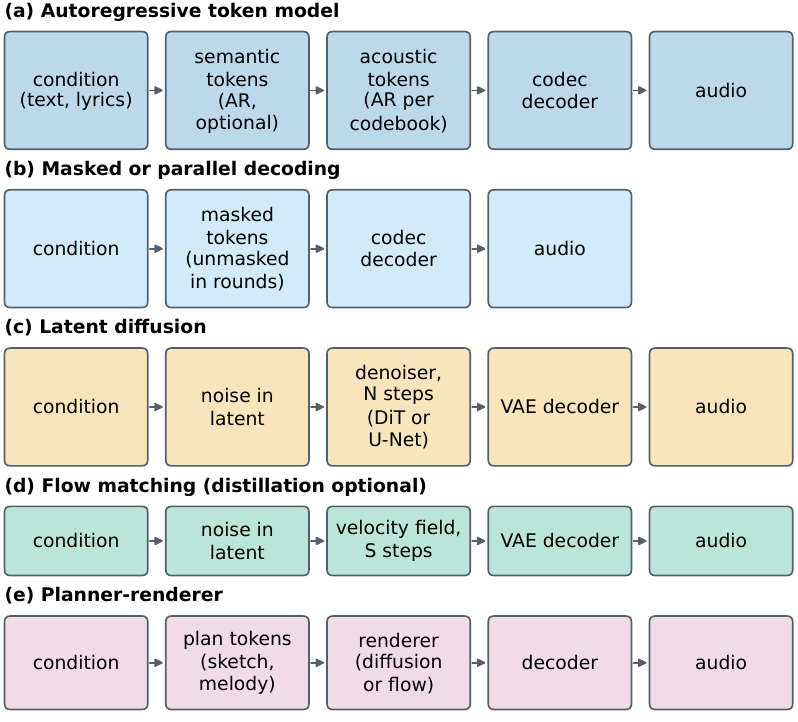}
  \caption{Data paths of the five generative paradigms (D2). Boxes are the stages a sample passes through, from the condition to the audio, and arrows give the order of computation. Notation follows Section~\ref{sec:notation}.}
  \label{fig:schematics}
\end{figure}

Every paradigm models $p(\cdot \mid c)$ over one of the representations above. Two time axes appear below and must be kept apart. The frames of a representation follow the playback time of the music. The time variable of a diffusion or flow sampler indexes the progress of sampling: each step updates a representation of the whole excerpt, so the number of frames sets the size of one update and the number of steps sets how many updates are made.

\begin{figure}[H]
  \centering
  \includegraphics[width=\linewidth]{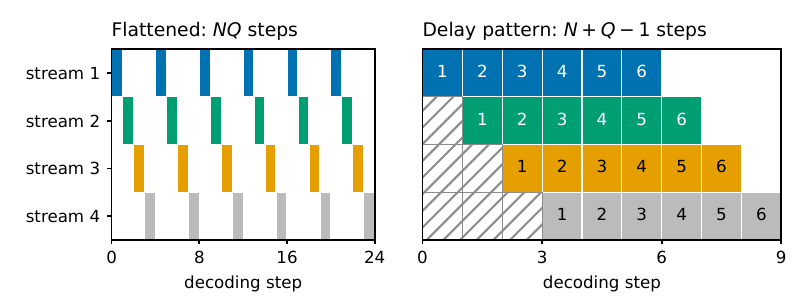}
  \caption{Two schedules for predicting four token streams of six frames with one autoregressive model. Left: the flattened schedule predicts the four tokens of each frame one after another, so a clip of $N$ frames costs $NQ$ steps (here 24). Right: the delay schedule predicts one token of every stream at each step, with stream $q$ delayed by $q-1$ frames, so the same clip costs $N + Q - 1$ steps (here 9); the hatched cells are the steps at which a delayed stream has nothing to predict yet, and the numbers are frame indices. At each step of the delay schedule the later streams see the earlier streams of the same frame, which is what makes the single pass possible at a small cost in quality.}
  \label{fig:delay}
\end{figure}

\paragraph{Example first.} MusicGen predicts its four token streams in one left-to-right pass by delaying stream $q$ by $q-1$ steps, so that at each step the model emits one token of each stream and the later streams can see the earlier ones of the same frame~\parencite{copet2023musicgen}. The pattern is a canon: each voice enters one step after the previous one and follows it. Figure~\ref{fig:delay} draws the two schedules.

 A 30-second clip at 50 frames per second therefore costs $1{,}500$ decoding steps plus three delay steps at the end, each a full transformer pass over the growing context, which is the figure its paper reports. Predicting the four streams one after another would cost $6{,}000$ steps for the same clip, and predicting a semantic stream first and the acoustic streams second, as MusicLM does, requires three stages of different lengths: semantic, coarse acoustic, and fine acoustic. The choice of pattern is therefore a choice of sequential step count. Wall time also depends on the cost of each step, which grows with the context even with key-value caching, and on model size, hardware, and codec decoding. No autoregressive token model in Table~\ref{tab:t1} carries the faster-than-real-time code. The sequential step count is one reason. Another is that the code records only what each paper measured, and these papers report no such measurement. The general form follows.

\paragraph{Autoregressive token language models.} An interleaving pattern assigns every token $z^{(q)}_n$ of the $Q \times N$ grid to one of $S$ decoding steps, and the model factorizes the joint distribution over steps, $p_\theta(z \mid c) = \prod_{s=1}^{S} \prod_{(q, n) \in P_s} p_\theta\big(z^{(q)}_n \mid z_{<s}, c\big)$, where $P_s$ is the set of tokens predicted at step $s$ and $z_{<s}$ are the tokens of the earlier steps. The tokens within one step are predicted in parallel from the same context. The model is trained by maximum likelihood, $\mathcal{L}_{\mathrm{AR}} = -\sum_s \sum_{(q,n) \in P_s} \log p_\theta(z^{(q)}_n \mid z_{<s}, c)$. When every step holds one token the factorization is the usual chain rule over a flattened sequence $u_{1:NQ}$ and $S = NQ$. The interleaving pattern decides which streams are predicted at each step: a hierarchy predicts the semantic stream first and the acoustic streams second, and the delay pattern of MusicGen offsets stream $q$ by $q-1$ steps so that all $Q$ streams are predicted in one pass with a controlled loss of quality~\parencite{copet2023musicgen}. Sampling takes one forward pass per step, so the number of sequential steps grows with the token rate and the output length. Wall time also depends on model size, hardware, and the cost of each pass. Key-value caching avoids recomputing earlier states, but each new query still reads a cache that grows with the context. Under the delay pattern $P_s$ holds one token of each stream that has started, so a clip of $N$ frames costs $S = N + Q - 1$ steps rather than $NQ$, which is the arithmetic of the example.

\paragraph{Masked and parallel decoding.} A masked model predicts a masked subset $z_{\mathcal{M}}$ from the visible tokens, $p_\theta(z_{\mathcal{M}} \mid z_{\bar{\mathcal{M}}}, c)$, trained with the cross-entropy on the masked positions for masks drawn from a schedule. Generation starts from a fully masked sequence and unmasks in $S$ iterations, keeping at each iteration the predictions of highest confidence and re-masking the rest, so that $S \ll N Q$ forward passes suffice~\parencite{chang2022maskgit,garcia2023vampnet}. Inpainting and outpainting need no new model: the mask $\mathcal{M}$ is chosen to cover the region to regenerate.

\paragraph{Diffusion.} With a latent $h_0 \sim p(h \mid c)$, the forward process adds Gaussian noise, $h_t = \alpha_t h_0 + \sigma_t \varepsilon$ with $\varepsilon \sim \mathcal{N}(0, I)$ and a noise schedule $(\alpha_t, \sigma_t)$, and a network $\varepsilon_\theta$ is trained to predict the noise,
\begin{equation}
\mathcal{L}_{\mathrm{DM}} = \mathbb{E}_{t, h_0, \varepsilon}\, \big\lVert \varepsilon - \varepsilon_\theta(h_t, t, c) \big\rVert_2^2,
\end{equation}
which is the denoising objective of DDPM~\parencite{ho2020ddpm}. The score-based view learns the score $\nabla_{h_t} \log p_t(h_t \mid c)$ (the unconditional case drops $c$), which equals $-\mathbb{E}[\varepsilon \mid h_t, c] / \sigma_t$ for this forward process, so the two parameterizations carry the same information. It samples by integrating a reverse-time stochastic or ordinary differential equation~\parencite{song2021score}. The EDM parameterization gives the preconditioning and schedule choices that several current audio latent diffusion models use~\parencite{karras2022edm}. Sampling integrates the reverse process over many steps, and each step is a full network evaluation, so the number of sequential steps of unaccelerated sampling is set by the schedule and does not grow with output length, although the computation and memory of each step grow with the latent sequence length, quadratically in the attention of a diffusion transformer, whereas autoregressive decoding takes one sequential step per token. Which of the two is slower for a given output depends on the step count, the token rate, the model size, and the hardware. The one comparison that holds data and architecture fixed found no dominant paradigm~\parencite{tal2025auto}. Conditioning is applied with classifier-free guidance~\parencite{ho2022cfg}: the network is trained with the condition dropped at random, and at sampling time the prediction is extrapolated, $\tilde{\varepsilon} = \varepsilon_\theta(h_t, t, \varnothing) + w\,(\varepsilon_\theta(h_t, t, c) - \varepsilon_\theta(h_t, t, \varnothing))$ with a guidance weight $w > 1$, which trades diversity for adherence to $c$.

\paragraph{Notation remark.} The three diffusion papers parameterize the same process differently. DDPM writes the forward process with a variance schedule $\beta_t$ and the cumulative product $\bar{\alpha}_t$, so that our $\alpha_t = \sqrt{\bar{\alpha}_t}$ and $\sigma_t = \sqrt{1 - \bar{\alpha}_t}$, with time running over integer steps $1$ to $T$. The score-based formulation uses continuous time and a score network $s_\theta$, related to ours by $s_\theta(h_t, t) = -\varepsilon_\theta(h_t, t) / \sigma_t$. EDM sets $\alpha_t = 1$ and indexes the process by the noise level $\sigma$ alone, a convention that several audio latent diffusion implementations follow. A paper's ``number of steps'' refers to the discretization of this continuous process at sampling time, which training leaves open.

\paragraph{Flow matching.} Flow matching replaces the noise-prediction objective by regression onto a velocity field along a prescribed path from noise to data. With the linear path $y_t = (1-t)\,\varepsilon + t\, h_0$ for $t \in [0, 1]$, which runs from the noise $y_0 = \varepsilon$ to the clean latent $y_1 = h_0$, the target velocity is $h_0 - \varepsilon$ and the objective is
\begin{equation}
\mathcal{L}_{\mathrm{FM}} = \mathbb{E}_{t, h_0, \varepsilon}\, \big\lVert v_\theta(y_t, t, c) - (h_0 - \varepsilon) \big\rVert_2^2,
\end{equation}
after which samples are drawn by integrating $\mathrm{d}y_t / \mathrm{d}t = v_\theta(y_t, t, c)$ from $t = 0$ to $1$~\parencite{lipman2023flow,liu2023rectified}. The path is a choice: flow matching admits a family of interpolations between noise and data that includes the diffusion path, and the linear interpolation above is the conditional optimal-transport path~\parencite{lipman2023flow}. A straight path between each sampled noise and data pair does not make the learned marginal field straight, so the number of solver steps a model needs depends on the learned geometry and on the solver. Reflow, which retrains on the model's own noise-to-sample pairs, straightens the learned paths and is one optional procedure for reducing the step count~\parencite{liu2023rectified}. Diffusion and flow matching are the same family in the sense that both learn a time-dependent vector field over the latent. They differ in the path and the parameterization, which is why the tables distinguish them as paradigms but Section~\ref{sec:instrumental} treats their systems together.

\paragraph{Notation remark.} The subscript $0$ marks the clean latent $h_0$ throughout this monograph, and the point on a path at time $t$ is written $h_t$ for diffusion and $y_t$ for flow matching, because the two run in opposite directions: diffusion's $h_t$ goes from data at $t = 0$ to noise, flow matching's $y_t$ from noise at $t = 0$ to data at $t = 1$. The flow-matching paper writes the data point $x_1$ and the vector field $u_t$. Rectified flow writes the endpoints $X_0$ (noise) and $X_1$ (data) and the velocity $X_1 - X_0$. These conventions correspond to the target velocity $h_0 - \varepsilon$ along the linear path used here.

\warning{flow matching, rectified flow, and consistency models are not one method} Flow matching is a training objective (regression onto a velocity along a chosen path). Rectified flow is that objective with the linear path. Reflow is an optional later procedure that retrains on the model's own noise-to-sample pairs to straighten the learned paths. Consistency models are a distillation or training method that maps any point on a trajectory to its endpoint in one step, and they can be built on a diffusion or a flow model. A system reporting ``few-step'' generation should therefore state both its training objective and any acceleration procedure. The post-training column of the tables records the procedures that a paper reports.

\paragraph{Planner-renderer hybrids.} A hybrid factorizes $p(x \mid c) = \sum_{\pi} p(\pi \mid c)\, p(x \mid \pi, c)$, where the plan $\pi$ is a semantic token sequence, a sketch, a melody, or a human-readable layout produced by an autoregressive language model, and the renderer is a diffusion or flow model over a continuous latent. The two stages are trained separately with the objectives above. At inference the plan is sampled first and the renderer is conditioned on it. The factorization moves long-range decisions, such as structure and lyrics, into the discrete stage, where an autoregressive model handles long contexts, and moves acoustic detail into the continuous stage, where parallel sampling is cheap.

\paragraph{How the condition enters.} Three mechanisms carry $c$ into the networks above, and they are the choices of slot 3. A global condition is encoded once, by a text encoder or a contrastive audio-text model, into one or a few vectors that every layer attends to through cross-attention, or that are added to the time embedding of a diffusion network. A time-varying condition (chords, beats, a melody curve, a reference stem) is encoded into a sequence aligned with the frames of the representation. It enters as extra input channels, as additional tokens in the context of a token model, or through a trained branch. A ControlNet-style branch copies part of the network, receives the control, and adds its outputs to the frozen main network. An adapter is a small module inserted into a frozen network and trained alone, so that the cost of a new control is the adapter's training run. Reference audio enters as the model's own representation of it, as a prefix the token model continues or as a latent the diffusion model is conditioned on.

\paragraph{Inference-time control and inversion.} Two further routes change what a trained diffusion or flow model produces without training it further. Inference-time optimization treats the initial noise $\varepsilon$ as the free variable: for a feature extractor $g$ and a target $y^\star$ (an intensity curve, a melody, a structure), it solves $\min_{\varepsilon} \lVert g(\mathrm{Sample}_\theta(\varepsilon, c)) - y^\star \rVert^2$ by gradient descent through the sampler, which is the DITTO formulation of Section~\ref{sec:control}. Each gradient step costs a sampling run and its backward pass, with the memory that entails, and a distilled sampler made the method cheap enough for interactive use in the cited work. Inversion runs the sampler backwards: the deterministic (ordinary-differential-equation) sampler maps noise to a latent, so integrating it in the reverse direction from a given recording's latent recovers a noise $\hat{\varepsilon}$ that regenerates the recording, and an edit is made by regenerating from $\hat{\varepsilon}$ under a changed condition or with the attention of the network steered. Errors accumulate along the reverse integration, which is the failure the editing methods of Section~\ref{sec:control} address, and inversion-free methods avoid the reverse pass by constructing the edit from the difference between the source-conditioned and target-conditioned predictions.

\section{Sampling Acceleration}
\label{sec:acceleration}

Sampling from a diffusion or flow model takes many network evaluations. Several acceleration methods recur in Sections~\ref{sec:instrumental} to~\ref{sec:realtime}. \emph{Step distillation} trains a student to reproduce in one step what the teacher reaches in many. The consistency formulation~\parencite{song2023consistency} trains $f_\theta(h_t, t)$ to map any point on a sampling trajectory to its origin, with the consistency loss $\mathbb{E}\, d\big(f_\theta(h_{t_{n+1}}, t_{n+1}),\, f_{\theta^-}(\hat{h}_{t_n}, t_n)\big)$ between adjacent points, where $\hat{h}_{t_n}$ is obtained from $h_{t_{n+1}}$ by one solver step and $\theta^-$ is a slowly updated copy of $\theta$. The parameterization fixes $f_\theta(h_{t_{\min}}, t_{\min}) = h_{t_{\min}}$ at the smallest noise level $t_{\min}$, which rules out the constant function that the agreement loss alone would admit, and the gradient is stopped through the target branch. When the solver step uses a pretrained teacher the procedure is consistency distillation, and when it is estimated from the data alone it is consistency training~\parencite{song2023consistency}. \emph{Adversarial post-training} adds a discriminator loss on few-step samples so that the student's outputs match the data distribution rather than the teacher's trajectory. \emph{Layer distillation} removes layers from the network and fine-tunes the rest. The tables record the resulting runtime class and whether a streaming procedure exists, because a fast batch sampler is not a streaming one: a streaming system must emit audio for the first chunk before the condition for later chunks is known. Consider a model that computes each output chunk of duration $\tau$ in one pass, after it has received the input it may condition on, and plays the chunks in order. Its future visibility $\phi$ is the offset between the latest input a chunk may condition on and the start of the output interval the chunk accompanies~\parencite{wu2025streaming}. Under this schedule the output stream lags the input it accompanies by at least $\phi$ plus the computation time of a chunk, and a new input is heard only when the next chunk plays, up to $\tau$ later. Computing each chunk in less than $\tau$ keeps the stream continuous. That is a throughput condition, separate from the delay. Section~\ref{sec:realtime} reports the trade-off.

\warning{real-time factor, latency, and future visibility measure three different things} The real-time factor is the generation time divided by the duration of the audio generated. A value below one says that a system produces audio faster than the audio plays, in a batch, once the whole condition is known. It says nothing about how soon the first sound arrives. Latency is the time from an input to the audio that responds to it. For a chunked streaming model it has two parts: the output stream lags the input it accompanies by at least the future visibility $\phi$ plus the computation time of a chunk, and a new input is heard only at the next chunk boundary, up to the chunk duration $\tau$ later. Future visibility is a design choice: a positive $\phi$ gives the model input look-ahead at the price of a later stream, and a negative $\phi$ makes it predict ahead to absorb system delay~\parencite{wu2025streaming}. The tables record the real-time factor where a paper reports one and mark streaming separately, because live use requires output while the input is still arriving, a separate property from batch speed.

\section{Preference Alignment}

Preference alignment is a post-training stage that can be applied to a model of any paradigm, so we state it once. Given a reward $r(x, c)$ from human ratings or a learned judge, reinforcement learning post-trains a generator $p_\theta$ initialized from a reference $p_{\mathrm{ref}}$ by maximizing
\begin{equation}
\mathbb{E}_{c}\Big[\mathbb{E}_{x \sim p_\theta(\cdot \mid c)}\big[r(x, c)\big] - \beta\, \mathrm{KL}\big(p_\theta(\cdot \mid c)\,\|\,p_{\mathrm{ref}}(\cdot \mid c)\big)\Big],
\end{equation}
where $\beta$ controls how far the policy may move from the reference. MusicRL applied this to a token language model with policy-gradient optimization~\parencite{cideron2024musicrl,schulman2017ppo}. Direct preference optimization (DPO)~\parencite{rafailov2023dpo} removes the explicit reward: for a pair $(x^+, x^-)$ preferred and dispreferred under $c$, the loss is
\begin{equation}
\mathcal{L}_{\mathrm{DPO}} = -\log \sigma\!\left(\beta \log \frac{p_\theta(x^+ \mid c)}{p_{\mathrm{ref}}(x^+ \mid c)} - \beta \log \frac{p_\theta(x^- \mid c)}{p_{\mathrm{ref}}(x^- \mid c)}\right),
\end{equation}
which diffusion models cannot evaluate directly, because their likelihoods are not available in closed form. Diffusion-DPO derives a surrogate from the evidence lower bound of the diffusion likelihood, which reduces to a difference of denoising losses on the preferred and the dispreferred sample~\parencite{wallace2023diffusion}, and text-to-audio alignment uses that construction~\parencite{majumder2024tango2}. Preference methods for flow models use constructions of their own. The post-training column of the tables records which of these stages a system reports, and the evaluation section returns to the circularity that arises when the judge that supplies $r$ is also the metric.

\paragraph{Notation remark.} The $\beta$ of the reinforcement objective and the $\beta$ of the DPO loss play the same role, the weight of the pull toward the reference model, and DPO uses the analytic relationship between a reward and the optimal KL-regularized policy to write the pairwise preference likelihood in terms of policy ratios, which gives a training objective without an explicit reward model. The reference $p_{\mathrm{ref}}$ is the model before alignment, in most music systems the supervised model of the previous training stage. Papers write it $\pi_{\mathrm{ref}}$ when they borrow the language-model notation in which the generator is a policy $\pi$, a symbol this monograph reserves for the plan.

\warning{preference optimization and adversarial post-training are different stages with similar names} Both are applied after the main training run and both change what the model produces, but they answer different questions. Preference optimization moves the model toward outputs a judge or a listener prefers, and needs ratings or a reward model. Adversarial post-training moves a few-step sampler toward the data distribution, needs only data and a discriminator, and is a speed method. The tables mark them with different codes, and a system can report both, one, or neither.

\section{Distribution-Level Metrics}

The default automatic metric of the period, the Fr\'echet Audio Distance~\parencite{kilgour2019fad}, embeds generated and reference clips with a fixed audio encoder, fits a Gaussian to each set, and reports the Fr\'echet distance between the two Gaussians,
\begin{equation}
\mathrm{FAD} = \lVert \mu_g - \mu_r \rVert_2^2 + \mathrm{tr}\big(\Sigma_g + \Sigma_r - 2 (\Sigma_g \Sigma_r)^{1/2}\big),
\end{equation}
so it depends on the encoder, assumes Gaussian embeddings, and its estimate from a small set is biased, generally upward, and noisy. The sample covariances themselves are unbiased, but the distance is a nonlinear function of the estimated means and covariances, and that nonlinearity introduces the bias. The kernel alternative uses the maximum mean discrepancy: for a kernel $k$ and the generated and reference embedding distributions $P_g$ and $P_r$, $\mathrm{MMD}^2(P_g, P_r) = \mathbb{E}\,k(g, g') + \mathbb{E}\,k(r, r') - 2\,\mathbb{E}\,k(g, r)$, and its unbiased estimator replaces each expectation by a U-statistic, an average over distinct pairs that is unbiased for the expectation, over the two sets~\parencite{gretton2012kernel}. On a finite sample this unbiased estimate can be negative. The Kernel Audio Distance (KAD) is this estimator over audio embeddings~\parencite{chung2025kad}, unbiased in expectation, without the Gaussian assumption, and still subject to sampling variance. Section~\ref{sec:evaluation} reviews the evidence on how these distances correlate with human judgments.

\paragraph{Interpreting a distribution distance.} Two consequences of the definitions decide how a comparison should be read. First, both distances are computed in an embedding space, so two papers that use different encoders, or the same encoder on different reference sets, report numbers that cannot be compared. The tables of Section~\ref{sec:evaluation} therefore name the encoder wherever a paper does. Second, the Fr\'echet distance is a nonlinear function of the means and covariances estimated from the samples, so with few generated clips the estimate is biased, generally upward, and a system evaluated on a small set can be penalized for the set size rather than its quality. The unbiased MMD estimator removes that bias but not the variance, so a small evaluation set still gives a noisy number. A reader who sees a distribution distance reported without the encoder, the reference set, and the number of clips has three questions to ask before believing a ranking.

\begin{takeaways}
\begin{itemize}
  \item Every system fills four slots: a representation, a factorization, a route for the condition, and a sampler. The comparison framework's first three dimensions correspond to the representation, generative method, and conditioning route. The fourth and fifth dimensions, the application class and the artifact attributes, describe use and release.
  \item The five paradigms differ in how the number of sequential steps scales: with the token count for autoregressive models, with the sampler's step count for diffusion and flow models, and with both for a planner-renderer pair. Wall time also depends on the cost of each step and on the hardware.
  \item A distribution distance depends on its encoder, its reference set, and its sample size, and the plug-in estimate of the Fr\'echet distance is biased at finite sample sizes.
\end{itemize}
\end{takeaways}

\chapter{Comparing Music Generation Systems}
\label{sec:taxonomy}

Two music generators can accept the same prompt and still differ in what they generate internally, how they produce it, and what control they give the user. Their published results may also describe different tasks, hardware, or release conditions. A comparison has to keep these differences visible. We compare systems through five questions: what representation is generated (D1), how it is generated (D2), what conditions guide it (D3), which musical task is evaluated (D4), and what is documented about the released system (D5). The first three describe the model's design. The last two establish the context in which its capabilities and reported results should be read. The questions are complementary, and the answers interact: a planner-renderer system, for example, uses different representations and generative methods at different stages. Sections~\ref{sec:building-blocks} and~\ref{sec:notation} explained the mechanisms. This section defines the comparison categories, resolves the cases that could fit more than one category, and explains the tables used in the application sections. The codes D1 to D5 and T1 to T6 are reference labels for the tables, and the discussion uses the corresponding terms. Figure~\ref{fig:paradigm-tier} shows which method-by-class combinations the reviewed systems occupy.

For MusicGen, the generated representation is a sequence of discrete codec tokens, the generative method is autoregressive, and the main condition is a text description, with melody conditioning in one variant. Its class in this review is instrumental text-to-music. Its duration, output format, runtime evidence, and release terms are recorded separately, because none of them follows from the words ``autoregressive model''. Table~\ref{tab:t1} holds the resulting row.

Practitioners and product descriptions more often group systems by task, as an input-to-output pair: text-prompt instrumental generation, lyrics-to-song, audio-to-audio generation (accompaniment, continuation, and style transfer from an audio input), music editing, and video-to-music. Table~\ref{tab:functional} therefore provides a task-based entry point to the review. It maps each functional category to the conditioning class (D3) and application class (D4) that define it and to the section that reviews it. The remaining dimensions describe how a system in that category is built (D1, D2) and what it is as a released artifact (D5).

\begin{table}[H]
  \caption{Functional (input-to-output) categories of music generation systems and their place in the taxonomy.}
  \label{tab:functional}
  \footnotesize
  \begin{tabular}{@{}p{2.4cm}p{2.2cm}p{0.75cm}p{5.1cm}@{}}
    \toprule
    Functional category & Input (D3) & Class (D4) & Examples and section \\
    \midrule
    Text-prompt instrumental & global text & T1 & MusicGen~\parencite{copet2023musicgen}, Stable Audio~\parencite{evans2024fast}; Section~\ref{sec:instrumental} \\
    Lyrics-to-song & lyrics, structure tags, text & T2 & YuE~\parencite{yuan2025yue}, DiffRhythm~\parencite{ning2025diffrhythm}; Section~\ref{sec:song} \\
    Audio-to-audio & reference or live audio & T2, T3, T4 & SingSong~\parencite{donahue2023singsong} (vocals to accompaniment, tabulated with the accompaniment systems of T3), LiveBand~\parencite{pasini2026liveband}; Sections~\ref{sec:song}, \ref{sec:control}, \ref{sec:realtime} \\
    Music editing & audio and an instruction & T3 & InstructME~\parencite{han2023instructme}, MusicMagus~\parencite{zhang2024musicmagus}; Section~\ref{sec:control} \\
    Video-to-music & video or image & T5 & VidMuse~\parencite{tian2024vidmuse}, V2M-Zero~\parencite{lin2026vm}; Section~\ref{sec:crossmodal-symbolic} \\
    Understanding and generation & dialogue and audio & T6 & M2UGen~\parencite{hussain2023m2ugen}, MusicAgent~\parencite{yu2023musicagent}; Section~\ref{sec:crossmodal-symbolic} \\
    \bottomrule
  \end{tabular}
\end{table}

\section{Audio Representations}
\label{sec:d1}

Models generate over one of four classes of representation. \emph{Symbolic} representations (MIDI, ABC) separate composition from rendering~\parencite{lu2023musecoco,wang2025notagen}. \emph{Discrete acoustic} representations are codec token sequences, typically from RVQ codecs~\parencite{zeghidour2022soundstream,defossez2023encodec,kumar2023dac}, often arranged in a semantic-then-acoustic hierarchy~\parencite{borsos2023audiolm,agostinelli2023musiclm}. The cost of stacked codebooks has motivated single-codebook, two-dimensional, semantically supervised, and multi-scale tokenizers~\parencite{ji2025wavtokenizer,cheng2026bandtok,ye2024codec,siuzdak2024snac}. \emph{Continuous acoustic} representations are VAE latents, mel spectrograms, or waveforms, consumed by diffusion and flow models~\parencite{liu2023audioldm,evans2024fast}. Recent autoencoders are designed to be simultaneously compact, accurate in reconstruction, and semantically structured~\parencite{evans2026stableaudio3}. \emph{Hybrid or sketch} representations insert an intermediate plan or learned sketch between the condition and the audio, which symbolic layouts and programs can make interpretable: a trajectory of audio-text embedding tokens~\parencite{lam2025musicot}, a coarse semantic sketch~\parencite{yang2025songbloom}, a melody token sequence~\parencite{qwen2026music}, or a human-readable structural layout or program~\parencite{li2026musiclayout,wang2025musicweaver}.

\section{Generative Methods}
\label{sec:d2}

The generative method is the factorization of the distribution over the representation. Five principal paradigms account for most of the literature (Figure~\ref{fig:schematics}). The paradigm code of a system follows the training objective its paper states: a diffusion path with a noise or v-prediction target is coded as diffusion, and the linear noise-to-data path with the target $h_0 - \varepsilon$ as flow matching, whichever word the paper uses for its family. Both targets are called velocities in some papers, although they differ. The post-training stages and the order of generation are recorded separately, in the post-training, latency, and streaming cells. The tables add four smaller groups (inference-time optimization, language-model agents, adversarial continuous generators, and autoencoder instruments), each described where it occurs. The paradigms differ in how they factor the distribution over a piece of music, and the factorization constrains what a model can do cheaply. \emph{Autoregressive token language models} predict successive tokens or groups of tokens from earlier predictions, which makes continuation and streaming natural and makes cost grow with the token count~\parencite{agostinelli2023musiclm,copet2023musicgen,yang2024uniaudio,yuan2025yue}. \emph{Masked or parallel decoding} predicts many tokens at once and refines the result over a few rounds, trading some quality for large latency reductions~\parencite{garcia2023vampnet,ziv2024magnet,parker2024stemgen}. \emph{Latent and cascaded diffusion} starts from noise and removes it in steps over a continuous representation, so that every step sees the whole piece, which suits editing and inference-time control~\parencite{huang2023noise2music,liu2023audioldm,evans2024longform,li2025qamdt}. \emph{Flow matching and rectified flow} regress a velocity field along a chosen path from noise to data, and the optimal-transport and rectified-flow variants used in music aim at straight paths, so that fewer solver steps suffice~\parencite{prajwal2024musicflow,tal2024jasco,fei2024fluxmusic}, and combine with consistency training, step and layer distillation, and adversarial post-training to reach faster-than-real-time generation~\parencite{liu2024audiolcm,novack2025presto,evans2026stableaudio3}. Block-wise flow matching keeps the objective and generates a song block by block, each block conditioned on the ones before it~\parencite{jiang2025diffrhythm2}. \emph{Planner-renderer hybrids} split the task: a language model produces a plan, a sketch, a melody, or a score, and a diffusion or flow model renders audio from it~\parencite{lam2023efficient,yang2025songbloom,bai2024seedmusic,zhang2025inspiremusic,qwen2026music,dai2026pushing}.

Orthogonal to all five, \emph{preference alignment} is an increasingly common post-training stage, from reinforcement learning on human feedback~\parencite{cideron2024musicrl} to direct preference optimization adapted to the denoising objective~\parencite{wallace2023diffusion,majumder2024tango2}, distributional reward optimization~\parencite{bai2025dragon}, multi-axis preference alignment~\parencite{lei2025levo}, multi-reward optimization for flow matching~\parencite{ziv2025mr}, cross-pair schemes that avoid model merging~\parencite{jiang2025diffrhythm2}, and diversity-preserving distillation of guidance~\parencite{cideron2024diversity}. A survey of alignment methods for generative music summarizes the options and their open problems~\parencite{herremans2025aligning}. A controlled comparison of the autoregressive and flow-matching paradigms with data and architecture held fixed finds trade-offs rather than a winner~\parencite{tal2025auto}.

Several systems reviewed below combine paradigms to use their different strengths. Autoregression contributes streaming and variable-length generation. Diffusion contributes parallel refinement, editing by inpainting, and inference-time control~\parencite{novack2024ditto,evans2026stableaudio3}. Block-wise and chunked schemes now give diffusion models autoregressive interfaces, including key-value caching, which keeps the attention states of earlier chunks so they are not recomputed, for streaming inference~\parencite{novack2026lmdm}, and planner-renderer systems give diffusion renderers the long-range structure of a language-model plan~\parencite{li2026beyond}.

\section{Conditioning and Control}
\label{sec:d3}

A condition can specify a whole piece in a few words or constrain every beat, and the systems reviewed here span that range. The main distinction is how the condition enters the model: a global condition applies to the whole output and is represented by one vector or by a sequence of embeddings that every step of generation attends to, a time-varying condition is a sequence aligned with the audio, and a reference recording enters either as an embedding or as audio the model continues. \emph{Global text} is the default interface~\parencite{agostinelli2023musiclm,copet2023musicgen}. \emph{Lyrics and structure tags} specify full songs~\parencite{yuan2025yue,ning2025diffrhythm,yang2026heartmula}, and segment-level prompts allow different instructions at different points in a song~\parencite{wang2026segtune}. \emph{Reference audio} conveys melody, style, or voice, including in-context learning from short excerpts~\parencite{copet2023musicgen,yuan2025yue,tsai2024apadapter} and the melody of an existing song for cover generation~\parencite{li2026songecho}. \emph{Time-varying signals} such as chords, rhythm, dynamics, intensity, and emotion curves give bar-level and beat-level control~\parencite{wu2024musiccontrolnet,lin2023cocomulla,lan2024musicongen,tal2024jasco,melechovsky2024mustango,mei2025lara}, with adapter and conditioner designs reducing the cost of adding controls~\parencite{baker2025lilac,tsai2025musecontrollite}, inference-time optimization avoiding retraining altogether~\parencite{novack2024ditto,novack2024ditto2}, and gestural interfaces accepting hummed or sketched shapes~\parencite{garcia2025sketch2sound}. \emph{Structural plans}, human-readable layouts or programs, control form directly~\parencite{li2026musiclayout,wang2025musicweaver}. \emph{Multimodal} conditioning maps video or images to music~\parencite{tian2024vidmuse,li2024muvi,lin2026vm}. \emph{Interactive and streaming} control operates while audio is being generated, in live music models~\parencite{caillon2025live,novack2026lmdm} and causal accompaniment to live input~\parencite{pasini2026liveband,wu2025streaming}.

\section{Application Classes}
\label{sec:d4}

We distinguish six application classes, labeled T1 to T6 for brevity. The labels index the classes and do not order them: T4 is not above T1, and the numbering follows the order in which the classes are reviewed. \emph{T1, instrumental text-to-music}: clip- to track-level instrumental generation~\parencite{copet2023musicgen,evans2024longform,li2025qamdt}. \emph{T2, lyrics-to-song}: full songs with vocals~\parencite{suno2026,udio2026,yuan2025yue,gong2025acestep,lei2025levo,yang2026heartmula,lei2026levo2,qwen2026music}. \emph{T3, production co-creation}: stems, accompaniment, inpainting, editing, and variations~\parencite{parker2024stemgen,rouard2025musicgenstem,yao2025jen1composer,han2023instructme,zhang2024musicmagus,zhang2024instructmusicgen,mariani2024msdm,nistal2024diffariff,donahue2023singsong}. \emph{T4, real-time and live interaction}: streaming generation and accompaniment as an instrument~\parencite{caillon2025live,novack2026lmdm,pasini2026liveband,novack2025presto}. \emph{T5, cross-modal scoring}: music for video, images, and games~\parencite{tian2024vidmuse,su2023vmeow,ji2025comprehensive}. \emph{T6, unified understanding and generation}: systems that analyze, discuss, and generate music in one interface, including audio-language models and agents that control a generative backend~\parencite{hussain2023m2ugen,yang2024uniaudio,yu2023musicagent,liu2024mumu}.

The classes are defined by the task a system is built and evaluated for, and a system can belong to more than one. Three rules fix the placement. First, a system is tabulated once, in its primary class, which is the task of its main evaluation. Secondary memberships are stated in the text. Second, the boundaries follow the output and the condition rather than the architecture. A model that generates vocals from lyrics is T2 whatever its paradigm. A model prompted with audio rather than text is T3 even when it was built as a general generator. A model is T4 only if it emits audio under a stated latency bound while receiving input. Third, where two classes apply equally, the class with the stricter constraint wins (T4 over T3, T2 over T1), so that the table records the result under the more specific constraint. The hard cases show the rules at work. SingSong generates accompaniment for a given vocal, so its main evaluation is an audio-to-audio task and it is tabulated with the stem and accompaniment systems of T3, although Section~\ref{sec:song} keeps it as the first decomposition of the song problem. VampNet is a general masked generator prompted with audio, so it is tabulated with the editing systems of T3 rather than with T1. Shao is trained with lyrics and generates vocal songs, so it is tabulated in T2 although its design question, a deep token hierarchy with super-resolution in the same token space, belongs to T1. Presto is a distillation method whose paper evaluates offline text-to-music generation, so it is tabulated in T1 although Section~\ref{sec:realtime} cites it for the speed that live use needs. The six classes were fixed before the structured searches of Section~\ref{sec:method} were run, on the basis of the landmark systems and the functional categories of Table~\ref{tab:functional}, and the boundary rules above were added after the first screening pass exposed the cases just listed. No class was added or merged afterwards. Most new systems addressed T1 in 2023 and T2 by 2025. Figure~\ref{fig:paradigm-tier} and Tables~\ref{tab:t1} to~\ref{tab:t3} show the distribution of the reviewed systems across classes and methods.

\section{Reported Attributes and Release Conditions}
\label{sec:d5}

Two systems with the same architecture can differ as released artifacts in ways that decide whether a reader can use or reproduce them, and the tables record these differences as release attributes. Tables~\ref{tab:t1}, \ref{tab:t2}, and~\ref{tab:t3} tabulate seven of them per system for the four application classes covered by these tables (T1, T2, and T3 with T4):
\begin{itemize}
  \item reported duration, distinguishing context windows, evaluation clips, demonstrated outputs, and streaming;
  \item output format (sample rate and channels);
  \item latency class (offline, faster than real time, streaming), with the runtime of the measurement;
  \item openness of weights, code, and training data, with the license and the backbone's weights where a system builds on another;
  \item data provenance (licensed, Creative Commons, undisclosed)~\parencite{evans2025stableaudioopen,nistal2024diffariff,evans2026stableaudio3};
  \item vocals;
  \item the type of post-training stage (preference, reinforcement, adversarial, distillation).
\end{itemize}
Cells the system's paper or model card does not support are marked as unknown rather than inferred. The openness cells follow the dated artifact checks against model cards and repositories documented in Appendix~\ref{app:screening}, because openness changes after publication (Section~\ref{sec:trends}). Three further attributes are discussed where they arise and are not tabulated. Model scale is omitted because it is not reported consistently across the systems. The evaluation reported is treated by instrument in Section~\ref{sec:evaluation}. Whether a memorization measurement or a watermark accompanies the release~\parencite{sanroman2024audioseal,yuan2025yue} is documented for ten of the 105 systems, and the remaining papers were not searched for it, so the attribute is not tabulated. The tables report output sample rate and channel count, which describe format rather than perceived quality, and they do not score output quality. Section~\ref{sec:evaluation} assesses the evaluation instruments themselves.

\begin{guidelines}{Reporting a released system}
A system report or model card that lets a reader place the system in the tables states the seven attributes above. It also states the measurement hardware, the objective or feedback used in each post-training stage, model scale, and any memorization test or watermark.
\end{guidelines}

\section{Reading the Comparison Tables}
\label{sec:map}

Table~\ref{tab:map} links the common comparison questions to the sections that answer them: for each value a system can take on D1 to D5, it names the section in which the mechanism is explained and the section in which systems with that value are reviewed. Given a system report or a product description, a reader can identify the relevant design choices and follow the corresponding rows. Starting from a goal instead, the application class of D4 and the constraints of D5 lead backward to the design choices that meet them. The reading paths in Section~\ref{sec:intro} provide different routes through this table, and the closing subsection of each of Sections~\ref{sec:instrumental} to~\ref{sec:realtime}, titled Lessons and Open Trade-offs, states which design choices the evidence supports and which remain unresolved.

\begin{table}[H]
  \caption{Comparison questions and the sections that answer them. For each value of the five dimensions, the section that explains the mechanism and the section that reviews the systems that use it.}
  \label{tab:map}
  \footnotesize
  \begin{tabular}{@{}p{0.7cm}p{3.4cm}p{2.8cm}p{3.5cm}@{}}
    \toprule
    Dim. & Value & Mechanism explained in & Systems reviewed in \\
    \midrule
    D1 & Discrete codec tokens & Sections~\ref{sec:codecs} and~\ref{sec:notation} & Sections~\ref{sec:t1-ar}, \ref{sec:t1-masked}, and~\ref{sec:song} \\
    D1 & Continuous latents & Sections~\ref{sec:codecs} and~\ref{sec:notation} & Sections~\ref{sec:t1-diffusion} and~\ref{sec:song} \\
    D1 & Hybrid plan or sketch & Section~\ref{sec:d1} & Sections~\ref{sec:song-convergence} and~\ref{sec:control-timevarying} \\
    D1 & Symbolic & Section~\ref{sec:d1} & Section~\ref{sec:symbolic} \\
    D2 & Autoregressive & Section~\ref{sec:notation} & Section~\ref{sec:t1-ar} \\
    D2 & Masked and parallel & Section~\ref{sec:notation} & Section~\ref{sec:t1-masked} \\
    D2 & Diffusion and flow matching & Section~\ref{sec:notation} & Section~\ref{sec:t1-diffusion} \\
    D2 & Planner-renderer hybrid & Section~\ref{sec:notation} & Section~\ref{sec:song-convergence} \\
    D5 & Few-step and distilled sampling (post-training) & Section~\ref{sec:notation} & Sections~\ref{sec:t1-diffusion} and~\ref{sec:streaming} \\
    D5 & Preference alignment (post-training) & Section~\ref{sec:notation} & Section~\ref{sec:song-convergence}; Table~\ref{tab:t2} \\
    D3 & Global text & Section~\ref{sec:d3} & Section~\ref{sec:instrumental} \\
    D3 & Lyrics and structure tags & Section~\ref{sec:d3} & Section~\ref{sec:song} \\
    D3 & Time-varying signals & Section~\ref{sec:d3} & Section~\ref{sec:control-timevarying} \\
    D3 & Reference audio and instructions & Section~\ref{sec:d3} & Sections~\ref{sec:control-editing} and~\ref{sec:control-stems} \\
    D3 & Live audio and gesture & Section~\ref{sec:d3} & Section~\ref{sec:realtime} \\
    D3 & Video, image, dialogue & Section~\ref{sec:d3} & Section~\ref{sec:crossmodal-symbolic} \\
    D4 & T1 to T6 & Section~\ref{sec:d4} & Sections~\ref{sec:instrumental} to~\ref{sec:crossmodal-symbolic}, one per class \\
    D5 & Duration, format, latency & Section~\ref{sec:d5} & Tables~\ref{tab:t1} to~\ref{tab:t3}; Section~\ref{sec:realtime} for latency \\
    D5 & Openness and provenance & Section~\ref{sec:d5} & Tables~\ref{tab:t1} to~\ref{tab:t3}; Section~\ref{sec:data} \\
    -- & Evaluation (by instrument, not tabulated per system) & Section~\ref{sec:evaluation} & Table~\ref{tab:eval} \\
    \bottomrule
  \end{tabular}
\end{table}

\section{Relation to Earlier Surveys}
\label{sec:prior-taxonomies}

Several earlier taxonomies emphasized symbolic generation. We relate their organizing questions to the audio systems reviewed here. The five-dimension analysis of Briot, Hadjeres, and Pachet asks what is generated, how it is represented, which architecture generates it, and what challenges and strategies apply~\parencite{briot2017deep}. Our D1 and D2 correspond to its representation and architecture dimensions, and D3 to D5 replace its challenge and strategy dimensions with the axes that distinguish released systems: what a system can be told, what it is for, and what was released. The functional taxonomy of Herremans, Chuan, and Chew indexes systems by the musical dimension they generate (melody, harmony, rhythm, timbre, interaction, narrative, difficulty)~\parencite{herremans2018functional}. That index remains useful for systems that generate particular musical dimensions, and its interaction category anticipates our class T4. Full-mixture audio generators often combine melody, harmony, rhythm, and timbre in one output, and we distinguish them by input, paradigm, and artifact. The three-level organization of Ji, Luo, and Yang (score, performance, audio) places all of our systems in its audio level~\parencite{ji2020comprehensive}. The foundation-model survey of Ma et al.\ organizes by pretraining paradigm and downstream task across understanding and generation~\parencite{ma2024foundation}. Our D2 refines its generative paradigms and our D4 and D5 add the application and artifact axes.

The modality-centered survey of Li et al.\ organizes systems by guidance modality, single, cross, and multi-modal, across symbolic and audio output, and records for each representative method its modeling approach, modal encoding, fusion, and architecture~\parencite{li2025survey}. Our D3 covers modality as one of five dimensions, and Tables~\ref{tab:t1} to~\ref{tab:t3} record the released artifact (D5) rather than the fusion design. The video-to-music survey of Ji et al.\ decomposes that branch into conditioning input construction, conditioning mechanism, and generation framework, and tabulates demos, generation lengths, model sizes, and training resources for its systems~\parencite{ji2025comprehensive}. The three components map onto our D3 and D2, and its table is the branch-level counterpart of ours. The evaluation survey of Lerch et al.\ groups evaluation targets into system output and model use and reviews subjective, objective, and HCI methodologies~\parencite{lerch2025survey}. Section~\ref{sec:evaluation} points to it for methodology and covers only the instruments the reviewed systems report. The functional grouping of Table~\ref{tab:functional} is the bridge between these organizations and ours.

\begin{takeaways}
\begin{itemize}
  \item The application classes are unordered labels. A system is tabulated once, in the class of its main evaluation, and the boundary rules of Section~\ref{sec:d4} decide the hard cases.
  \item Artifact attributes are taken from the system's own paper or model card, and the openness cells follow the dated artifact checks documented in Appendix~\ref{app:screening}. Cells without a source are marked as unknown. Section~\ref{sec:d5} explains which attributes the tables omit and why.
  \item Table~\ref{tab:map} links a system's design choices to the sections that explain them.
\end{itemize}
\end{takeaways}

\chapter{Instrumental Text-to-Music}
\label{sec:instrumental}

\begin{figure}[H]
  \centering
  \includegraphics[width=\linewidth,alt={Timeline of landmark systems of the period reviewed arranged by paradigm; colored by application class; with filled; hollow; and grey markers for open; closed; and partial or withdrawn weights.}]{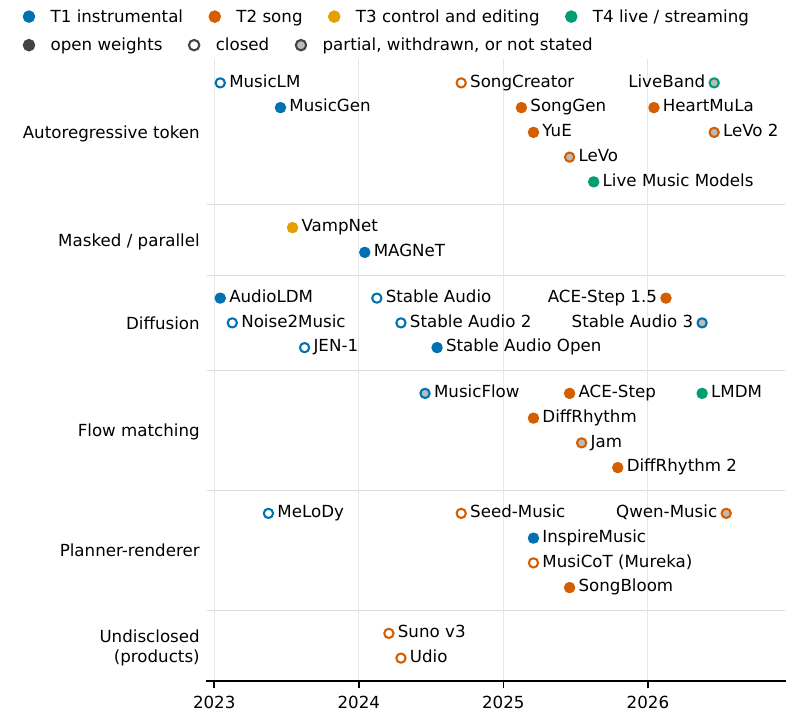}
  \caption{Landmark music generation systems of the period reviewed, by generative paradigm (rows), application class (color), and openness (filled: open weights; hollow: closed; grey: partial release, withdrawn release, or release not stated in the paper). Dates are first public release, placed to the month.}
  \label{fig:timeline}
\end{figure}

Instrumental text-to-music (class T1) was the task addressed by the 2023 systems, and it is the class in which the paradigms can be compared without the added problems of vocals and lyrics. A description specifies the music and the system produces an excerpt, so the comparison asks how each method handles sequence length, sampling cost, and control. Reported quality also depends on training data and evaluation, and cannot be attributed to the model family alone. Table~\ref{tab:t1} lists the systems reviewed in this section with their taxonomy placement and released-artifact attributes. Figure~\ref{fig:timeline} places the landmark systems of this and the next class on a common time axis.

\section{Autoregressive Token Language Models}
\label{sec:t1-ar}

MusicLM established a hierarchical codec-token approach to text-to-music generation, and the token-based systems of Tables~\ref{tab:t1} and~\ref{tab:t2} keep its main elements~\parencite{agostinelli2023musiclm}. A MuLan embedding of the text prompt conditions a first language model that predicts semantic tokens (in the original, units from w2v-BERT, a self-supervised speech encoder), a second model predicts the coarse SoundStream codebooks given the semantic tokens, and a third predicts the fine codebooks. The SoundStream decoder renders audio. The hierarchy separates long-range structure, which the semantic stream carries, from acoustic detail, and it lets each stage run at its own sequence length. MusicLM also released MusicCaps, the caption set that became the default text-to-music benchmark. The model was not released.

MusicGen then showed that the three-stage hierarchy can be replaced by a single model~\parencite{copet2023musicgen}. A single transformer predicts EnCodec tokens under one of several codebook interleaving patterns. The delay pattern, which offsets each codebook by one step, lets one autoregressive pass predict all codebooks with a small quality cost relative to fully sequential decoding. The model adds melody conditioning through a chromagram (a per-frame profile of energy over the twelve pitch classes) of a reference recording, was released with weights and code, and is the most common open baseline among the papers reviewed here: it appears as a baseline or a backbone in the control, editing, stem, and adaptation work of Section~\ref{sec:control}~\parencite{lin2023cocomulla,lan2024musicongen,zhang2024instructmusicgen,rouard2025musicgenstem,lin2024arrange,mehta2025exploring}.

\paragraph{Running example.} MusicGen is the system this monograph returns to, because its paper states every slot of the generic problem and because so much later work is built on it. Slot 1 is the EnCodec codec of Section~\ref{sec:notation}, 50 frames per second and four streams. Slot 2 is a single transformer over the delay pattern, 1,503 decoding steps for 30 seconds. Slot 3 is a global text embedding from a frozen text encoder, with an optional chromagram of a reference recording added as a time-aligned condition for melody. Slot 4 is plain autoregressive sampling, with no acceleration, which is why the table lists it as offline. Each later section says what happened to one of these slots.

\begin{figure}[H]
  \centering
  \includegraphics[width=0.8\linewidth,alt={Bubble chart of the number of reviewed systems per generative paradigm and application class.}]{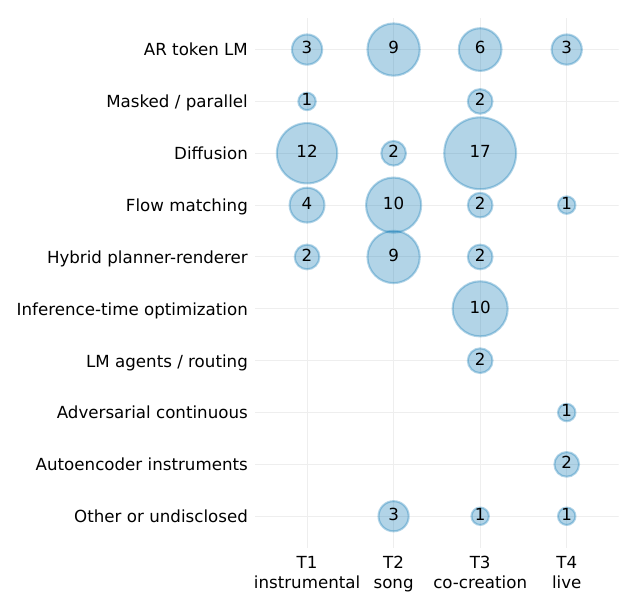}
  \caption{Coverage of the reviewed systems across generative paradigm (D2) and application class (D4), for the classes tabulated in Tables~\ref{tab:t1} to~\ref{tab:t3}. A system that spans classes is counted once, under its table row (M2UGen and the agent systems under T3, production co-creation, for their editing and co-creation use); the cross-modal (T5) and unified (T6) systems of Section~\ref{sec:crossmodal-symbolic} without a table row do not appear. The last row counts the rows whose paradigm is undisclosed (the two commercial products) or outside the nine groups, so that the figure covers all 105 rows. Bubble area and number give the count of systems per cell; assignments are the table rows. Empty cells mark combinations with no tabulated system. Counts are derived from the rows of the tables.}
  \label{fig:paradigm-tier}
\end{figure}

The systems discussed next illustrate two approaches to acoustic detail, and the choice between them is the planning question Q4 in its earliest form. The first keeps the language model for what it does well, long-range structure, and delegates acoustic detail to a separate renderer. MeLoDy reduced sampling cost by rendering semantic tokens with a dual-path diffusion model~\parencite{lam2023efficient}, and InspireMusic pairs an autoregressive model with a flow-matching super-resolution stage for long-form 48\,kHz output~\parencite{zhang2025inspiremusic}. MusiCoT has the model generate an analyzable structural sketch, a trajectory of audio-text embedding tokens, before the audio tokens, which its authors report improves global structure. The method is the basis of a commercial system~\parencite{lam2025musicot}. The second direction keeps everything in one acoustic token space and scales it, accepting long sequences for the sake of a single model. UniAudio includes music in a single language model with universal token input and output across audio tasks~\parencite{yang2024uniaudio}. Shao builds a deep RVQ hierarchy and generates coarse tokens for the whole track before a super-resolution model completes the finer codebooks within the same token space. It is trained with lyrics and generates vocal songs, so it appears in Table~\ref{tab:t2}, but its design question is the one of this section~\parencite{liu2026shao}. Codec design constrains what a token model can express: codecs trained for reconstruction alone may represent semantic information in a form that language models predict inefficiently, which motivated semantically supervised codecs (Section~\ref{sec:codecs})~\parencite{ye2024codec}, and codecs with frequency-ordered or multi-scale codebooks make truncating the codebook stack a controlled loss~\parencite{yang2026harp,siuzdak2024snac}. A controlled comparison that holds data and architecture fixed and varies only the modeling paradigm found that autoregressive and flow-matching models trade off differently on quality, speed, and controllability, so the choice between them depends on the deployment target rather than on one being dominant~\parencite{tal2025auto}.

\section{Masked and Parallel Decoding}
\label{sec:t1-masked}

Masked token modeling generates all positions iteratively rather than left to right. VampNet showed that one masked model covers generation, inpainting, outpainting, and variations by changing the mask~\parencite{garcia2023vampnet}. Because it is prompted with audio rather than text, it is tabulated with the editing systems in Table~\ref{tab:t3}. MAGNeT reached near-autoregressive quality at substantially lower latency with a pure non-autoregressive schedule over EnCodec codebooks~\parencite{ziv2024magnet}. The approach was extended to accompaniment, where StemGen generates a stem conditioned on an existing mix~\parencite{parker2024stemgen}. Masked decoding has not been used for full songs in the systems reviewed here (Figure~\ref{fig:paradigm-tier}). The song systems that need parallel generation use diffusion or flow matching over continuous latents instead.

\section{Diffusion and Flow Matching}
\label{sec:t1-diffusion}

Diffusion models reached music by two routes, waveform cascades at industrial scale and compressed latent representations, which reduced the computation needed and supported several open implementations. Noise2Music demonstrated text-to-music with cascaded waveform diffusion at industrial scale, with pseudo-captions generated by a language model~\parencite{huang2023noise2music}. AudioLDM diffuses over mel-VAE latents with CLAP conditioning and renders with a vocoder~\parencite{liu2023audioldm}. Other latent models released in this period included Make-An-Audio with prompt enhancement~\parencite{huang2023makeanaudio}, ERNIE-Music with weakly supervised text-music pairs~\parencite{zhu2023ernie}, AudioLDM~2 with one latent for speech, sound, and music~\parencite{liu2024audioldm2}, Mo\^usai with long contexts under strong compression~\parencite{schneider2024mousai}, MusicLDM with beat-synchronous mixup~\parencite{chen2024musicldm}, and JEN-1 at 48\,kHz stereo~\parencite{li2024jen1}.

The Stable Audio series exemplifies the design that several later systems use: a diffusion transformer over a highly compressed continuous latent, with the same or a derived autoencoder~\parencite{ning2025diffrhythm,wang2026segtune,chen2026wansong}. Timing conditioning informs the model of its position within a track, enabling structured 95-second stereo pieces~\parencite{evans2024fast}. A diffusion transformer over a compressed latent extended this to full-length tracks with coherent long-range structure~\parencite{evans2024longform}. Stable Audio Open is an open-weights reference model with documented Creative Commons training sources~\parencite{evans2025stableaudioopen}. Stable Audio 3 added a semantic-acoustic autoencoder, variable-length generation, inpainting-based editing, and adversarial post-training for few-step inference, with open small and medium weights and a commercial tier~\parencite{evans2026stableaudio3,stability2026stableaudio}. Orthogonal improvements target data quality, with quality-aware masked diffusion transformers steerable toward high-quality output from noisy corpora~\parencite{li2025qamdt}. The latent autoencoder is a separate design decision: decoders at high compression lose high frequencies, phase coherence, and stereo image, and complex-spectral autoencoders with per-band correction address these failures~\parencite{wang2026fourier}.

Flow matching replaces the diffusion objective with a velocity regression along a chosen path, and the variants used in music choose the linear path and reach few solver steps with reflow or distillation. MusicFlow reached comparable quality with smaller, faster cascaded models and native infilling~\parencite{prajwal2024musicflow}, FluxMusic adapted the rectified-flow transformer design from image generation~\parencite{fei2024fluxmusic}, and JASCO used flow matching for temporally precise joint conditioning~\parencite{tal2024jasco}. Sampling cost falls further with consistency and distillation methods: consistency training reaches few-step music generation~\parencite{fei2024music}, latent consistency models do the same for text-to-audio~\parencite{liu2024audiolcm}, flow matching with preference optimization gives fast and faithful text-to-audio generation~\parencite{hung2024tangoflux}, and Presto distills both sampling steps and network layers, reaching faster-than-real-time latent diffusion~\parencite{novack2025presto}. Instrument-level synthesis with explicit playing techniques~\parencite{tian2026violet} and unified anything-to-audio generation~\parencite{tian2025audiox} are neighboring tasks that share these architectures.

\section{Lessons and Open Trade-offs}
\label{sec:t1-lessons}

This class poses the simplest form of the generation problem: produce seconds to minutes of instrumental audio from a global text condition in the format of a studio recording. It is the class in which the representation and cost questions are answered most directly. The tabulated answers are autoregressive language models over codec tokens, masked token models, and diffusion or flow matching over continuous latents, with planner-renderer designs as a fourth once MeLoDy delegated rendering to a diffusion model~\parencite{lam2023efficient}. The evidence that separates them is uneven. Reported runtime gives the most concrete distinction, although the reported values depend on the hardware and configuration of each measurement. Six rows of Table~\ref{tab:t1} carry the faster-than-real-time code: four diffusion or flow models, with or without distillation~\parencite{evans2024fast,evans2024longform,novack2025presto,evans2026stableaudio3}, one planner-renderer design with a diffusion renderer~\parencite{lam2023efficient}, and one masked token model, MAGNeT, whose paper gives its speed relative to autoregressive decoding~\parencite{ziv2024magnet}. No autoregressive token model reports such a measurement. On control the reviewed corpus shows a pattern rather than a general result: adapters were built on token models, largely because MusicGen's weights were available, which is why much of the control literature of Section~\ref{sec:control} was built on it~\parencite{copet2023musicgen}, and gradient-based control at inference time was built on diffusion models~\parencite{novack2024ditto}. Neither route is exclusive to its paradigm. On quality the evidence is weakest: the one study that holds data and architecture fixed and varies only the paradigm found no dominant choice~\parencite{tal2025auto}, and every other comparison in this section confounds the paradigm with the training set, the codec or autoencoder, the model size, and the metric, none of which is held constant across papers. Output quality and the effect of training data cannot be compared from the table at all. The systems report different metrics over different embeddings (Section~\ref{sec:evaluation}), and five of the 22 rows carry only the data code U, with no named source and no license, so a difference between two rows may be a difference in data.

A reader who starts from a practical constraint can take three rules from the table. For an open backbone to fine-tune or adapt, MusicGen remains the common choice because the adapters and controls that exist were built for it. For long-form stereo output with a documented training set, the Stable Audio reports provide examples with documented training-source statements, from the Creative Commons model~\parencite{evans2025stableaudioopen} to the licensed and screened corpus of its third generation~\parencite{evans2026stableaudio3}. For few-step or faster-than-real-time generation, several diffusion or flow systems reach these speeds through distillation or adversarial post-training. The table also includes faster-than-real-time masked and hybrid systems. The open trade-off of the class is the autoencoder. Compression sets the latent sequence length and with it part of the sampling cost, and the losses at high compression, in high frequencies, phase coherence, and stereo image, are documented~\parencite{wang2026fourier} but have not been traded against sampling cost in a controlled study. The choice between a codec with semantic supervision~\parencite{ye2024codec} and one with frequency-ordered codebooks~\parencite{yang2026harp} is likewise made per system and not compared.

\noindent{\small Legend for Tables~\ref{tab:t1} to~\ref{tab:t3}. D1 representation: Tok, discrete codec tokens; Lat, continuous latent or spectrogram; Wav, waveform; Hyb, plan or sketch plus acoustic representation; Sym, symbolic. Duration: the reported figure with its kind: \textsc{ctx}, training or context window; \textsc{eval}, evaluated clip length; \textsc{gen}, demonstrated or single-pass output; \textsc{stream}, unbounded; a training window is not a maximum. Str.\ (streaming): S, the paper describes a streaming or live generation procedure; --, batch generation. Run (measured runtime class for batch generation): O, slower than real time or not timed; F, faster than real time for the reported configuration, task, and hardware; M, faster for some reported tasks and slower for others; --, streaming system (no batch runtime reported); speed does not imply streaming. Open (weights): \checkmark, released; S, some sizes; R, a released checkpoint that differs from the paper's configuration (retrained or later version); C, code only; A, announced; W, released and later withdrawn; $\times$, not released; ?, not stated in paper or model card. For inference-time methods a second mark gives the backbone's weights (method / backbone). Data (training data of the listed method): L, licensed; CC, Creative Commons; P, public research datasets; Syn, synthetic; U, undisclosed, in-house, or web-collected; hours as reported; \textsc{mix}, the corpus is general audio, not music only; \textsc{coll}, the figure is the collection before filtering; --, none (inference-time method on a pretrained backbone). The data codes summarize what the sources state. They do not certify rights clearance or the compensation of creators. Post-training: \textsc{pref}, preference optimization or reinforcement learning; \textsc{adv}, adversarial training or adversarial distillation; \textsc{distill}, distillation without an adversarial loss; \textsc{sft}, supervised fine-tuning; --, not reported. D1 describes the representation the acoustic generator produces; a system whose plan or sketch precedes that generator is Hyb. A question mark means the value is not stated in the system's paper or model card (each such cell was checked against the paper or card), and such cells are left unfilled rather than inferred. Rows are grouped by generative paradigm (D2) and ordered by year of first public release, which can precede the year of the cited version.}

\begin{table}[H]
  \caption{Instrumental text-to-music systems (T1). Coverage and source dates are defined in Section~\ref{sec:method} and Appendix~\ref{app:search}.}
  \label{tab:t1}
  \footnotesize
  \setlength{\tabcolsep}{2.5pt}
  \begin{tabular}{@{}>{\raggedright\arraybackslash}p{2.90cm}>{\raggedright\arraybackslash}p{8.50cm}@{}}
    \toprule
    System, D1 & Attributes (D3, D5) \\
    \midrule
    \multicolumn{2}{@{}l}{\emph{Autoregressive token language models}} \\
    \textbf{MusicGen}\newline \parencite{copet2023musicgen}\newline D1 Tok & \textit{Inputs} text; melody; \textit{Duration} 30 s window; longer by windowed continuation (official code) \textsc{ctx}; \textit{Output} 32 kHz mono; stereo models by fine-tuning (paper Section 4.3), released as separate checkpoints; \textit{Latency} batch, offline; \textit{Open} \checkmark; \textit{Data} L 20k\,h lic.; \textit{Post-training} --. \\
    \textbf{MusicLM}\newline \parencite{agostinelli2023musiclm}\newline D1 Tok & \textit{Inputs} text; melody (hummed); \textit{Duration} 5 min demonstrated \textsc{gen}; \textit{Output} 24 kHz mono; \textit{Latency} batch, offline; \textit{Open} $\times$; \textit{Data} U 280k\,h; \textit{Post-training} --. \\
    \textbf{UniAudio}\newline \parencite{yang2024uniaudio}\newline D1 Tok & \textit{Inputs} text and task tokens; \textit{Duration} ?; \textit{Output} ?; \textit{Latency} batch, offline; \textit{Open} R; \textit{Data} P 165k\,h \textsc{mix} (all-audio multi-task corpus); \textit{Post-training} --. \\
    \multicolumn{2}{@{}l}{\emph{Masked and parallel decoding}} \\
    \textbf{MAGNeT}\newline \parencite{ziv2024magnet}\newline D1 Tok & \textit{Inputs} text; \textit{Duration} 30 s \textsc{ctx}; \textit{Output} 32 kHz; \textit{Latency} batch, faster than real time; \textit{Open} \checkmark; \textit{Data} L 20k\,h lic.; \textit{Post-training} --. \\
    \multicolumn{2}{@{}l}{\emph{Latent and cascaded diffusion}} \\
    \textbf{AudioLDM}\newline \parencite{liu2023audioldm}\newline D1 Lat & \textit{Inputs} text (CLAP); \textit{Duration} 10 s \textsc{gen}; \textit{Output} 16 kHz mono; \textit{Latency} batch, offline; \textit{Open} \checkmark; \textit{Data} P \textsc{mix} (L-Full: AudioSet, AudioCaps, Freesound, BBC); \textit{Post-training} --. \\
    \bottomrule
  \end{tabular}
\end{table}

\begin{table}[H]
  \addtocounter{table}{-1}
  \caption{Instrumental text-to-music systems (T1) (continued).}
  \footnotesize
  \setlength{\tabcolsep}{2.5pt}
  \begin{tabular}{@{}>{\raggedright\arraybackslash}p{2.90cm}>{\raggedright\arraybackslash}p{8.50cm}@{}}
    \toprule
    System, D1 & Attributes (D3, D5) \\
    \midrule
    \textbf{Make-An-Audio}\newline \parencite{huang2023makeanaudio}\newline D1 Lat & \textit{Inputs} text (prompt enhancement); \textit{Duration} 10 s \textsc{gen}; \textit{Output} 16 kHz; \textit{Latency} batch, offline; \textit{Open} \checkmark; \textit{Data} P 3k\,h \textsc{mix} (AudioSet, AudioCaps, Clotho, ESC-50, BBC and others); \textit{Post-training} --. \\
    \textbf{Noise2Music}\newline \parencite{huang2023noise2music}\newline D1 Wav & \textit{Inputs} text (pseudo-captions); \textit{Duration} 30 s \textsc{gen}; \textit{Output} 24 kHz; \textit{Latency} batch, offline; \textit{Open} $\times$; \textit{Data} U 340k\,h; \textit{Post-training} --. \\
    \textbf{AudioLDM 2}\newline \parencite{liu2024audioldm2}\newline D1 Lat & \textit{Inputs} text; \textit{Duration} 10 s \textsc{gen}; \textit{Output} 16 kHz (paper); 48 kHz checkpoint in the repository; \textit{Latency} batch, offline; \textit{Open} \checkmark; \textit{Data} P 29.5k\,h \textsc{mix} (includes speech and sound effects); \textit{Post-training} --. \\
    \textbf{JEN-1}\newline \parencite{li2024jen1}\newline D1 Lat & \textit{Inputs} text; audio (continuation, inpainting); \textit{Duration} ?; \textit{Output} 48 kHz stereo; \textit{Latency} batch, offline; \textit{Open} $\times$; \textit{Data} L 15k tracks; \textit{Post-training} --. \\
    \textbf{Moûsai}\newline \parencite{schneider2024mousai}\newline D1 Lat & \textit{Inputs} text; \textit{Duration} about 44 s per evaluated window; minutes by chaining \textsc{eval}; \textit{Output} 48 kHz stereo; \textit{Latency} batch, offline; \textit{Open} \checkmark; \textit{Data} U 2.5k\,h coll.; \textit{Post-training} --. \\
    \textbf{MusicLDM}\newline \parencite{chen2024musicldm}\newline D1 Lat & \textit{Inputs} text; \textit{Duration} 10 s \textsc{gen}; \textit{Output} 16 kHz; \textit{Latency} batch, offline; \textit{Open} \checkmark; \textit{Data} L 455.6\,h \textsc{coll} (Audiostock); \textit{Post-training} --. \\
    \textbf{Stable Audio}\newline \parencite{evans2024fast}\newline D1 Lat & \textit{Inputs} text; timing; \textit{Duration} 95 s \textsc{gen}; \textit{Output} 44.1 kHz stereo; \textit{Latency} batch, faster than real time; \textit{Open} $\times$; \textit{Data} L 19.5k\,h lic.; \textit{Post-training} --. \\
    \textbf{Stable Audio 2 (long-form)}\newline \parencite{evans2024longform}\newline D1 Lat & \textit{Inputs} text; timing; \textit{Duration} 4 min 45 s (research model) \textsc{gen}; \textit{Output} 44.1 kHz stereo; \textit{Latency} batch, faster than real time; \textit{Open} $\times$; \textit{Data} L 19,500\,h; \textit{Post-training} --. \\
    \textbf{Presto}\newline \parencite{novack2025presto}\newline D1 Lat & \textit{Inputs} text; \textit{Duration} 32 s \textsc{gen}; \textit{Output} 44.1 kHz mono; stereo through the additional MusicHiFi stage; \textit{Latency} batch, faster than real time; \textit{Open} $\times$; \textit{Data} L 3.6k\,h lic.; \textit{Post-training} \textsc{adv}: GAN-based step distillation (DMD) and layer distillation, alone or combined (variant-dependent). \\
    \textbf{QA-MDT}\newline \parencite{li2025qamdt}\newline D1 Lat & \textit{Inputs} text; quality token; \textit{Duration} 10 s \textsc{gen}; \textit{Output} 16 kHz; \textit{Latency} batch, offline; \textit{Open} \checkmark; \textit{Data} CC/P 12.5k\,h; \textit{Post-training} --. \\
    \bottomrule
  \end{tabular}
\end{table}

\begin{table}[H]
  \addtocounter{table}{-1}
  \caption{Instrumental text-to-music systems (T1) (continued).}
  \footnotesize
  \setlength{\tabcolsep}{2.5pt}
  \begin{tabular}{@{}>{\raggedright\arraybackslash}p{2.90cm}>{\raggedright\arraybackslash}p{8.50cm}@{}}
    \toprule
    System, D1 & Attributes (D3, D5) \\
    \midrule
    \textbf{Stable Audio Open}\newline \parencite{evans2025stableaudioopen}\newline D1 Lat & \textit{Inputs} text; timing; \textit{Duration} 47 s \textsc{gen}; \textit{Output} 44.1 kHz stereo; \textit{Latency} batch, offline; \textit{Open} \checkmark; \textit{Data} CC/P 7.3k\,h \textsc{mix} (Freesound, FMA); \textit{Post-training} --. \\
    \multicolumn{2}{@{}l}{\emph{Flow matching and rectified flow}} \\
    \textbf{FluxMusic}\newline \parencite{fei2024fluxmusic}\newline D1 Lat & \textit{Inputs} text; \textit{Duration} 10 s \textsc{gen}; \textit{Output} 16 kHz; \textit{Latency} batch, offline; \textit{Open} \checkmark; \textit{Data} P/U 22k\,h (ASM, MTT, MSD, FMA, Music4All plus private); \textit{Post-training} --. \\
    \textbf{JASCO}\newline \parencite{tal2024jasco}\newline D1 Lat & \textit{Inputs} text; chords; melody; drums; audio; \textit{Duration} about 10 s \textsc{gen}; \textit{Output} 32 kHz; \textit{Latency} batch, offline; \textit{Open} \checkmark; \textit{Data} L 16k\,h lic. (Meta Music Initiative, Shutterstock, Pond5 per the model card); \textit{Post-training} --. \\
    \textbf{MusicFlow}\newline \parencite{prajwal2024musicflow}\newline D1 Lat & \textit{Inputs} text; audio (continuation); \textit{Duration} 10 s training chunks \textsc{ctx}; \textit{Output} 32 kHz mono; \textit{Latency} batch, offline; \textit{Open} A; \textit{Data} U 20k\,h; \textit{Post-training} --. \\
    \textbf{Stable Audio 3}\newline \parencite{evans2026stableaudio3}\newline D1 Lat & \textit{Inputs} text; inpainting; variable length; \textit{Duration} 2 min (small); 6 min 20 s (medium, large) \textsc{gen}; \textit{Output} 44.1 kHz stereo; \textit{Latency} batch, faster than real time; \textit{Open} S; \textit{Data} L/CC; \textit{Post-training} \textsc{adv}: adversarial post-training. \\
    \multicolumn{2}{@{}l}{\emph{Planner-renderer hybrids}} \\
    \textbf{MeLoDy}\newline \parencite{lam2023efficient}\newline D1 Hyb & \textit{Inputs} text; audio (continuation); \textit{Duration} 10 to 30 s evaluated \textsc{eval}; \textit{Output} 24 kHz; \textit{Latency} batch, faster than real time; \textit{Open} $\times$; \textit{Data} U 257k\,h int.; \textit{Post-training} --. \\
    \textbf{InspireMusic}\newline \parencite{zhang2025inspiremusic}\newline D1 Hyb & \textit{Inputs} text; audio prompt (continuation); \textit{Duration} up to 8 min \textsc{gen}; \textit{Output} 48 kHz output (24 kHz tokenizer); \textit{Latency} batch, offline; \textit{Open} \checkmark; \textit{Data} P/U 100k\,h; \textit{Post-training} --. \\
    \bottomrule
  \end{tabular}
\end{table}

\FloatBarrier

\chapter{Song Generation}
\label{sec:song}

Full songs with vocals (class T2) became the most active class of this collection in 2024 and 2025. The task adds lyric-to-vocal alignment and coordination between vocals and accompaniment. These relationships must remain coherent over a full song, often several minutes long. The words must align with the singing, the voice must stay consistent, and the accompaniment and the sections must support the same song, so longer output is only part of the problem. The systems differ in how they divide these responsibilities between representations, generation stages, and conditioning. Table~\ref{tab:t2} lists the systems reviewed here.

\section{Commercial Systems and Compositional Precursors}

The early commercial song generators disclosed little about their architectures. Suno and Udio demonstrated full-song generation without technical reports~\parencite{suno2026,udio2026}. Seed-Music documented an industrial framework that mixes token language models and diffusion and adds vocal cloning and note-level editing~\parencite{bai2024seedmusic}. Mureka offers the chain-of-thought planning of MusiCoT as a product~\parencite{lam2025musicot}. Academic work reached songs by a different route: it decomposed the song into parts that existing models could handle, and the decompositions are instructive because later single-model systems had to solve the same sub-problems internally. One decomposition took the vocals as given and generated the accompaniment: SingSong did this with an AudioLM-style token stack~\parencite{donahue2023singsong}, and FastSAG replaced its autoregressive stages with non-autoregressive diffusion over mel spectrograms to make it fast~\parencite{chen2024fastsag}. Another generated the vocals first and the accompaniment second. Two-stage text-to-song did so with a tri-tower contrastive pretraining that supplies the text representation for the accompaniment stage~\parencite{hong2024texttosong}. MelodyLM put a MIDI melody before the vocals, so that the text controls an explicit melody representation before audio rendering, and rendered the accompaniment with latent diffusion~\parencite{li2024accompanied}. The Chord-Conditioned Song Generator fed extracted chords into an end-to-end model through a cross-attention mechanism robust to extraction errors~\parencite{gao2024end}. The decomposition then moved inside one model: SongCreator modeled vocals and accompaniment as two interacting token sequences in one language model, covering generation, continuation, and editing~\parencite{lei2024songcreator}, and SongEditor added segment-wise and track-wise editing of lyrics, vocals, and accompaniment to a language-model song generator~\parencite{yang2024songeditor}.

\section{Open Systems}

Open models reached the same task in 2025, within a year of the commercial products, and some of them provided inspectable and extensible baselines. They split along the two lineages of Section~\ref{sec:instrumental}. On the token side, YuE scaled a LLaMA-style language model to five-minute songs with track-decoupled next-token prediction, in which vocals and accompaniment are predicted as separate token tracks rather than as one mixed track, and added in-context style transfer and a memorization analysis. Its authors present it as the first open lyrics-to-song foundation model~\parencite{yuan2025yue}. SongGen reached single-stage generation by interleaving vocal and accompaniment tokens, with mixed and dual-track output modes~\parencite{liu2025songgen}. On the continuous side, DiffRhythm generated full songs end to end in seconds with flow matching over a continuous latent, which is possible because every sampling step treats the whole song at once~\parencite{ning2025diffrhythm}. DiffRhythm+ added style control and preference optimization~\parencite{chen2025diffrhythmplus}. ACE-Step released an open flow-matching model for songs, which its successor in Section~\ref{sec:song-convergence} extends~\parencite{gong2025acestep}. VersBand decoupled the vocal and accompaniment models under prompt-based control~\parencite{zhang2025versatile}. LeVo showed what post-training adds: it aligned a token language model to several human preference axes with direct preference optimization~\parencite{lei2025levo}, and its successor targets stability and melodic quality through hierarchical representation modeling and progressive post-training~\parencite{lei2026levo2}. Several projects also make reproducibility an explicit goal: Muse releases a song generator together with a licensed synthetic training set and the training and evaluation pipeline~\parencite{jiang2026muse}, and the ICME 2026 grand challenge on academic text-to-music generation requires training from scratch on a Creative Commons subset of MTG-Jamendo, for ten-second instrumental clips~\parencite{hsieh2026academic}.

\section{Planning, Scale, and Post-Training}
\label{sec:song-convergence}

The two lineages, token language models and latent diffusion, were combined most often in this class, and the combinations show what each side contributes: the token model supplies a sketch of the whole song, and the diffusion or flow model supplies the acoustic detail that is costly to predict in long token sequences. SongBloom interleaves autoregressive sketching with patch-wise diffusion refinement, so that each sketched segment is rendered before the next is sketched~\parencite{yang2025songbloom}. DiffRhythm~2 is semi-autoregressive block flow matching: the song is generated block by block, each block by flow matching conditioned on the blocks before it, which recovers lyric alignment without external labels. The same work introduces cross-pair optimization for multi-preference alignment~\parencite{jiang2025diffrhythm2}. SketchSong plans a song-level sketch along time and models the parts separately along the track axis~\parencite{duan2026sketchsong}. Two results show the range of scale and openness: Jam shows that a small flow-matching model suffices for full songs with word-level lyric control~\parencite{liu2025jam}, and HeartMuLa releases its model, codec, and training procedure, with multilingual conditioning and a reinforcement-refined release~\parencite{yang2026heartmula}. The YuE group released YuE2 on 9 September 2026, followed by a technical report on 27 September~\parencite{map2026yue2,yuan2026yue2}. Its autoregressive and non-autoregressive mixture-of-transformers backbone generates an editable score and semantic tokens, which condition a flow-matching renderer for 48 kHz stereo audio. The weights were released under CC BY-NC 4.0.

\begin{figure}[H]
  \centering
  \includegraphics[width=\linewidth]{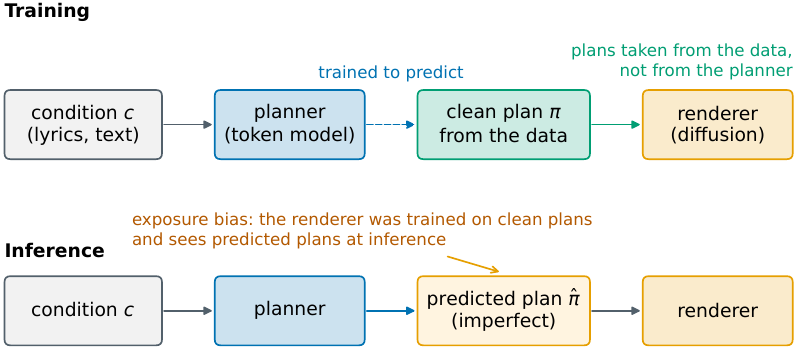}
  \caption{The planner-renderer data path and where its training and inference differ. In training (top), the planner learns to predict a plan $\pi$ (semantic tokens, a melody, or a score) from the condition, and the renderer learns to produce audio from plans taken from the data, which are clean. At inference (bottom), the renderer receives the planner's own prediction $\hat\pi$, which contains errors the renderer never saw in training. This mismatch is the exposure bias that the full-context renderer of Section~\ref{sec:song-convergence} is trained against. It is the cost of the separation that gives the design its long-range structure.}
  \label{fig:planner}
\end{figure}

Among the 2026 industrial reports in Table~\ref{tab:t2}, ACE-Step 1.5, Qwen-Music, and the eight-codebook report describe a planner and a renderer, and WanSong describes a single flow model without a planner. The three planner-renderer systems differ in the plan, the renderer, and the training procedure, and Qwen-Music reports the largest training corpus, in hours, of any row in Table~\ref{tab:t2}. ACE-Step 1.5 extended the open ACE-Step line and put a language-model planner in front of its generator~\parencite{gong2026acestep15}. Qwen-Music pairs a 3B semantic-token language model with a diffusion-transformer renderer, trained over more than five million hours of multilingual audio. Its addition is a melody-token chain of thought, a melody sequence the planner writes before the song, which plans the vocal contour and also serves as an interface for melody cloning and cover generation. The authors report blind expert tests in which it matches the strongest closed commercial systems~\parencite{qwen2026music}. A second report describes a semantic-aware eight-codebook tokenizer, a hybrid language-model planner, and a FullDiT renderer covering lyrics-to-song, instrumental, and cover generation~\parencite{dai2026pushing}. Its renderer is described separately, and that description states the central difficulty of the planner-renderer design (Figure~\ref{fig:planner}): in training the renderer sees clean token plans, at inference it sees the planner's imperfect ones, and a renderer trained without regard to this exposure bias degrades. The proposed remedy is to train the renderer for full-context generation from imperfect plans~\parencite{li2026beyond}. WanSong instead generates the song directly with one rectified-flow model and produces multilingual songs of several minutes with separate vocal and accompaniment stems in one pass. It relies on step distillation for speed~\parencite{chen2026wansong}.

Once a system can generate a whole song, the next problem is controlling what happens inside it, and much of the recent work on this class addresses that problem with the conditioning forms of Section~\ref{sec:d3} applied along the song's timeline. SegTune broadcasts segment-level prompts to their time windows, so that a verse and a chorus can receive different descriptions, and predicts sentence-level lyric timestamps with a language model~\parencite{wang2026segtune}. Cover-song generation conditions on the melody of an existing song and regenerates vocals and accompaniment~\parencite{li2026songecho}, with explicit phoneme conditioning reducing lyric errors~\parencite{lee2026mpecho}. MIDI-informed accompaniment lets songwriters retain direct control over the melody they supply by conditioning on vocal MIDI, chords, and structure~\parencite{tsai2026midi}, and UniSinger unifies speaker-cloning song generation with singing voice conversion in one multimodal diffusion transformer~\parencite{zhang2026unified}. Score-first pipelines through editable bar-level notation~\parencite{wang2025score} and video-synchronized song generation~\parencite{wang2025audiogen} lie at the boundary of the class.

Preference alignment has become common in this class: 11 of the 25 song systems of Table~\ref{tab:t2} first released in 2025 and 2026 report such a stage (the tables date a system by its first public release, which can precede the year of the cited version). The stage suits songs because errors in lyrics, unstable vocals, and unsatisfactory melodies can be easier for listeners to recognize than to specify in a training loss. Of the systems in Table~\ref{tab:t2}, DiffRhythm+, LeVo, LeVo~2, DiffRhythm~2, Jam, SegTune, HeartMuLa, WanSong, Qwen-Music, the Dai et al.\ framework, and ACE-Step~1.5 report a preference or reinforcement stage, as does the 2024 Seed-Music framework, which is outside the count of 11, while YuE, SongGen, DiffRhythm, ACE-Step, SongBloom, SketchSong, Muse, and Shao report none (the table marks adversarial post-training separately). Several reports combine supervised initialization, offline preference optimization, and online policy optimization in sequence~\parencite{qwen2026music,lei2026levo2,dai2026pushing}. Alignment targets have widened from overall preference to specific failure modes: phoneme-error-based preference data against lyric hallucination~\parencite{zhang2025hallucination}, multiple off-the-shelf rewards for text alignment, production quality, and semantic consistency in flow-matching generators~\parencite{ziv2025mr}, distributional rewards that match a target distribution rather than a scalar~\parencite{bai2025dragon}, and open pairwise preference metrics built for alignment~\parencite{kim2026tunejury}. Alignment can also reduce output diversity, which makes diversity preservation a separate design concern~\parencite{cideron2024diversity}.

\section{Lessons and Open Trade-offs}
\label{sec:t2-lessons}

Song generation adds constraints to the instrumental problem: the vocals must follow the lyrics, the accompaniment and vocals must remain coordinated, and these relations must hold over a full song. Three answers are tabulated:
\begin{itemize}
  \item a language model over interleaved or decoupled vocal and accompaniment tokens~\parencite{yuan2025yue,liu2025songgen};
  \item end-to-end diffusion or flow over the whole song~\parencite{ning2025diffrhythm,gong2025acestep,chen2026wansong}, within which step distillation is a sampling choice rather than a different model~\parencite{chen2026wansong}, and semi-autoregressive block generation a variant that decodes the song in order~\parencite{jiang2025diffrhythm2};
  \item planner-renderer hybrids in which a token model writes a sketch, a melody, or a score and a diffusion or flow model renders it~\parencite{yang2025songbloom,qwen2026music,dai2026pushing,map2026yue2}.
\end{itemize}
Errors in lyric alignment are a useful point of comparison between the paradigms: token models carry the lyrics in the sequence, diffusion models need the alignment supplied or recovered, DiffRhythm~2 recovers it without labels~\parencite{jiang2025diffrhythm2}, and lyric hallucination has become a target of preference data in both lineages~\parencite{zhang2025hallucination}. Sampling cost separates them as in Section~\ref{sec:instrumental}: the flow-matching systems report full songs in seconds~\parencite{ning2025diffrhythm,gong2025acestep}, and the sketch-then-render system that reports a real-time factor is slower than real time at full size~\parencite{yang2025songbloom}. Structure separates the planner-renderer designs from the rest, and it is the main response of this class to the planning question: several of the 2026 industrial reports use an explicit plan~\parencite{qwen2026music,dai2026pushing}, and the exposure bias between the clean plans of training and the predicted plans of inference is the cost that one of them trains against~\parencite{dai2026pushing}. The limits of cross-system comparison are especially pronounced for songs. The industrial systems are trained on undisclosed corpora of five million hours and more and the academic systems on tens of thousands of hours or on a released synthetic corpus~\parencite{jiang2026muse}, so a quality difference between rows may partly reflect a data difference. The evaluation batteries differ by report. Differences in evaluation protocols further limit independent comparison. For ten of the 33 rows of Table~\ref{tab:t2} the paper states nothing about released weights.

The practical choices follow the constraint a reader starts from. For training from scratch at academic scale, Muse releases a song pipeline and a corpus~\parencite{jiang2026muse}. The ICME challenge fixes a comparable setting for ten-second instrumental clips~\parencite{hsieh2026academic}. Fine-tuning an open backbone has ACE-Step and YuE as the common bases and HeartMuLa as the family that releases codec and training procedure as well as weights~\parencite{gong2025acestep,yuan2025yue,yang2026heartmula}. The commercial systems of Table~\ref{tab:t2} are closed, and the tables do not compare their quality with that of the open ones. Preference alignment can also reduce output diversity, making diversity preservation a separate design concern~\parencite{cideron2024diversity}. Which signals the alignment reward should contain remains an open question. The contents of the plan are chosen per system, a melody token stream~\parencite{qwen2026music}, an editable score~\parencite{map2026yue2}, or a semantic sketch~\parencite{yang2025songbloom}, without a comparison of what each gives the renderer and the user.

\begin{table}[H]
  \caption{Song-generation systems (T2). Coverage and source dates are defined in Section~\ref{sec:method} and Appendix~\ref{app:search}. Legend as in Table~\ref{tab:t1}.}
  \label{tab:t2}
  \footnotesize
  \setlength{\tabcolsep}{2.5pt}
  \begin{tabular}{@{}>{\raggedright\arraybackslash}p{2.90cm}>{\raggedright\arraybackslash}p{8.50cm}@{}}
    \toprule
    System, D1 & Attributes (D3, D5) \\
    \midrule
    \multicolumn{2}{@{}l}{\emph{Autoregressive token language models}} \\
    \textbf{Chord-Conditioned Song Generator}\newline \parencite{gao2024end}\newline D1 Tok & \textit{Inputs} lyrics; chords; \textit{Duration} 12 s segments \textsc{gen}; \textit{Output} ?; \textit{Latency} batch, offline; \textit{Open} ?; \textit{Data} U 554k songs; \textit{Post-training} --. \\
    \textbf{SongCreator}\newline \parencite{lei2024songcreator}\newline D1 Tok & \textit{Inputs} lyrics; prompts; audio; \textit{Duration} up to 30 s \textsc{gen}; \textit{Output} 24 kHz tokens, 44.1 kHz render; \textit{Latency} batch, offline; \textit{Open} $\times$; \textit{Data} U 8.5k\,h; \textit{Post-training} --. \\
    \textbf{LeVo}\newline \parencite{lei2025levo}\newline D1 Tok & \textit{Inputs} lyrics; text; audio prompt; \textit{Duration} ?; \textit{Output} 48 kHz; \textit{Latency} batch, offline; \textit{Open} W; \textit{Data} P/U 110k\,h; \textit{Post-training} \textsc{pref}: multi-preference DPO. \\
    \textbf{SongGen}\newline \parencite{liu2025songgen}\newline D1 Tok & \textit{Inputs} lyrics; text; voice prompt; \textit{Duration} up to 30 s \textsc{gen}; \textit{Output} 16 kHz (X-Codec); \textit{Latency} batch, offline; \textit{Open} \checkmark; \textit{Data} P 2k\,h; \textit{Post-training} --. \\
    \textbf{YuE}\newline \parencite{yuan2025yue}\newline D1 Tok & \textit{Inputs} lyrics; structure tags; in-context audio; \textit{Duration} up to 5 min generated; about 163 s context window \textsc{gen}; \textit{Output} 16 kHz, upsampled to 44.1 kHz; \textit{Latency} batch, offline; \textit{Open} \checkmark; \textit{Data} U; \textit{Post-training} --. \\
    \bottomrule
  \end{tabular}
\end{table}

\begin{table}[H]
  \addtocounter{table}{-1}
  \caption{Song-generation systems (T2) (continued).}
  \footnotesize
  \setlength{\tabcolsep}{2.5pt}
  \begin{tabular}{@{}>{\raggedright\arraybackslash}p{2.90cm}>{\raggedright\arraybackslash}p{8.50cm}@{}}
    \toprule
    System, D1 & Attributes (D3, D5) \\
    \midrule
    \textbf{HeartMuLa}\newline \parencite{yang2026heartmula}\newline D1 Tok & \textit{Inputs} lyrics; tags (multilingual); \textit{Duration} up to 6 min \textsc{gen}; \textit{Output} 48 kHz stereo; \textit{Latency} batch, offline; \textit{Open} \checkmark; \textit{Data} U 100k\,h; \textit{Post-training} \textsc{pref}: DPO. \\
    \textbf{LeVo 2}\newline \parencite{lei2026levo2}\newline D1 Tok & \textit{Inputs} lyrics; text; \textit{Duration} full-length songs (length not stated) \textsc{gen}; \textit{Output} 48 kHz; \textit{Latency} batch, offline; \textit{Open} W; \textit{Data} U 500k\,h; \textit{Post-training} \textsc{sft/pref}: SFT, offline DPO, semi-online DPO. \\
    \textbf{Muse}\newline \parencite{jiang2026muse}\newline D1 Tok & \textit{Inputs} lyrics; style descriptions; \textit{Duration} about 4 min songs \textsc{gen}; \textit{Output} MuCodec tokens (rate not stated); \textit{Latency} batch, offline; \textit{Open} \checkmark; \textit{Data} L/Syn 7.8k\,h lic.; \textit{Post-training} --. \\
    \textbf{Shao}\newline \parencite{liu2026shao}\newline D1 Tok & \textit{Inputs} lyrics; duration; text; \textit{Duration} about 6 min (381 s at 21.5 Hz token rate) \textsc{gen}; \textit{Output} 44.1 kHz; \textit{Latency} batch, offline; \textit{Open} \checkmark; \textit{Data} U 1.2M\,h int.; \textit{Post-training} --. \\
    \multicolumn{2}{@{}l}{\emph{Latent and cascaded diffusion}} \\
    \textbf{MIDI-SAG}\newline \parencite{tsai2026midi}\newline D1 Lat & \textit{Inputs} vocal MIDI; chords; structure; \textit{Duration} 47 s single pass; 90 to 120 s with continuation \textsc{gen}; \textit{Output} ?; \textit{Latency} batch, offline; \textit{Open} \checkmark; \textit{Data} U 2.5k\,h; \textit{Post-training} --. \\
    \textbf{SongEcho}\newline \parencite{li2026songecho}\newline D1 Lat & \textit{Inputs} reference vocal melody; text; \textit{Duration} up to 240 s \textsc{gen}; \textit{Output} ?; \textit{Latency} batch, offline; \textit{Open} C; \textit{Data} Syn 3k\,h syn.; \textit{Post-training} --. \\
    \multicolumn{2}{@{}l}{\emph{Flow matching and rectified flow}} \\
    \textbf{ACE-Step}\newline \parencite{gong2025acestep}\newline D1 Lat & \textit{Inputs} lyrics; text; \textit{Duration} up to 4 min \textsc{gen}; \textit{Output} 44.1 kHz (32 kHz mono vocoder); \textit{Latency} batch, faster than real time; \textit{Open} \checkmark; \textit{Data} U 100k\,h; \textit{Post-training} --. \\
    \textbf{DiffRhythm}\newline \parencite{ning2025diffrhythm}\newline D1 Lat & \textit{Inputs} lyrics; style prompt; \textit{Duration} 4 min 45 s \textsc{gen}; \textit{Output} 44.1 kHz stereo; \textit{Latency} batch, faster than real time; \textit{Open} \checkmark; \textit{Data} U 60k\,h; \textit{Post-training} --. \\
    \textbf{DiffRhythm 2}\newline \parencite{jiang2025diffrhythm2}\newline D1 Lat & \textit{Inputs} lyrics; style; \textit{Duration} up to 210 s \textsc{gen}; \textit{Output} 48 kHz output (mono VAE); \textit{Latency} batch, faster than real time; \textit{Open} \checkmark; \textit{Data} U 70k\,h; \textit{Post-training} \textsc{pref}: cross-pair DPO. \\
    \textbf{DiffRhythm+}\newline \parencite{chen2025diffrhythmplus}\newline D1 Lat & \textit{Inputs} lyrics; style; \textit{Duration} over 4 min \textsc{gen}; \textit{Output} 44.1 kHz stereo; \textit{Latency} batch, faster than real time; \textit{Open} C; \textit{Data} U 120k\,h train.; 300k\,h coll.; \textit{Post-training} \textsc{pref}: DPO. \\
    \bottomrule
  \end{tabular}
\end{table}

\begin{table}[H]
  \addtocounter{table}{-1}
  \caption{Song-generation systems (T2) (continued).}
  \footnotesize
  \setlength{\tabcolsep}{2.5pt}
  \begin{tabular}{@{}>{\raggedright\arraybackslash}p{2.90cm}>{\raggedright\arraybackslash}p{8.50cm}@{}}
    \toprule
    System, D1 & Attributes (D3, D5) \\
    \midrule
    \textbf{Jam}\newline \parencite{liu2025jam}\newline D1 Lat & \textit{Inputs} lyrics (word-level timing); text; \textit{Duration} up to 3 min 50 s \textsc{gen}; \textit{Output} 44.1 kHz stereo; \textit{Latency} batch, offline; \textit{Open} ?; \textit{Data} U 54k\,h; \textit{Post-training} \textsc{pref}: DPO with SongEval scores. \\
    \textbf{VersBand}\newline \parencite{zhang2025versatile}\newline D1 Lat & \textit{Inputs} prompts (text, audio); \textit{Duration} phrases up to 20 s \textsc{gen}; \textit{Output} 24 kHz; \textit{Latency} batch, offline; \textit{Open} C; \textit{Data} P/U 1k\,h; \textit{Post-training} --. \\
    \textbf{MPEcho}\newline \parencite{lee2026mpecho}\newline D1 Lat & \textit{Inputs} reference melody; phonemes; text; \textit{Duration} ?; \textit{Output} ?; \textit{Latency} batch, offline; \textit{Open} \checkmark; \textit{Data} P 1.4k\,h; \textit{Post-training} --. \\
    \textbf{SegTune}\newline \parencite{wang2026segtune}\newline D1 Lat & \textit{Inputs} lyrics; segment prompts; \textit{Duration} 30 s to 6 min training data range; generation maximum not stated \textsc{ctx}; \textit{Output} 44 kHz (as written in the paper); \textit{Latency} batch, offline; \textit{Open} ?; \textit{Data} U 27k\,h pretrain.; 4k\,h ft.; \textit{Post-training} \textsc{pref}: iterative DPO. \\
    \textbf{UniSinger}\newline \parencite{zhang2026unified}\newline D1 Lat & \textit{Inputs} lyrics; speaker prompt; \textit{Duration} 15 to 20 s evaluation clips; generation length not stated \textsc{eval}; \textit{Output} 44.1 kHz; \textit{Latency} batch, offline; \textit{Open} ?; \textit{Data} U 30k\,h coll.; 20k\,h train.; 5k\,h accomp.; \textit{Post-training} --. \\
    \textbf{WanSong}\newline \parencite{chen2026wansong}\newline D1 Lat & \textit{Inputs} lyrics; text; \textit{Duration} up to 5 min \textsc{gen}; \textit{Output} 44.1 kHz stereo; \textit{Latency} batch, faster than real time; \textit{Open} ?; \textit{Data} U; \textit{Post-training} \textsc{pref}: RLHF (DPO and RL with reward models). \\
    \multicolumn{2}{@{}l}{\emph{Planner-renderer hybrids}} \\
    \textbf{MelodyLM}\newline \parencite{li2024accompanied}\newline D1 Hyb & \textit{Inputs} text; vocal prompt; \textit{Duration} sentences up to 30 s \textsc{gen}; \textit{Output} 16 kHz vocals, 24 kHz output; \textit{Latency} batch, offline; \textit{Open} $\times$; \textit{Data} P/U 1k\,h; \textit{Post-training} --. \\
    \textbf{Seed-Music}\newline \parencite{bai2024seedmusic}\newline D1 Hyb & \textit{Inputs} lyrics; style; reference vocal; notes; \textit{Duration} short clips and full-length tracks (length not stated) \textsc{gen}; \textit{Output} 44.1 kHz stereo; \textit{Latency} streaming; \textit{Open} $\times$; \textit{Data} U; \textit{Post-training} \textsc{pref}: RL post-training. \\
    \textbf{SongEditor}\newline \parencite{yang2024songeditor}\newline D1 Hyb & \textit{Inputs} lyrics; segment masks; 10 s acoustic prompt; source audio for editing; \textit{Duration} up to 2 min \textsc{gen}; \textit{Output} ?; \textit{Latency} batch, offline; \textit{Open} ?; \textit{Data} U 50k\,h; \textit{Post-training} --. \\
    \textbf{MusiCoT}\newline \parencite{lam2025musicot}\newline D1 Hyb & \textit{Inputs} lyrics; musical tags; text (chain-of-musical-thought sketch); \textit{Duration} ?; \textit{Output} 44.1 kHz (stereo latents); \textit{Latency} batch, offline; \textit{Open} $\times$; \textit{Data} U 10M songs; \textit{Post-training} --. \\
    \bottomrule
  \end{tabular}
\end{table}

\begin{table}[H]
  \addtocounter{table}{-1}
  \caption{Song-generation systems (T2) (continued).}
  \footnotesize
  \setlength{\tabcolsep}{2.5pt}
  \begin{tabular}{@{}>{\raggedright\arraybackslash}p{2.90cm}>{\raggedright\arraybackslash}p{8.50cm}@{}}
    \toprule
    System, D1 & Attributes (D3, D5) \\
    \midrule
    \textbf{SongBloom}\newline \parencite{yang2025songbloom}\newline D1 Hyb & \textit{Inputs} lyrics (structured); 10 s reference audio; text; \textit{Duration} up to 150 s \textsc{gen}; \textit{Output} 48 kHz stereo; \textit{Latency} batch, offline; \textit{Open} \checkmark; \textit{Data} U 100k\,h; \textit{Post-training} --. \\
    \textbf{ACE-Step 1.5}\newline \parencite{gong2026acestep15}\newline D1 Hyb & \textit{Inputs} lyrics; text; \textit{Duration} up to 10 min \textsc{gen}; \textit{Output} 48 kHz stereo; \textit{Latency} batch, faster than real time; \textit{Open} \checkmark; \textit{Data} U; \textit{Post-training} \textsc{pref}: intrinsic RL (GRPO, DiffusionNTF). \\
    \textbf{Dai et al.\ full-song framework}\newline \parencite{dai2026pushing}\newline D1 Hyb & \textit{Inputs} lyrics; text; attributes; cover; \textit{Duration} up to about 328 s \textsc{gen}; \textit{Output} 48 kHz stereo; \textit{Latency} batch, offline; \textit{Open} ?; \textit{Data} U; \textit{Post-training} \textsc{sft/pref}: SFT, DPO, GRPO, on-policy distillation. \\
    \textbf{Qwen-Music}\newline \parencite{qwen2026music}\newline D1 Hyb & \textit{Inputs} lyrics; text; melody tokens; \textit{Duration} renderer training samples up to 6 min; generation maximum not stated \textsc{ctx}; \textit{Output} 48 kHz stereo; \textit{Latency} batch, offline; \textit{Open} ?; \textit{Data} U; \textit{Post-training} \textsc{sft/pref}: SFT, offline DPO, online GSPO. \\
    \textbf{SketchSong}\newline \parencite{duan2026sketchsong}\newline D1 Hyb & \textit{Inputs} lyrics; text; \textit{Duration} song-level (length not stated) \textsc{gen}; \textit{Output} LeVo codec pipeline (rate not stated); \textit{Latency} batch, offline; \textit{Open} ?; \textit{Data} U 54k\,h; \textit{Post-training} --. \\
    \multicolumn{2}{@{}l}{\emph{Undisclosed architecture (products)}} \\
    \textbf{Suno (v3 2024-03 to v6 2026-09; observed 2026-09-14)}\newline \parencite{suno2026}\newline D1 ? & \textit{Inputs} lyrics; style text; audio, image, and video references, editing and stem operations (v6, product-reported); \textit{Duration} not documented per version (product); \textit{Output} not documented; \textit{Latency} batch, offline; \textit{Open} $\times$; \textit{Data} U; \textit{Post-training} --. \\
    \textbf{Udio (v1 2024-04 to v1.5 Allegro 2025-03; observed 2026-09-14)}\newline \parencite{udio2026}\newline D1 ? & \textit{Inputs} lyrics; style text; \textit{Duration} about 30 s per generation, extended in 30 s steps (product docs) \textsc{gen}; \textit{Output} not documented; \textit{Latency} batch, offline; \textit{Open} $\times$; \textit{Data} U; \textit{Post-training} --. \\
    \multicolumn{2}{@{}l}{\emph{Other or not stated}} \\
    \textbf{Two-stage text-to-song}\newline \parencite{hong2024texttosong}\newline D1 Tok & \textit{Inputs} text; lyrics; \textit{Duration} 10 s \textsc{gen}; \textit{Output} 16 kHz mono; \textit{Latency} batch, offline; \textit{Open} ?; \textit{Data} P/U 180\,h; \textit{Post-training} --. \\
    \bottomrule
  \end{tabular}
\end{table}

\FloatBarrier

\chapter{Control, Editing, and Co-Creation}
\label{sec:control}

Production use (class T3) requires more than a prompt: time-varying control, targeted edits, and stem-level operations. A request such as ``change the instrument but keep the melody'' specifies a change and a constraint, and a useful method must satisfy both. This section compares how controls enter a generator, how an existing recording is represented during editing, and where the cost is paid: in additional training, in per-edit computation, or in limits on what the model can change. Table~\ref{tab:t3} lists the control, editing, stem, and live systems of this section and of Section~\ref{sec:realtime}.

\section{Time-Varying Control}
\label{sec:control-timevarying}

Time-varying controls specify how a piece should change over time: a chord sequence, a melody line, or a dynamics curve. A generator trained with text conditioning aligns such a signal with its audio output through an additional mechanism, and the first approach learns that mechanism during training. Music ControlNet adapted the ControlNet design to music: a control branch receives the time-varying signal and feeds the diffusion generator, giving melody, dynamics, and rhythm controls~\parencite{wu2024musiccontrolnet}. For token language models, Coco-Mulla adds chord and drum conditioning with a parameter-efficient adapter on a frozen backbone~\parencite{lin2023cocomulla}, MusiConGen reaches practical rhythm and chord control by fine-tuning MusicGen~\parencite{lan2024musicongen}, and a heterogeneous adapter with a masking scheme gives MusicGen inpainting together with frame-level content controls~\parencite{lin2024arrange}. JASCO conditions flow matching jointly on symbolic signals and audio stems with temporal precision~\parencite{tal2024jasco}, and Mustango takes a different route to the same end, providing control through the prompt itself by predicting chords, beats, tempo, and key from rich captions~\parencite{melechovsky2024mustango}. Several designs reduce the additional training needed to support a control. LiLAC replaces ControlNet-style branches with lightweight modules~\parencite{baker2025lilac}, and MuseControlLite reaches multiple conditions with small conditioners built on positionally enhanced cross attention~\parencite{tsai2025musecontrollite}. AP-Adapter adds audio-prompt control to a frozen diffusion model~\parencite{tsai2024apadapter}, and adapter studies compare architecture, placement, and size for low-resource adaptation~\parencite{mehta2025exploring,zou2024teadapter}.

The second approach keeps the pretrained generator fixed: DITTO optimizes the initial diffusion noise at inference time, by gradient descent through the sampler, to improve agreement with an intensity, melody, or structure target~\parencite{novack2024ditto}, and DITTO-2 reduces the cost of this optimization by separately training a distilled sampler~\parencite{novack2024ditto2}. Sketch2Sound extends the interface to vocal imitations and sketched control curves~\parencite{garcia2025sketch2sound}, and LARA-Gen adds continuous valence-arousal emotion control by aligning hidden states with a music understanding model~\parencite{mei2025lara}. A third approach modifies generation without updating the generator's weights or optimizing each output: once a concept direction has been identified inside a pretrained generator, activation steering applies it directly, although finding the direction may itself require training a sparse autoencoder or an activation-patching analysis~\parencite{singh2025discovering,facchiano2025activation,narashiman2025genre,zhao2025steering}.

\paragraph{Running example.} Several MusicGen variants illustrate the different costs of adding control. Coco-Mulla keeps MusicGen frozen and trains an adapter that reads chords and drums~\parencite{lin2023cocomulla}. MusiConGen fine-tunes the backbone itself for rhythm and chords~\parencite{lan2024musicongen}. Instruct-MusicGen adds instruction following to it~\parencite{zhang2024instructmusicgen}. A heterogeneous adapter gives it inpainting together with frame-level controls~\parencite{lin2024arrange}. MusicGen's melody-conditioned variant accepts a chromagram as a control input. The variants differ in where they introduce the control sequence and in which parameters they update: the adapter methods keep MusicGen frozen, and the adapter itself still needs a training run, which is the cost the later lightweight conditioners of this section reduce.

A global text prompt can ask for a chorus or a returning motif but gives little direct control over their timing, and two recent systems provide more explicit structural control by generating a plan before the audio. MusicLayout generates a time-aligned layout of sections, textures, repetitions, and arrangement before the audio~\parencite{li2026musiclayout}, and MusicWeaver predicts a multi-level song program encoding form, motif recurrence, and bar-level attributes, with edits formalized as typed operations on the program~\parencite{wang2025musicweaver}. Both are planner-renderer systems in the sense of Section~\ref{sec:d2}, with a human-readable plan.

\section{Editing and Instruction Following}
\label{sec:control-editing}

Editing an existing recording, rather than generating a new one, is approached by three families of methods, which differ in how they obtain the requested change while preserving the source recording. Some learn editing from examples or train additional modules. Others use a pretrained generator to reconstruct or transform the input at inference, by inversion into the model's generation process, by a directly constructed editing trajectory, or by an intervention on attention. The routes overlap, so the comparison below states what is trained, how the source enters, and what is changed during generation. The first family learns editing from examples: instruction-tuned models are trained on triplets of a recording, an instruction, and the edited result. AUDIT introduced the formulation for general audio~\parencite{wang2023audit}, InstructME frames editing and remixing as instruction following over latent diffusion, covering stem addition, removal, and replacement~\parencite{han2023instructme}, and Instruct-MusicGen brings instruction tuning to token language models~\parencite{zhang2024instructmusicgen}.

A second family uses inversion. The recording is mapped back to the latent or the noise from which a pretrained generator would have produced it. Generation is then steered toward the requested edit. These methods can use a pretrained generator without task-specific editor training. MusicMagus changes instrument, genre, or mood by steering latents~\parencite{zhang2024musicmagus}. DDPM inversion gives text-based and unsupervised editing directions~\parencite{manor2024zero}. MEDIC disentangles the inversion to reduce the error that accumulates when a recording is inverted and regenerated~\parencite{liu2024medic}. MelodyFlow adapts latent inversion to flow matching over a stereo latent codec~\parencite{lan2024high}. MusRec and FlowSonic invert real recordings with rectified-flow diffusion transformers~\parencite{boudaghi2025musrec,boudaghi2026flowsonic}. SteerMusic uses score distillation for consistency~\parencite{niu2025steermusic}. Melodia and AnchorSteer intervene on attention maps and on self-discovered concept vectors to preserve rhythm and melody while changing attributes~\parencite{yang2025melodia,chang2026anchorsteer}.

A third family edits without updating the generator's weights or first inverting the recording. Inversion-free methods construct the editing path directly from the difference between source- and target-conditioned predictions~\parencite{fu2026echoedit,ge2026directaudioedit}, and EditGen steers the cross-attention of a frozen autoregressive model~\parencite{sioros2025editgen}. Two capabilities cut across the families. Inpainting-capable generators cover targeted regeneration and continuation directly, by regenerating only a masked region~\parencite{garcia2023vampnet,evans2026stableaudio3}, and similarity-guided posterior sampling extends inpainting to multi-second gaps~\parencite{turland2025similarity}. What an edit should leave unchanged has its own evaluation framework~\parencite{vishe2025evaluating}, and instruction-based editing across audio modalities has a benchmark~\parencite{ma2026mmae}.

\section{Stems and Accompaniment}
\label{sec:control-stems}

A stem is an audio track containing one instrument, voice, or group of parts from a mix. Stem-level generation asks a model to produce one part given the others, or all parts so that they can be mixed separately. Early examples include SingSong's vocal-conditioned accompaniment~\parencite{donahue2023singsong} and StemGen's context-aware stems~\parencite{parker2024stemgen}. Later systems approach the task in three ways. The first design generates and edits named stems: MusicGen-Stem extends one autoregressive model to drums, bass, and other stems~\parencite{rouard2025musicgenstem}, and JEN-1 Composer generates or completes multiple tracks jointly on a diffusion stack~\parencite{yao2025jen1composer}. The second treats the stems as parallel channels of one distribution, so that generation, separation, and accompaniment are all the same conditional inference. Multi-source diffusion unifies generation, source separation, and partial accompaniment over parallel stems~\parencite{mariani2024msdm}, and Multi-Track MusicLDM extends a latent diffusion model to arrangement-aware multi-track generation~\parencite{karchkhadze2024multi}. MGE-LDM learns a joint latent distribution over mixtures, submixtures, and stems, so that generation, imputation, and text-conditioned extraction are all conditional inpainting~\parencite{chae2025mge}. The third design generates one accompaniment track conditioned on the remaining mix, as in Diff-A-Riff~\parencite{nistal2024diffariff}. Diff-A-Riff is trained on licensed data~\parencite{nistal2024diffariff}. The three designs differ in the data they need. In Table~\ref{tab:t3}, the systems that treat stems as parallel channels are trained on public research datasets, and the named-stem and mix-conditioned systems on in-house or licensed catalogs, so the design a reader can reproduce depends on the multi-track data available.

Beyond generation, the production steps themselves are being modeled: Diff2Mix controls mixing with diffusion models and differentiable effects~\parencite{zong2026diffmix}, and RIME formalizes agentic post-production, in which individual aspects of a song are targeted and refined in sequence~\parencite{schaffer2026rime}. All stems of a track can also be generated at once with flexible conditioning~\parencite{wu2026stemphonic}, hierarchical autoregressive foundation models target accompaniment generation as a task in its own right~\parencite{zhu2026hafm}, and MIDI-to-music synthesis with optional prompt MIDI unifies rendering and local editing~\parencite{jing2026p}.

A language model can also sit above these tools and call them in sequence on the user's behalf. M2UGen connects music encoders and generators to a large language model for chat-style understanding and generation~\parencite{hussain2023m2ugen}, the T6 pattern of Section~\ref{sec:taxonomy}. MusiChat separates lyric-aligned structure generation from audio generation so that a conversation can evolve a piece rather than regenerate it~\parencite{liao2026musichat}. Loop Copilot uses a language model to sequence specialized models in an iterative generate-and-edit loop~\parencite{zhang2023loop,zhang2024improving}. Section~\ref{sec:crossmodal-symbolic} returns to these systems as the unified class.

\section{Lessons and Open Trade-offs}
\label{sec:t3-lessons}

Adding control requires choices about what to train, what to compute for each edit, and how closely to preserve the source. Five routes are reviewed above:
\begin{itemize}
  \item conditioning added at training time, through a control branch~\parencite{wu2024musiccontrolnet}, an adapter on a frozen backbone~\parencite{lin2023cocomulla}, fine-tuning~\parencite{lan2024musicongen}, or instruction triplets~\parencite{han2023instructme,zhang2024instructmusicgen};
  \item optimization at inference time against a target~\parencite{novack2024ditto};
  \item inversion of a recording into the generator's latent~\parencite{zhang2024musicmagus,liu2024medic}, or an editing path that needs no inversion~\parencite{fu2026echoedit,ge2026directaudioedit};
  \item an explicit plan between intent and audio~\parencite{li2026musiclayout,wang2025musicweaver};
  \item steering of concepts found inside the model~\parencite{singh2025discovering,facchiano2025activation}.
\end{itemize}
The routes trade additional training against inference cost and against preservation of the source. Control branches and adapters learn a new conditioning signal, and some keep the generator fixed while others update it. The cost of this route has fallen from a full control branch to lightweight modules~\parencite{baker2025lilac,tsai2025musecontrollite}. Inference-time optimization avoids parameter updates and adds computation to every output, which distillation made affordable~\parencite{novack2024ditto2}. Inversion-based and inversion-free editors differ in how they retain the source. Later methods seek to reduce errors introduced during inversion and regeneration~\parencite{liu2024medic}. Explicit plans make structure over minutes easier to specify and revise. The papers report different training costs, inference costs, and preservation measures, and their protocols do not support a ranking: one framework defines preservation~\parencite{vishe2025evaluating}, one benchmark covers instruction-based editing across audio modalities~\parencite{ma2026mmae}, and otherwise each paper measures its own control against its own target. For a reader with a constraint, the routes sort as follows. An adapter adds a control to a token generator with its weights fixed, although the adapter itself is trained. A control branch or inference-time optimization adds one to a diffusion generator. Inversion-based or inversion-free editing uses a pretrained generator without additional training for the edit. A plan is useful when the user must specify or revise structure over minutes. The five routes have not been compared under one protocol, and Section~\ref{sec:discussion} returns to this gap. How best to support multiple controls within a common training and evaluation framework remains open.

\begin{table}[H]
  \caption{Control, editing, stem, and live systems (T3 and T4). Coverage and source dates are defined in Section~\ref{sec:method} and Appendix~\ref{app:search}. Legend as in Table~\ref{tab:t1}.}
  \label{tab:t3}
  \footnotesize
  \setlength{\tabcolsep}{2.5pt}
  \begin{tabular}{@{}>{\raggedright\arraybackslash}p{2.90cm}>{\raggedright\arraybackslash}p{8.50cm}@{}}
    \toprule
    System, D1 & Attributes (D3, D5) \\
    \midrule
    \multicolumn{2}{@{}l}{\emph{Autoregressive token language models}} \\
    \textbf{Coco-Mulla}\newline \parencite{lin2023cocomulla}\newline D1 Tok & \textit{Inputs} chords; drums; \textit{Duration} 20 s \textsc{gen}; \textit{Output} 32 kHz (MusicGen EnCodec backbone); \textit{Latency} batch, offline; \textit{Open} C; \textit{Data} U 17\,h; \textit{Post-training} --. \\
    \textbf{SingSong}\newline \parencite{donahue2023singsong}\newline D1 Tok & \textit{Inputs} input vocals; \textit{Duration} 10 s clips \textsc{eval}; \textit{Output} 16 kHz; \textit{Latency} batch, offline; \textit{Open} $\times$; \textit{Data} U 46k\,h; \textit{Post-training} --. \\
    \textbf{Arrange-Inpaint-Refine}\newline \parencite{lin2024arrange}\newline D1 Tok & \textit{Inputs} frame-level content controls; score; \textit{Duration} 15 s \textsc{gen}; \textit{Output} 32 kHz (MusicGen EnCodec backbone); \textit{Latency} batch, offline; \textit{Open} ?; \textit{Data} P 1,289 songs; \textit{Post-training} --. \\
    \textbf{Instruct-MusicGen}\newline \parencite{zhang2024instructmusicgen}\newline D1 Tok & \textit{Inputs} instruction; audio; \textit{Duration} 5 s \textsc{gen}; \textit{Output} 32 kHz (MusicGen EnCodec backbone); \textit{Latency} batch, offline; \textit{Open} R; \textit{Data} P/Syn; \textit{Post-training} --. \\
    \textbf{MusiConGen}\newline \parencite{lan2024musicongen}\newline D1 Tok & \textit{Inputs} rhythm; chords; \textit{Duration} 30 s \textsc{gen}; \textit{Output} 32 kHz (EnCodec 32k); \textit{Latency} batch, offline; \textit{Open} \checkmark; \textit{Data} U 250\,h; \textit{Post-training} --. \\
    \bottomrule
  \end{tabular}
\end{table}

\begin{table}[H]
  \addtocounter{table}{-1}
  \caption{Control, editing, stem, and live systems (T3 and T4) (continued).}
  \footnotesize
  \setlength{\tabcolsep}{2.5pt}
  \begin{tabular}{@{}>{\raggedright\arraybackslash}p{2.90cm}>{\raggedright\arraybackslash}p{8.50cm}@{}}
    \toprule
    System, D1 & Attributes (D3, D5) \\
    \midrule
    \textbf{Lyria RealTime}\newline \parencite{caillon2025live}\newline D1 Tok & \textit{Inputs} text; audio prompts, continuously; \textit{Duration} arbitrary; 2 s chunks, 10 s context \textsc{stream}; \textit{Output} 48 kHz stereo (SpectroStream); \textit{Latency} streaming (generates at least as fast as playback); \textit{Open} $\times$; \textit{Data} ?; \textit{Post-training} --. \\
    \textbf{Magenta RealTime}\newline \parencite{caillon2025live}\newline D1 Tok & \textit{Inputs} text; audio prompts, continuously; \textit{Duration} arbitrary; 2 s chunks, 10 s context \textsc{stream}; \textit{Output} 48 kHz stereo (SpectroStream); \textit{Latency} streaming (generates at least as fast as playback); \textit{Open} \checkmark; \textit{Data} U 190k\,h; \textit{Post-training} --. \\
    \textbf{MusicGen-Stem}\newline \parencite{rouard2025musicgenstem}\newline D1 Tok & \textit{Inputs} text; stems; \textit{Duration} 30 s \textsc{gen}; \textit{Output} 32 kHz mono per stem; \textit{Latency} batch, offline; \textit{Open} \checkmark; \textit{Data} U 17k\,h; \textit{Post-training} --. \\
    \textbf{Streaming accompaniment (Wu et al.)}\newline \parencite{wu2025streaming}\newline D1 Tok & \textit{Inputs} live audio input; \textit{Duration} streaming; 10 s training windows \textsc{ctx}; \textit{Output} 32 kHz (DAC); \textit{Latency} streaming (future-visibility and chunk trade-off); \textit{Open} \checkmark; \textit{Data} P 145\,h; \textit{Post-training} --. \\
    \multicolumn{2}{@{}l}{\emph{Masked and parallel decoding}} \\
    \textbf{VampNet}\newline \parencite{garcia2023vampnet}\newline D1 Tok & \textit{Inputs} audio prompt; mask pattern (no text); \textit{Duration} 10 s samples \textsc{eval}; \textit{Output} 44.1 kHz; \textit{Latency} batch, faster than real time; \textit{Open} \checkmark; \textit{Data} U 797k tracks; \textit{Post-training} --. \\
    \textbf{StemGen}\newline \parencite{parker2024stemgen}\newline D1 Tok & \textit{Inputs} context mix; instrument category; \textit{Duration} 20 s training crops from 1 min segments \textsc{ctx}; \textit{Output} 32 kHz; \textit{Latency} batch, offline; \textit{Open} $\times$; \textit{Data} L/P/Syn/U 145\,h syn.; 500\,h lic.; \textit{Post-training} --. \\
    \multicolumn{2}{@{}l}{\emph{Latent and cascaded diffusion}} \\
    \textbf{AUDIT}\newline \parencite{wang2023audit}\newline D1 Lat & \textit{Inputs} source audio; instruction; \textit{Duration} 10 s \textsc{gen}; \textit{Output} 16 kHz; \textit{Latency} batch, offline; \textit{Open} ?; \textit{Data} P; \textit{Post-training} --. \\
    \textbf{InstructME}\newline \parencite{han2023instructme}\newline D1 Lat & \textit{Inputs} instruction; stems; \textit{Duration} 10 s (30 to 60 s tested) \textsc{eval}; \textit{Output} 24 kHz; \textit{Latency} batch, offline; \textit{Open} $\times$; \textit{Data} U 417\,h int.; \textit{Post-training} --. \\
    \textbf{AP-Adapter}\newline \parencite{tsai2024apadapter}\newline D1 Lat & \textit{Inputs} audio prompt; \textit{Duration} 10 s \textsc{gen}; \textit{Output} 16 kHz (AudioLDM 2); \textit{Latency} batch, offline; \textit{Open} \checkmark; \textit{Data} P 500\,h; \textit{Post-training} --. \\
    \textbf{Control and style transfer (Demerlé et al.)}\newline \parencite{demerle2024combining}\newline D1 Lat & \textit{Inputs} audio structure; reference audio (timbre, style); \textit{Duration} ?; \textit{Output} RAVE-based codec (rate not stated); \textit{Latency} batch, faster than real time; \textit{Open} ?; \textit{Data} P/Syn 400\,h; 4\,h; \textit{Post-training} --. \\
    \bottomrule
  \end{tabular}
\end{table}

\begin{table}[H]
  \addtocounter{table}{-1}
  \caption{Control, editing, stem, and live systems (T3 and T4) (continued).}
  \footnotesize
  \setlength{\tabcolsep}{2.5pt}
  \begin{tabular}{@{}>{\raggedright\arraybackslash}p{2.90cm}>{\raggedright\arraybackslash}p{8.50cm}@{}}
    \toprule
    System, D1 & Attributes (D3, D5) \\
    \midrule
    \textbf{Diff-A-Riff}\newline \parencite{nistal2024diffariff}\newline D1 Lat & \textit{Inputs} mix context; text; \textit{Duration} 10 s windows; 95 s generated in 8 s (A100) \textsc{gen}; \textit{Output} 48 kHz pseudo-stereo; \textit{Latency} batch, faster than real time; \textit{Open} $\times$; \textit{Data} L 12k recordings; \textit{Post-training} --. \\
    \textbf{Diff-A-Riff 2}\newline \parencite{nistal2024improving}\newline D1 Lat & \textit{Inputs} context mix; text via embedding translation; reference audio; \textit{Duration} 10 s training windows \textsc{ctx}; \textit{Output} 48 kHz stereo; \textit{Latency} batch, faster than real time; \textit{Open} ?; \textit{Data} U 20,000 recordings; \textit{Post-training} --. \\
    \textbf{FastSAG}\newline \parencite{chen2024fastsag}\newline D1 Lat & \textit{Inputs} input vocals; \textit{Duration} 10 s \textsc{gen}; \textit{Output} 24 kHz; \textit{Latency} batch, faster than real time; \textit{Open} C; \textit{Data} U 3k\,h; \textit{Post-training} --. \\
    \textbf{MSDM}\newline \parencite{mariani2024msdm}\newline D1 Wav & \textit{Inputs} stems (separation, imputation); \textit{Duration} about 12 s \textsc{gen}; \textit{Output} 22 kHz (44 kHz variant); \textit{Latency} batch, offline; \textit{Open} \checkmark; \textit{Data} P; \textit{Post-training} --. \\
    \textbf{Multi-Track MusicLDM}\newline \parencite{karchkhadze2024multi}\newline D1 Lat & \textit{Inputs} text; tracks; \textit{Duration} 10.24 s segments \textsc{gen}; \textit{Output} 16 kHz; \textit{Latency} batch, offline; \textit{Open} \checkmark; \textit{Data} P; \textit{Post-training} --. \\
    \textbf{Music ControlNet}\newline \parencite{wu2024musiccontrolnet}\newline D1 Lat & \textit{Inputs} melody; dynamics; rhythm; \textit{Duration} 6 s (up to 24 s) \textsc{gen}; \textit{Output} 22.05 kHz mel; \textit{Latency} batch, offline; \textit{Open} $\times$; \textit{Data} L 1.8k\,h lic.; \textit{Post-training} --. \\
    \textbf{MusicMagus}\newline \parencite{zhang2024musicmagus}\newline D1 Lat & \textit{Inputs} source audio; text edit; \textit{Duration} 5 s \textsc{gen}; \textit{Output} 16 kHz (AudioLDM 2); \textit{Latency} batch, offline; \textit{Open} \checkmark/\checkmark; \textit{Data} --; \textit{Post-training} --. \\
    \textbf{Mustango}\newline \parencite{melechovsky2024mustango}\newline D1 Lat & \textit{Inputs} text with chords, beats, tempo, key; \textit{Duration} 10 s \textsc{gen}; \textit{Output} 16 kHz; \textit{Latency} batch, offline; \textit{Open} \checkmark; \textit{Data} CC; \textit{Post-training} --. \\
    \textbf{JEN-1 Composer}\newline \parencite{yao2025jen1composer}\newline D1 Lat & \textit{Inputs} text; tracks; \textit{Duration} ?; \textit{Output} 48 kHz stereo; \textit{Latency} batch, offline; \textit{Open} ?; \textit{Data} U 640\,h; \textit{Post-training} --. \\
    \textbf{LiLAC}\newline \parencite{baker2025lilac}\newline D1 Lat & \textit{Inputs} time-varying controls; \textit{Duration} 10 s \textsc{gen}; \textit{Output} 48 kHz codec; \textit{Latency} batch, offline; \textit{Open} $\times$; \textit{Data} L/P; \textit{Post-training} --. \\
    \bottomrule
  \end{tabular}
\end{table}

\begin{table}[H]
  \addtocounter{table}{-1}
  \caption{Control, editing, stem, and live systems (T3 and T4) (continued).}
  \footnotesize
  \setlength{\tabcolsep}{2.5pt}
  \begin{tabular}{@{}>{\raggedright\arraybackslash}p{2.90cm}>{\raggedright\arraybackslash}p{8.50cm}@{}}
    \toprule
    System, D1 & Attributes (D3, D5) \\
    \midrule
    \textbf{MGE-LDM}\newline \parencite{chae2025mge}\newline D1 Lat & \textit{Inputs} text; stems; \textit{Duration} 10.24 s \textsc{gen}; \textit{Output} 16 kHz mono; \textit{Latency} batch, offline; \textit{Open} \checkmark; \textit{Data} P; \textit{Post-training} --. \\
    \textbf{MuseControlLite}\newline \parencite{tsai2025musecontrollite}\newline D1 Lat & \textit{Inputs} melody; dynamics; audio; \textit{Duration} 47 s \textsc{gen}; \textit{Output} 44.1 kHz stereo (Stable Audio Open); \textit{Latency} batch, offline; \textit{Open} \checkmark; \textit{Data} P; \textit{Post-training} --. \\
    \textbf{Sketch2Sound}\newline \parencite{garcia2025sketch2sound}\newline D1 Lat & \textit{Inputs} vocal imitation; control curves; \textit{Duration} up to 5 s \textsc{gen}; \textit{Output} 48 kHz mono; \textit{Latency} batch, offline; \textit{Open} ?; \textit{Data} L/CC/U; \textit{Post-training} --. \\
    \multicolumn{2}{@{}l}{\emph{Flow matching and rectified flow}} \\
    \textbf{MelodyFlow}\newline \parencite{lan2024high}\newline D1 Lat & \textit{Inputs} source audio; text edit (inversion); \textit{Duration} up to 30 s \textsc{gen}; \textit{Output} 48 kHz stereo; \textit{Latency} batch, faster than real time; \textit{Open} \checkmark; \textit{Data} L/U 20k\,h int.; \textit{Post-training} --. \\
    \textbf{MusRec}\newline \parencite{boudaghi2025musrec}\newline D1 Lat & \textit{Inputs} source audio; text edit; \textit{Duration} 10 s \textsc{gen}; \textit{Output} 16 kHz; \textit{Latency} batch, offline; \textit{Open} ?/\checkmark; \textit{Data} --; \textit{Post-training} --. \\
    \textbf{Live Music Diffusion Models}\newline \parencite{novack2026lmdm}\newline D1 Lat & \textit{Inputs} text, continuously; \textit{Duration} up to 2 min evaluated \textsc{eval}; \textit{Output} stereo (rate not stated); \textit{Latency} streaming (delay under 1 s (8 steps)); \textit{Open} \checkmark; \textit{Data} P; \textit{Post-training} \textsc{adv}: adversarial post-training (ARC-Forcing). \\
    \multicolumn{2}{@{}l}{\emph{Adversarial continuous generators}} \\
    \textbf{LiveBand}\newline \parencite{pasini2026liveband}\newline D1 Lat & \textit{Inputs} live audio input; \textit{Duration} streaming (12 s training crops) \textsc{stream}; \textit{Output} 44.1 kHz stereo; \textit{Latency} streaming (real time, about 93 ms anticipation); \textit{Open} ?; \textit{Data} P/U; \textit{Post-training} --. \\
    \multicolumn{2}{@{}l}{\emph{Planner-renderer hybrids}} \\
    \textbf{MusicWeaver}\newline \parencite{wang2025musicweaver}\newline D1 Hyb & \textit{Inputs} source audio; song program; edits; \textit{Duration} 60 to 180 s evaluated \textsc{eval}; \textit{Output} 24 kHz; \textit{Latency} batch, offline; \textit{Open} ?; \textit{Data} U 610\,h; \textit{Post-training} --. \\
    \textbf{MusicLayout}\newline \parencite{li2026musiclayout}\newline D1 Hyb & \textit{Inputs} text; layout; \textit{Duration} layouts up to 240 s; 20 s windows with continuation \textsc{ctx}; \textit{Output} 44.1 kHz mono; \textit{Latency} batch, offline; \textit{Open} C; \textit{Data} P; \textit{Post-training} --. \\
    \multicolumn{2}{@{}l}{\emph{Inference-time optimization and steering}} \\
    \textbf{DITTO}\newline \parencite{novack2024ditto}\newline D1 Lat & \textit{Inputs} intensity; melody; structure; \textit{Duration} 6 s (24 s variant) \textsc{gen}; \textit{Output} 22.05 kHz mel; \textit{Latency} batch, offline; \textit{Open} $\times$/$\times$; \textit{Data} --; \textit{Post-training} --. \\
    \bottomrule
  \end{tabular}
\end{table}

\begin{table}[H]
  \addtocounter{table}{-1}
  \caption{Control, editing, stem, and live systems (T3 and T4) (continued).}
  \footnotesize
  \setlength{\tabcolsep}{2.5pt}
  \begin{tabular}{@{}>{\raggedright\arraybackslash}p{2.90cm}>{\raggedright\arraybackslash}p{8.50cm}@{}}
    \toprule
    System, D1 & Attributes (D3, D5) \\
    \midrule
    \textbf{DITTO-2}\newline \parencite{novack2024ditto2}\newline D1 Lat & \textit{Inputs} as DITTO; \textit{Duration} 6 s \textsc{gen}; \textit{Output} 22.05 kHz mel; \textit{Latency} batch, mixed; \textit{Open} $\times$/$\times$; \textit{Data} L 1.8k\,h lic.; \textit{Post-training} \textsc{distill}: consistency and consistency-trajectory distillation without the GAN loss (32 GPU hours). \\
    \textbf{MEDIC}\newline \parencite{liu2024medic}\newline D1 Lat & \textit{Inputs} source audio; text edit; \textit{Duration} 10 s (ZoME-Bench clips) \textsc{eval}; \textit{Output} 16 kHz (benchmark clips); \textit{Latency} batch, offline; \textit{Open} C/\checkmark; \textit{Data} --; \textit{Post-training} --. \\
    \textbf{ZETA / ZEUS}\newline \parencite{manor2024zero}\newline D1 Lat & \textit{Inputs} source audio; text edit; unsupervised directions; \textit{Duration} 8 to 10 s \textsc{gen}; \textit{Output} 16 kHz (AudioLDM 2); \textit{Latency} batch, offline; \textit{Open} \checkmark/\checkmark; \textit{Data} --; \textit{Post-training} --. \\
    \textbf{Melodia}\newline \parencite{yang2025melodia}\newline D1 Lat & \textit{Inputs} source audio; text edit; \textit{Duration} ?; \textit{Output} 16 kHz (AudioLDM 2) or 44.1 kHz (Stable Audio Open); \textit{Latency} batch, offline; \textit{Open} ?/\checkmark; \textit{Data} --; \textit{Post-training} --. \\
    \textbf{SteerMusic}\newline \parencite{niu2025steermusic}\newline D1 Lat & \textit{Inputs} source audio; text edit; concept; \textit{Duration} 10 s \textsc{gen}; \textit{Output} 16 kHz (backbone); \textit{Latency} batch, offline; \textit{Open} C/\checkmark; \textit{Data} --; \textit{Post-training} --. \\
    \textbf{AnchorSteer}\newline \parencite{chang2026anchorsteer}\newline D1 Lat & \textit{Inputs} source audio; text edit; concept vectors; \textit{Duration} 47 s \textsc{gen}; \textit{Output} 44.1 kHz (Stable Audio Open); \textit{Latency} batch, offline; \textit{Open} \checkmark/\checkmark; \textit{Data} --; \textit{Post-training} --. \\
    \textbf{Diffusion timbre transfer (Lee et al.)}\newline \parencite{lee2026diffusion}\newline D1 Lat & \textit{Inputs} source audio; reference audio (timbre); \textit{Duration} 10 s evaluation clips \textsc{eval}; \textit{Output} 48 kHz stereo (Diff-A-Riff autoencoder, 11.7 Hz latent); \textit{Latency} batch, offline; \textit{Open} ?/$\times$; \textit{Data} --; \textit{Post-training} --. \\
    \textbf{EchoEdit}\newline \parencite{fu2026echoedit}\newline D1 Lat & \textit{Inputs} source audio; text edit; \textit{Duration} 30 s (15 to 120 s) \textsc{gen}; \textit{Output} 44.1 kHz (Stable Audio 3 medium); \textit{Latency} batch, offline; \textit{Open} C/\checkmark; \textit{Data} --; \textit{Post-training} --. \\
    \textbf{FlowSonic}\newline \parencite{boudaghi2026flowsonic}\newline D1 Lat & \textit{Inputs} source audio; text edit; \textit{Duration} 10 s \textsc{gen}; \textit{Output} 16 kHz; \textit{Latency} batch, offline; \textit{Open} \checkmark/\checkmark; \textit{Data} --; \textit{Post-training} --. \\
    \multicolumn{2}{@{}l}{\emph{Language-model agents and tool routing}} \\
    \textbf{M2UGen}\newline \parencite{hussain2023m2ugen}\newline D1 -- & \textit{Inputs} dialogue; images; video; \textit{Duration} ?; \textit{Output} ?; \textit{Latency} batch, offline; \textit{Open} \checkmark; \textit{Data} U 55.7\,h; \textit{Post-training} --. \\
    \bottomrule
  \end{tabular}
\end{table}

\begin{table}[H]
  \addtocounter{table}{-1}
  \caption{Control, editing, stem, and live systems (T3 and T4) (continued).}
  \footnotesize
  \setlength{\tabcolsep}{2.5pt}
  \begin{tabular}{@{}>{\raggedright\arraybackslash}p{2.90cm}>{\raggedright\arraybackslash}p{8.50cm}@{}}
    \toprule
    System, D1 & Attributes (D3, D5) \\
    \midrule
    \textbf{RIME}\newline \parencite{schaffer2026rime}\newline D1 ? & \textit{Inputs} natural-language refinement; \textit{Duration} --; \textit{Output} --; \textit{Latency} batch, offline; \textit{Open} ?; \textit{Data} U 37\,h train.; \textit{Post-training} --. \\
    \multicolumn{2}{@{}l}{\emph{Autoencoder instruments and latent-space synthesis}} \\
    \textbf{Continuous descriptor control (Devis et al.)}\newline \parencite{devis2023continuous}\newline D1 Lat & \textit{Inputs} continuous audio descriptors; \textit{Duration} --; \textit{Output} up to 48 kHz (RAVE); training sets at 16 and 48 kHz; \textit{Latency} streaming (real time; embedded in a hardware synthesizer); \textit{Open} C; \textit{Data} P/U 2\,h; \textit{Post-training} --. \\
    \textbf{Low-latency RAVE synthesizer (Caspe et al.)}\newline \parencite{caspe2025designing}\newline D1 Lat & \textit{Inputs} live audio input; \textit{Duration} --; \textit{Output} 44.1 kHz; \textit{Latency} streaming (real time; latency and jitter sources analyzed); \textit{Open} \checkmark; \textit{Data} P 2\,h; \textit{Post-training} --. \\
    \multicolumn{2}{@{}l}{\emph{Other or not stated}} \\
    \textbf{GAN post-training for live jamming}\newline \parencite{wu2025generative}\newline D1 Sym & \textit{Inputs} live interaction; \textit{Duration} --; \textit{Output} --; \textit{Latency} streaming (real time (user study with musicians)); \textit{Open} ?; \textit{Data} P; \textit{Post-training} \textsc{pref/adv}: RL with adversarial post-training. \\
    \textbf{LARA-Gen}\newline \parencite{mei2025lara}\newline D1 Tok & \textit{Inputs} valence-arousal; \textit{Duration} 30 s \textsc{gen}; \textit{Output} mono (rate not stated); \textit{Latency} batch, offline; \textit{Open} ?; \textit{Data} P 22k clips; \textit{Post-training} --. \\
    \bottomrule
  \end{tabular}
\end{table}

\FloatBarrier

\chapter{Real-Time and Interactive Generation}
\label{sec:realtime}

Whether a system can be played like an instrument depends on its latency, the time between an input and the audio that responds to it. A generator can produce a minute of audio in less than a minute and still respond too slowly for live use. A performer needs to know how much recent input the model can hear, when an output becomes available, and how soon a change in playing affects it. These timing questions separate fast batch generation, streaming generation, and participation in a live interaction. Offline generation serves production. Live use (class T4) requires streaming output under continuous control, which for most of the period only small models could attempt. An early survey of music generation in the context of interaction predates the audio systems reviewed here and frames the evaluation problem the class still has~\parencite{agchar2024survey}.

\section{Streaming Generation and Live Accompaniment}
\label{sec:streaming}

Interactive speed has been reached by two routes, one that makes a large offline model cheaper to sample and one that changes how it generates. The first reduces the computation needed for sampling: step and layer distillation makes latent diffusion faster than real time~\parencite{novack2025presto}, distilled inference-time control makes steering usable live~\parencite{novack2024ditto2}, adversarial post-training gives few-step, high-fidelity sampling in current open models~\parencite{evans2026stableaudio3}, and a dependency-free runtime runs such a model on CPUs and single-board computers with quantization~\parencite{spanio2026quantized}. The second is architectural streaming. Live music models generate continuously promptable audio with a fixed delay, built on autoregressive token modeling, and were designed with performance and jamming as the target use~\parencite{caillon2025live}. Live Music Diffusion Models then showed that open diffusion models can be repurposed into streaming generators: block-wise generation with key-value caching matches, and then improves on, the inference cost of the discrete autoregressive live models, on consumer hardware~\parencite{novack2026lmdm}.

\begin{figure}[H]
  \centering
  \includegraphics[width=\linewidth]{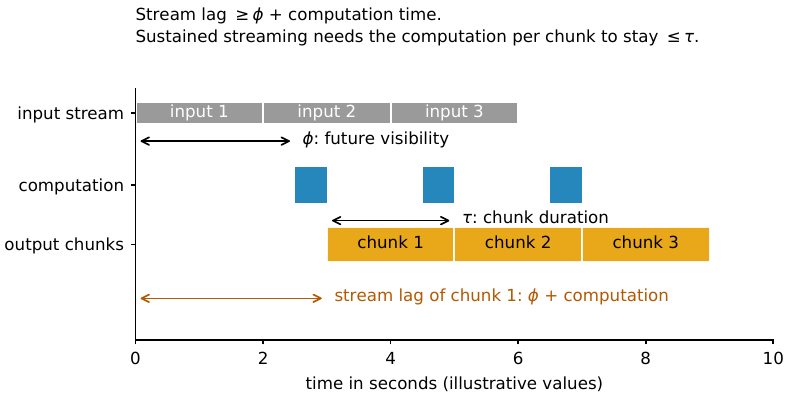}
  \caption{Illustrative chunked streaming schedule. Each output chunk has duration $\tau$. For a chunk that accompanies an interval beginning at time $t$, the latest input the chunk may use is at $t + \phi$, where $\phi$ is the future visibility. The chunk is computed after that input is available. Under the illustrated schedule, the output lags the input it accompanies by at least $\phi$ plus the computation time. Sustained streaming requires that computing a chunk take no longer than $\tau$. These quantities describe different constraints, so throughput alone does not determine response latency. The values shown are illustrative.}
  \label{fig:streaming}
\end{figure}

\paragraph{Running example.} MusicGen's arithmetic shows why offline models do not stream by default. At 50 frames per second under the delay pattern, playing in real time would require about 50 decoding steps every second, each a transformer pass over a growing context, before any computation time for the condition. The live token models of this section take the same route and make it work by fixing a chunk, bounding the context, and generating each chunk at least as fast as it plays. The block-wise diffusion models take the other route and generate a chunk at a time with a cache over the previous ones. Future visibility $\phi$ is the offset between the latest input a chunk may condition on and the start of the output interval it accompanies: a positive value gives the model look-ahead at the price of a later stream, and a negative value makes it predict ahead to absorb system delay~\parencite{wu2025streaming}. Figure~\ref{fig:streaming} draws the schedule.

Accompaniment to a live musician adds a causality constraint: the model hears an input stream and must emit a coherent output stream without seeing the future. LiveBand generates causal accompaniment to a musician's audio in real time, with training and inference computation matched by design~\parencite{pasini2026liveband}. A study of streaming accompaniment, evaluated in simulated streaming, varies future visibility and output chunk duration and examines their effects on perceived latency and coherence~\parencite{wu2025streaming}. ReaLchords learns adaptive chord accompaniment for live interaction with reinforcement learning~\parencite{wu2025adaptive}, and reinforcement-learning post-training for live jamming exposes a further problem, reward hacking toward repetitive but coherent output, which adversarial post-training mitigates~\parencite{wu2025generative}. System reports describe the architecture of real-time human-AI jamming systems~\parencite{liu2026lk} and low-latency gesture-driven generation~\parencite{jeyaraj2025gesturemusic}. Performance interfaces that map gesture to control signals for real-time symbolic generation show how such models are used in performance~\parencite{wu2026calliphony}, and a design space built from 184 live music agents across the literature and video organizes the field by usage context, interaction, technology, and ecosystem~\parencite{kim2026design}. In live coding, an agent generates TidalCycles patterns and steers them toward target affective states with reinforcement learning~\parencite{wilson2024tidal}.

\section{Neural Audio Synthesis as an Instrument}
\label{sec:instruments}

A second line of real-time work predates the text-to-music systems above and comes from computer music research. Here the goal is a playable instrument: a small model that a musician controls through a latent space, a synthesizer panel, or a live audio input. Sounds are represented by compact sets of numbers, and the model turns points in this latent space into audio. For a performer, the practical questions are what a gesture or a control movement changes, and how quickly the sound responds. The systems below are trained on hours rather than tens of thousands of hours of audio and run on a laptop CPU or on embedded hardware. Early contributions from IRCAM explored learned timbre spaces, controllable synthesis, and neural instruments~\parencite{esling2018generative,esling2019universal,caillon2021rave}. Generative timbre spaces regularize the latent space of a variational autoencoder with perceptual dissimilarity ratings, encouraging movements through the latent space to correspond to perceptually meaningful changes in timbre~\parencite{esling2018generative}. A modulated variant performs many-to-many timbre transfer with one encoder and one decoder~\parencite{bitton2018modulated}. FlowSynth maps such an organized latent audio space invertibly to synthesizer parameters with normalizing flows, so that one model gives both parameter inference from audio and macro-controls over the synthesizer~\parencite{esling2019universal}. A conditional autoencoder over drum spectrograms ships as a Max for Live plugin for real-time generation on a CPU~\parencite{aouameur2019neural}. Neural granular synthesis replaces the descriptor space over grains with a learned continuous latent~\parencite{bitton2020neural}. Structured lottery-ticket pruning makes deep generative audio models fit embedded hardware with real-time constraints~\parencite{esling2020diet}.

RAVE combined these elements into the model that most of the later instrument work builds on. It is a variational autoencoder with a two-stage training procedure, representation learning followed by adversarial fine-tuning, a multiband decomposition of the waveform, and a post-training analysis that compacts the latent space. The paper reports 48\,kHz synthesis running twenty times faster than real time on a laptop CPU~\parencite{caillon2021rave}. A companion method produces non-causal streaming models. It executes offline-trained convolutional models in the buffer-based computation of real-time audio software while preserving their non-causal context, at the cost of additional buffering latency~\parencite{caillon2022streamable}. Control then moved from latent traversal to explicit attributes: a lightweight model whose latent space is adversarially stripped of a salient descriptor turns that descriptor into a continuous control, small enough to embed in a hardware synthesizer~\parencite{devis2023continuous}. Demerlé et al.\ bring the latent-diffusion machinery of Sections~\ref{sec:instrumental} to~\ref{sec:control} into this setting: diffusion-autoencoder features separate local structure from global timbre, and one model serves both explicit control and example-based style transfer, at faster than real-time speed on a GPU~\parencite{demerle2024combining}. Two position papers state the goals of this line of work. \textcite{chemlaromeusantos2022challenges,chemlaromeusantos2022creative} argue that creative exploration can benefit from departures from the training distribution, which likelihood objectives and distribution-matching metrics may discourage. \textcite{douwes2021energy} argue that generative audio models should be scored on quality and energy consumption together. Later work extracts task-specific subnetworks from large audio models for consumer devices~\parencite{genova2025keep} and learns composable representations for diffusion-based synthesis~\parencite{bindi2024unsupervised}.

Other groups arrived at the same framing, a small controllable model that runs in real time, from different starting points. Differentiable digital signal processing starts from the synthesizer: it places interpretable components, oscillators, filters, and reverberation, inside the network, so that the network controls a synthesizer rather than producing samples, which gives controllable synthesis with small models~\parencite{engel2020ddsp}. It was extended to hierarchical note, expression, and synthesis modeling~\parencite{wu2021midi}, made lightweight through neural waveshaping~\parencite{hayes2021neural}, packaged as a real-time plugin~\parencite{ganis2021real}, and reviewed as a family~\parencite{hayes2023review}. Musika starts from the data budget: it trains on hundreds of hours with one consumer GPU and generates arbitrary-length music faster than real time on a CPU~\parencite{pasini2022musika}. The same group's later work connects the instrument line to the systems of Sections~\ref{sec:instrumental} and~\ref{sec:control}: Music2Latent is a consistency autoencoder with single-step reconstruction~\parencite{pasini2024musiclatent,pasini2025musiclatent}, noise augmentation lets continuous autoregressive models avoid error accumulation over long sequences~\parencite{pasini2024continuous}, and Diff-A-Riff 2 adds a stereo autoencoder, a diffusion transformer, and consistency training to accompaniment generation~\parencite{nistal2024improving,pasini2024bass}. Deployment has its own papers: the Neutone SDK wraps PyTorch models as real-time and offline plugins for digital audio workstations~\parencite{mitcheltree2025neutone}, and Caspe et al.\ analyze the sources of latency and jitter in interactive neural synthesizers, with RAVE-based timbre transfer as the case, and derive a design procedure for low-latency interaction~\parencite{caspe2025designing}. Related work covers novelty search over a RAVE generator~\parencite{guo2024lvns}, timbre transfer by inference-time editing of a pretrained latent diffusion model~\parencite{lee2026diffusion}, and an HCI framing of neural audio models as design materials for musicians~\parencite{zheng2026explainable}.

In the taxonomy these systems fall under audio and gesture control (D3) and classes T3 and T4 (D4). In the D5 attributes, these examples include compact models trained on relatively small corpora, released weights, and CPU or embedded deployment, although not every example has every property. Their evaluation runs through performance and practice, with little use of the distribution metrics of Section~\ref{sec:evaluation}. The two lines address the same latency budget in opposite orders. The instrument line starts with small models and adds control and quality. The streaming models of Section~\ref{sec:streaming} start from large offline models and reduce their cost. Demerlé et al.\ bring latent diffusion into the instrument setting~\parencite{demerle2024combining}, and the live models bring text conditioning to the instrument's latency budget~\parencite{caillon2025live,novack2026lmdm}.

The large text-conditioned streaming models are recent, and the interactive neural instruments have the longer history. Both must be evaluated on response latency and on the musical interaction as well as on output quality, and the available computation bounds the model that can meet a given latency budget~\parencite{novack2026lmdm}.

\section{Lessons and Open Trade-offs}
\label{sec:t4-lessons}

This class asks how to emit audio under the latency budget of Section~\ref{sec:notation} while receiving input, with whatever model capacity the bound allows. Live use requires the generator to meet a throughput budget and a response-delay budget at once. Two routes are reviewed above, and they approach the bound from opposite sides. The first reduces the cost of a large offline model, by step and layer distillation~\parencite{novack2025presto}, adversarial post-training~\parencite{evans2026stableaudio3}, or block-wise generation with key-value caching~\parencite{novack2026lmdm}, or builds a token model for continuous prompting~\parencite{caillon2025live}. The second builds a small model for the purpose: the autoencoder instruments of Section~\ref{sec:instruments}, which run on a laptop CPU~\parencite{caillon2021rave} or inside a hardware synthesizer~\parencite{devis2023continuous}. The evidence separates them on hardware and on input. The reviewed streaming text-to-music models typically run on a GPU and are prompted with text. The instruments typically run on a CPU and are played with audio or gesture. The one study that varies the streaming parameters shows that future visibility and chunk duration trade perceived latency against coherence inside the first route~\parencite{wu2025streaming}. Accompaniment to a live musician is a third problem, causal by construction~\parencite{pasini2026liveband}, and reinforcement post-training for it has a documented failure, reward hacking toward repetitive output~\parencite{wu2025generative}. The evaluations are difficult to compare. The instrument line is assessed through performance and practice, and the streaming models through distribution metrics and user studies. Latency figures are reported on different hardware without a shared procedure, although the sources of latency and jitter have been analyzed for one system class~\parencite{caspe2025designing}.

For CPU or embedded deployment with audio or gesture control, the neural-instrument literature provides the starting points and the deployment tools~\parencite{mitcheltree2025neutone}. Text-conditioned generation on a GPU has the live models and the block-wise models. Accompaniment to a live input needs a causal design. The remaining trade-offs concern latency, output quality, energy use, and hardware requirements, and this class reports embedded deployment and energy consumption where the other classes do not~\parencite{esling2020diet,douwes2021energy}. Evaluation also needs to examine the musical interaction as well as individual outputs.

\chapter{Boundaries and Adjacent Areas}
\label{sec:crossmodal-symbolic}

Neighboring areas intersect the audio-generation literature: music generated for video and images (class T5), systems that unify understanding and generation behind a language interface (class T6), and symbolic generation where it feeds audio pipelines or co-creation. This section marks where the scope of the monograph ends. We cover selected examples that show how each area connects to music audio generation. The structured searches of Section~\ref{sec:method} had no topic group for symbolic generation. The branch surveys cited in each subsection give fuller treatments.

\section{Cross-Modal Scoring}
\label{sec:t5}

Video-to-music generation conditions on visual content and must satisfy two alignments at once: semantic (mood, genre, setting) and temporal (cuts, motion, rhythm). Existing surveys cover the history, datasets, and metrics of video-to-music generation in greater detail~\parencite{ji2025comprehensive,wang2025vision}. An early system of the branch generated symbolic accompaniment, chords and MIDI, from a video's semantic, motion, and emotion features with an affective multimodal transformer and released the paired MuVi-Sync dataset~\parencite{kang2024video2music}. Within our scope the systems extend the conditioning dimension D3 without changing the paradigms. The examples below illustrate changes in conditioning and synchronization within the generative paradigms introduced earlier. V2Meow learned globally aligned video-acoustic signatures with a multi-stage autoregressive model over paired music videos~\parencite{su2023vmeow}. VidMuse combined local and global visual cues in a long-short-term model and released a large video-music dataset~\parencite{tian2024vidmuse}, MuVi added contrastive music-visual pretraining for rhythmic synchronization~\parencite{li2024muvi}, Diff-BGM brought diffusion and an annotated dataset with retrieval-based alignment metrics~\parencite{li2024diff}, and GVMGen used hierarchical spatial and temporal attention~\parencite{zuo2025gvmgen}. Recent systems address these alignments in different ways. V2M-Zero captures temporal structure through event curves computed within each modality and needs no paired video-music data at training time~\parencite{lin2026vm}. Video-Robin and VIBE combine autoregressive planning with diffusion synthesis, the first adding text conditioning for semantic control~\parencite{lokegaonkar2026video} and the second reward modeling for verifiable constraints such as tempo and key~\parencite{bhosale2026vibe}. Reproducibility is a stated concern here as elsewhere: a corpus of public-domain film clips replaces crawled video links and enables the study of dialogue as a conditioning signal~\parencite{kim2026dialogue}. Video-synchronized generation of audio, speech, and song in one diffusion transformer marks the boundary with general audio generation~\parencite{wang2025audiogen}.

Dance-to-music generation is the motion counterpart of the task, and its designs follow the same conditioning routes. One system inverts the rhythm and genre of a dance sequence into the text space of a text-to-music model~\parencite{li2024dance}. Others condition a latent diffusion model on positive and negative rhythmic cues from the video~\parencite{sun2025enhancing} or have a diffusion transformer extract genre-adaptive rhythm from joint motion and align it to the music latents~\parencite{wang2025gaca}. Visual features in place of single-dancer pose extend the task to multiple and non-human dancers~\parencite{im2026pf}, and beat-guided contrastive pretraining on unpaired motion and music reduces the need for paired data~\parencite{kimura2026dance}. Procedural music for games, where generation must react to play, has a survey of its own~\parencite{luo2025procedural}.

\section{Unified Understanding and Generation}
\label{sec:t6}

Systems in class T6 put a language model between the user and one or more generators. Two designs recur. In the first design, the language model routes requests to specialized tools. MusicAgent decomposes a request into tasks and calls specialized tools~\parencite{yu2023musicagent}, and M2UGen and MuMu-LLaMA connect music, image, and video encoders to a language model and route generation to MusicGen or AudioLDM~2~\parencite{hussain2023m2ugen,liu2024mumu}. WeaveMuse is an open multi-agent system covering understanding, symbolic composition, and audio synthesis~\parencite{karystinaios2025weavemuse}, and MusiChat uses the same pattern for iterative refinement of a composition~\parencite{liao2026musichat}. In the second, one model does both: UniAudio treated music as one task among many in a universal token language model~\parencite{yang2024uniaudio}, and UniAudio~2.0 factorizes its codec into text-aligned reasoning tokens and reconstruction tokens so that understanding and hierarchical generation share a representation~\parencite{yang2026uniaudio}. AudioChat unifies storytelling, editing, and understanding for general audio in one model~\parencite{chen2026audiochat}, and any-to-any multimodal language models include music among their output modalities~\parencite{wu2023next,zhan2024anygpt}.

A third design distributes composition itself over several agents, so far mostly in the symbolic domain. ComposerX assigns roles to language-model agents and reports that the multi-agent arrangement improves the compositions of a single model~\parencite{deng2024composerx}. CoComposer builds five agents on the stages of a composition workflow, evaluates them with the Audiobox aesthetics predictor, and reports better MusicLM performance under its evaluation protocol~\parencite{xing2025cocomposer}. Specialized agents generate higher-voice harmony for a given melody~\parencite{ganapathy2025agent}. MusicSwarm removes the central coordinator. Identical frozen foundation models act as bar-wise agents that exchange harmonic, rhythmic, and structural cues, adapt a short-term memory, and reach consensus without updating model weights. In the reported experiments, the decentralized swarm outperformed a centralized multi-agent system with a global critic on quality, diversity, and structural variety, in symbolic, audio, and graph-theoretic analyses~\parencite{buehler2025musicswarm}. Libretto gives such agents an inspectable object to work on, a symbolic grammar of onset slots, voices, and bars with a corpus-calibrated statistical space for diagnosis and self-revision~\parencite{xu2026libretto}, and a self-supervised world model over piano rolls is proposed as the representation co-creation agents need~\parencite{hawley2026helping}.

On the audio side, training-free multi-agent frameworks orchestrate existing generators for complex scenes: AudioGenie pairs a generation team with a supervisor team for multimodal-to-multiaudio generation~\parencite{rong2025audiogenie}, and Audio-Oscar coordinates specialist agents for voice design, speech, timeline planning, model selection, non-speech generation, and post-production, with a benchmark of scene descriptions~\parencite{duan2026audio}. A language model can also analyze a listener's preferences and write prompts for a commercial generator~\parencite{pandey2025tunegenie}, and an agent-built music-theory lexicon is reported to improve symbolic and audio generators~\parencite{hu2025complex}. In the taxonomy these systems are language-model agents (D2) whose generation step is symbolic or delegated to the generators of Sections~\ref{sec:instrumental} to~\ref{sec:control}. The claim that coordination rules can substitute for parameter updates has been tested on symbolic output and remains open for audio.

Music-language models also support captioning and question answering, the understanding side of the class: MU-LLaMA performs both from MERT features~\parencite{liu2023music}, and Music Flamingo and the Audio Flamingo family extend song and sound understanding in audio-language models~\parencite{ghosh2025music,ghosh2026audio}. Benchmarks test what these models know: multiple-choice music understanding~\parencite{weck2024muchomusic}, factual comprehension~\parencite{lin2025assessing}, multi-step reasoning in symbolic music~\parencite{zhou2024can}, and competence outside Western tonal music~\parencite{kader2026exploring}. Surveys of foundation models for understanding~\parencite{li2024survey} and of large audio models~\parencite{latif2023sparks} cover this class's understanding side in depth.

\section{Symbolic Generation at the Boundary}
\label{sec:symbolic}

Symbolic generation remains relevant to audio pipelines for two reasons. First, as an intermediate control layer: MuseCoco maps text to musical attributes to symbolic music, letting users steer composition before rendering~\parencite{lu2023musecoco}. Symbolic signals such as chords and melody are the control inputs of several audio systems~\parencite{tal2024jasco,lin2023cocomulla}. MIDI intermediates carry the melody in text-to-song systems~\parencite{li2024accompanied,tsai2026midi}. Text-to-MIDI generation with adapted or purpose-built language models can produce such intermediate representations~\parencite{wu2025midi,bhandari2024textmidi,ding2024songcomposer,yuan2024chatmusician,qu2024mupt}, supported by captioned MIDI datasets~\parencite{melechovsky2024midicaps}. Second, as an area of active co-creation research: anticipatory music transformers model asynchronous human-machine composition with strong infilling~\parencite{thickstun2024anticipatory}, Composer's Assistant brings multi-track infilling into a DAW-style workflow~\parencite{malandro2023composers}, NotaGen applies the full language-model pipeline of pretraining, fine-tuning, and reinforcement learning to sheet music~\parencite{wang2025notagen}, and hierarchical orchestral generation with a controllable harmony skeleton extends the approach to large ensembles~\parencite{he2026symphonygen}. The symbolic and audio threads are converging through hybrid sketch representations (Section~\ref{sec:d1}), where an intermediate plan or learned sketch is placed between the condition and the sound~\parencite{lam2025musicot,yang2025songbloom,li2026musiclayout}. A survey of language-model methods for symbolic music covers this branch in full~\parencite{le2024natural}.

\begin{takeaways}
\begin{itemize}
  \item Video-to-music systems extend the conditioning route (slot 3) with a second alignment, temporal as well as semantic, and leave the paradigms unchanged. The branch surveys cited in Section~\ref{sec:t5} cover them in depth.
  \item Unified systems come in two designs, a language model that routes to generators and a single model that understands and generates, with multi-agent composition as a third pattern that so far lives mostly in the symbolic domain.
  \item Symbolic generation matters to audio systems as a control layer and as the place where co-creation interfaces are studied. The two threads meet in the sketch representations of D1.
\end{itemize}
\end{takeaways}

\chapter{Evaluation}
\label{sec:evaluation}

\section{Distribution-Level Metrics}

For most of the period, the default automatic metric was the Fr\'echet Audio Distance (FAD), a distribution-level distance in an embedding space~\parencite{kilgour2019fad}. It was adopted for three reasons. It needs no paired reference for each generated clip, it is cheap to compute, and the original paper showed that it tracks human ratings of distorted music better than the signal-based measures it replaced. Two lines of evidence then showed its limits as a measure of generated music. First, FAD scores vary substantially with the choice of embedding model and reference set, so results are not comparable across papers~\parencite{gui2024adapting}. The correlation of FAD with human ratings depends on the embedding even for environmental audio~\parencite{tailleur2024correlation}, and the training task of the encoder shapes which acoustic properties the metric can detect~\parencite{jeong2026empirical}. Second, FAD correlates weakly with human preferences on modern text-to-music output~\parencite{huang2025aligning}. The same study released MusicPrefs, a corpus of pairwise human preferences over generated music, and proposed the MAUVE Audio Divergence (MAD), which computes the MAUVE divergence over self-supervised MERT embeddings. The metric and the embedding were selected with synthetic degradations and then validated against MusicPrefs, where MAD's system rankings agreed with listeners better than FAD's~\parencite{huang2025aligning}. Kernel-based alternatives avoid the Gaussian approximation~\parencite{chung2025kad}. Squared MMD admits an unbiased estimator, although estimates remain variable at finite sample sizes~\parencite{chung2025kad}. A comparative study of five systems found inconsistencies across these metrics and human ratings~\parencite{zhang2025aesthetics}, and a study on codec embeddings found FAD ahead of MMD against listening tests~\parencite{biswas2025evaluating}. Reference-free metrics remove the reference set altogether by reading quality from a generative backbone~\parencite{ziv2026invflowfd}. Complementary text-audio alignment scores based on contrastive embeddings~\parencite{wu2023clap} remain in use and measure prompt adherence rather than quality. A counterfactual protocol shows that prompted attribute agreement overstates control, because a requested attribute may appear because it is common in the model's outputs~\parencite{wang2026text}.

\paragraph{Running example.} The MusicGen paper used four forms of evaluation: the Fr\'echet Audio Distance with the VGGish encoder, a Kullback--Leibler divergence between label distributions predicted on generated and reference clips, a CLAP score for prompt adherence, and a listening test with ratings of overall quality and of relevance to the prompt~\parencite{copet2023musicgen}. Read with the evidence of this section, the first number depends on an encoder trained on general audio, the third shares the blind spots of the embedding it uses, and the listening test is the result a reader should weigh most, provided the report states the raters and the items. The protocol of Section~\ref{sec:listening} is the list of what such a test should state.

Evaluation begins with the claim being made. A distribution metric asks whether collections of generated and reference recordings resemble one another in a chosen representation, a prompt-adherence score asks whether an output matches its description, and listening tests and task-specific measures ask more particular questions about lyrics, structure, edits, or preference. These instruments give different evidence and are not interchangeable measures of quality. The expansion of generation capabilities over the period reviewed has not been matched by widely adopted evaluation protocols for full songs, editing, and live interaction, and much of the recent evaluation literature revisits the automatic metrics of the earlier system reports. Table~\ref{tab:eval} lists the metrics, judges, benchmarks, and protocols in use, what each measures, and whether the paper that introduced or studied it reports evidence of agreement with human judgments. Dedicated evaluation surveys cover methodology across subjective and objective methods~\parencite{xiong2023comprehensive}, metrics for audio and symbolic generation~\parencite{kader2025survey}, and the evaluation of generative models in music as a whole, including listening studies and their validity~\parencite{lerch2025survey}. This section covers what a reader needs to interpret the claims of the systems in Sections~\ref{sec:instrumental} to~\ref{sec:crossmodal-symbolic}. For each instrument those systems report, it says what the instrument measures and whether human-agreement evidence stands behind it.

\begin{figure}[H]
  \centering
  \includegraphics[width=\linewidth,alt={Timeline of evaluation instruments for generated music from 2019 to 2026.}]{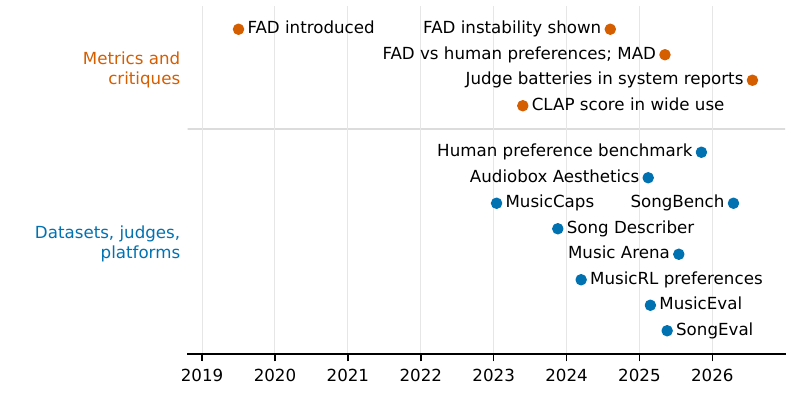}
  \caption{The evaluation literature for generated music. Top: metrics and critiques~\parencite{kilgour2019fad,wu2023clap,gui2024adapting,huang2025aligning}. Bottom: datasets, learned judges, and platforms~\parencite{agostinelli2023musiclm,manco2023songdescriber,cideron2024musicrl,tjandra2025aesthetics,liu2025musiceval,yao2025songeval,kim2025musicarena,grotschla2025benchmarking,wu2026songbench}. Positions within a year are approximate.}
  \label{fig:eval-timeline}
\end{figure}

\section{Human Evaluation and Learned Judges}

Human judgments are being collected at larger scale and in more settings than the listening tests of individual papers. Live arena platforms collect pairwise votes from real users across systems, providing continually updated preference data and a public leaderboard~\parencite{kim2025musicarena}. Large controlled preference studies compare models directly and test which automatic metrics track human judgments~\parencite{grotschla2025benchmarking}. User studies with musicians and producers measure workflow fit~\parencite{ronchini2024paguri,ronchini2025ai}, and a musician-centered framework scores adaptability, usability, and controllability~\parencite{ibanezmartinez2026musgu}. Co-design studies with music therapists, including a songwriting tool for Deaf and hard-of-hearing users, describe requirements that clip-level evaluation does not capture~\parencite{sun2024understanding,choi2026designing}. Listener studies of emotion conveyance test a specific promise of text conditioning~\parencite{go2025aimoclips}.

\FloatBarrier
\begin{table}[H]
\caption{Evaluation instruments for generated music. The evidence column states only what the cited work reports about agreement with human judgments.}
  \label{tab:eval}
  \footnotesize
  \setlength{\tabcolsep}{3pt}
  \begin{tabular}{@{}p{2.1cm}p{2.9cm}p{3.5cm}p{2.6cm}@{}}
    \toprule
    Instrument & What it measures & Human-agreement evidence & Known limits \\
    \midrule
    Fr\'echet Audio Distance~\parencite{kilgour2019fad} & Distance between embedding distributions of generated and reference sets & Weak correlation with human preference for music~\parencite{huang2025aligning}; correlation depends on the embedding~\parencite{tailleur2024correlation} & Scores vary with embedding and reference set~\parencite{gui2024adapting,jeong2026empirical}; high variance at small sample sizes \\
    Kernel Audio Distance and MMD distances~\parencite{chung2025kad,zhang2025aesthetics,biswas2025evaluating} & Distribution distance without the Gaussian assumption; KAD uses the unbiased estimator of squared MMD (unbiased in expectation, not free of small-sample variance) & Proposed as a replacement for FAD~\parencite{chung2025kad}; compared against MUSHRA and preference data in the cited studies; FAD outperformed MMD on codec embeddings~\parencite{biswas2025evaluating} & Same embedding dependence as FAD \\
    CLAP score~\parencite{wu2023clap} & Text-audio similarity of a generated clip and its prompt & Measures prompt adherence; no evidence of agreement on quality & Insensitive to musical quality; shared encoder with many training pipelines \\
    MAD~\parencite{huang2025aligning} & MAUVE divergence over self-supervised MERT embeddings & Higher correlation with pairwise human preferences than FAD, reported on MusicPrefs~\parencite{huang2025aligning} & Validated on one preference corpus (its systems, prompts, and listeners) \\
    InvFlowFD~\parencite{ziv2026invflowfd} & Reference-free and background-set-free quality from flow-matching inversion & Ranks generation models consistently with human judgments in the authors' evaluation & Depends on the inverted backbone \\
    \bottomrule
  \end{tabular}
\end{table}

\begin{table}[H]
  \addtocounter{table}{-1}
  \caption{Evaluation instruments for generated music (continued).}
  \footnotesize
  \setlength{\tabcolsep}{3pt}
  \begin{tabular}{@{}p{2.1cm}p{2.9cm}p{3.5cm}p{2.6cm}@{}}
    \toprule
    Instrument & What it measures & Human-agreement evidence & Known limits \\
    \midrule
    Aesthetics predictors~\parencite{tjandra2025aesthetics,zhang2026madb} & Predicted per-axis aesthetic scores (production quality, enjoyment, usefulness) & Trained on human ratings; agreement reported on held-out ratings & Also used as training rewards, so metric and objective coincide (Section~\ref{sec:circularity}) \\
    Song judges~\parencite{yao2025songeval,wu2026songbench,lv2026song} & Multi-aspect quality of full songs (melody, arrangement, vocals, mixing, structure) & Trained on expert annotations; SongBench reports Spearman 0.824 (per song) and 0.893 (per system) against expert structure scores~\parencite{wu2026songbench} & Genre shortcut learning documented~\parencite{zhang2026genre}; small rater pools per aspect \\
    Reward models~\parencite{zhuang2026musecritic,ma2026cmi} & Pairwise or scalar preference under text, lyric, and audio conditions & Benchmarked on human preference sets~\parencite{ma2026cmi} & Reward hacking when used for post-training~\parencite{wu2025generative} \\
    MOS predictors~\parencite{emon2025whisq,wang2026dera} & Predicted mean opinion scores for quality and text alignment & Trained and tested on MOS data & Distribution shift from natural to generated audio \\
    Live arenas~\parencite{kim2025musicarena} & Pairwise human votes across systems & Direct human data & Votes reflect the arena's users and prompts \\
    Controlled preference studies~\parencite{grotschla2025benchmarking,zhang2025aesthetics} & Human preference between systems under a fixed protocol; test of which metrics track it & Direct human data; metric inconsistencies reported & Costly to repeat; few systems, prompts, and raters \\
    \bottomrule
  \end{tabular}
\end{table}

\begin{table}[H]
  \addtocounter{table}{-1}
  \caption{Evaluation instruments for generated music (continued).}
  \footnotesize
  \setlength{\tabcolsep}{3pt}
  \begin{tabular}{@{}p{2.1cm}p{2.9cm}p{3.5cm}p{2.6cm}@{}}
    \toprule
    Instrument & What it measures & Human-agreement evidence & Known limits \\
    \midrule
    Counterfactual controllability~\parencite{wang2026text} & Instruction-attributable control, separated from attribute base rates & Not a human-agreement measure & Applied to key and beat grouping so far \\
    Emotion conveyance~\parencite{go2025aimoclips} & Rated valence and arousal against the intended emotion & Direct human data & One benchmark \\
    Editing preservation~\parencite{vishe2025evaluating} & What an edit left unchanged, per musical facet & Objective validation reported & Editing-specific \\
    Musician-centered frameworks~\parencite{ibanezmartinez2026musgu,ronchini2024paguri,ronchini2025ai} & Adaptability, usability, controllability & User studies and interviews & Small samples; qualitative \\
    Caption sets~\parencite{agostinelli2023musiclm,manco2023songdescriber} & Prompt sets for text-to-music evaluation & Human-written captions & Short clips; instrumental bias \\
    Academic challenge~\parencite{hsieh2026academic} & Fixed data and compute protocol for from-scratch systems & Protocol metrics plus human ratings & Instrumental only \\
    \bottomrule
  \end{tabular}
\end{table}
\FloatBarrier

In parallel, expert-annotated corpora support the training and testing of learned evaluators, models that predict a rating from the audio so that a system can be scored without a new listening test. The corpora differ in what was rated and by whom: ratings of generated music~\parencite{liu2025musiceval}, full-song aesthetics annotations built for the lyrics-to-song era~\parencite{yao2025songeval}, a large aesthetics dataset rated across ten perceptual dimensions by trained annotators~\parencite{zhang2026madb}, fine-grained multi-aspect song-quality benchmarks scoring melody, arrangement, vocals, mixing, and structure separately~\parencite{wu2026songbench}, and stem-aware song judges~\parencite{lv2026song}. General audio aesthetics predictors trained on such ratings are also used as reward models and data filters inside training pipelines~\parencite{tjandra2025aesthetics}, the dual use that Section~\ref{sec:circularity} returns to. Reward models for songs now produce a written critique before a score~\parencite{zhuang2026musecritic}, and a benchmark tests reward models under compositional text, lyric, and audio instructions~\parencite{ma2026cmi}. Mean-opinion-score predictors learn quality and text alignment directly~\parencite{emon2025whisq,wang2026dera}. The judges have known failure modes, and a reader should expect them: aesthetic scorers trained on SongEval learn genre as a shortcut for quality and overestimate pop music~\parencite{zhang2026genre}, and popularity on generative platforms is predictable from aesthetic features, which reflects the platforms' audiences as well as the music~\parencite{husain2026apex}. Cleaner caption datasets improve the text side of text-to-music evaluation~\parencite{manco2023songdescriber}, representation benchmarks support comparisons of the encoders that metrics are built on~\parencite{yuan2023marble}, and an academic challenge with fixed data and compute makes systems comparable at the protocol level~\parencite{hsieh2026academic}.

\section{A Minimum Listening-Test Protocol}
\label{sec:listening}

Listening tests need enough detail for readers to assess the comparison. The following protocol lists the information we recommend reporting. Each item addresses a source of uncertainty or bias discussed above.
\begin{guidelines}{A minimum listening-test protocol}
\begin{enumerate}
  \item \emph{Prompt and item sampling.} State how the prompts were chosen and how many outputs each system generated per prompt. Author-written prompts and outputs selected by hand can introduce selection bias. State the sampling procedure and any selection criteria, and fix the sampling seed.
  \item \emph{Listeners.} State the number of raters, how they were recruited, their musical training, and whether any rater is an author. Report expert and lay raters separately where both are present.
  \item \emph{Randomization and counterbalancing.} Present the items in an order randomized per rater, with the system identity hidden and the order of systems counterbalanced across raters, to reduce the influence of order and fatigue on the comparison.
  \item \emph{Repeated measures.} Have each rater judge several items per system and each item judged by several raters, so that rater effects and item effects can be separated.
  \item \emph{Intervals.} Report a confidence or credible interval for the contrast between systems, from a model that accounts for raters and items, rather than the standard deviation of the ratings alone. State the interval level and the primary contrast. If the interval for the contrast includes zero, the data do not establish the direction of the difference, and such an interval does not establish equivalence either.
  \item \emph{Rater effects.} Account for the dependence among ratings from the same listener and ratings of the same item, with a model or an analysis that fits the design. Report agreement or preference heterogeneity as a separate descriptive result. Low agreement can reflect differences in listeners' preferences and does not by itself invalidate a system comparison.
  \item \emph{Multiple comparisons.} When several systems or several aspects are compared, correct for the number of comparisons or state that no correction was made.
  \item \emph{Playback.} State the playback device, the loudness normalization, the sample rate and duration of the excerpts, and whether excerpts were cut from longer outputs. State whether the study compares systems in their native output formats or standardizes the format to isolate other differences, and use the same playback chain and loudness procedure across conditions.
  \item \emph{Independent validation.} If the system was post-trained against a learned judge, report the human test on the same items as the judge's score and never the judge's score alone. Otherwise the judge's score measures the optimization target rather than listeners' judgment (Section~\ref{sec:circularity}).
\end{enumerate}
\end{guidelines}
A multi-stimulus test with a hidden reference and anchors (MUSHRA), the format standardized for audio of intermediate quality~\parencite{itu2015mushra}, supports items 3, 4, and 8 when it is run with the sampling and reporting of the list above, and it is an appropriate format when a meaningful common reference exists, as in reconstruction or processing tasks. Unconstrained generation supplies no such reference. A preference test, pairwise or in an arena~\parencite{kim2025musicarena}, suits the comparison of unconstrained outputs on musical quality, where no reference exists, because it asks which of two outputs a listener prefers. The arena platforms~\parencite{kim2025musicarena} and the controlled preference studies~\parencite{grotschla2025benchmarking} provide examples of evaluation at scale. A smaller study can still provide useful evidence if its design and uncertainty are reported, and the nine items above state what such a study should report.

\section{Gaps and a Reporting Recommendation}
\label{sec:circularity}

Gaps remain after these developments, and each limits what a system report can claim. There is no accepted protocol for full songs, where lyric intelligibility, vocal quality, arrangement, and structure each need assessment. Practice is converging on batteries of learned judges reported side by side, as in the 2026 industrial reports~\parencite{qwen2026music} and in song benchmarks~\parencite{wu2026songbench}, but no battery is yet standard and each judge is validated against small rater pools. Long-range structure has no standard metric: system reports inspect it through self-similarity matrices of generated and real music~\parencite{evans2024longform}, a song judge reports rank correlation with expert structure scores~\parencite{wu2026songbench}, but no accepted measure distinguishes a coherent three-minute form from locally plausible drift, and structure-preserving editing has an evaluation framework~\parencite{vishe2025evaluating} that generation does not. Symbolic generation has its own distribution metrics~\parencite{retkowski2024frechet}, which audio systems with a symbolic plan could reuse. And the use of learned aesthetics predictors as both training rewards and evaluation metrics creates a circularity risk: a system post-trained against a judge can score well on that judge without listeners agreeing. Reward hacking of this kind has been observed in live interaction models~\parencite{wu2025generative}.

\begin{guidelines}{Reporting an evaluation}
Given the evidence in Table~\ref{tab:eval}, a system report should include:
\begin{itemize}
  \item at least one distribution-level metric, with the encoder, the reference set, and the number of clips named;
  \item a prompt-adherence score, read as adherence rather than as quality;
  \item a listening test that states its items, raters, and intervals (Section~\ref{sec:listening});
  \item for songs, per-aspect judge scores from a judge the system was not trained against.
\end{itemize} Report when a model was optimized using a learned judge, and provide an independent human comparison on matched items. For distribution-level scores, report the encoder, reference set, and sample count so that readers can assess whether comparisons are meaningful.
\end{guidelines}

\begin{takeaways}
\begin{itemize}
  \item A distribution distance depends on its encoder, its reference set, and its sample size. The plug-in FAD estimate is biased at finite sample sizes, an unbiased MMD estimate avoids that bias and remains variable, and agreement with listeners depends on the metric, the embedding, and the evaluation population. Report those choices with the number.
  \item Table~\ref{tab:eval} distinguishes the measures for which human-agreement evidence is reported and states the population and conditions to which that evidence applies.
  \item A judge used as a training reward cannot provide independent evidence of improvement. Its scores may be reported alongside an independent human evaluation, with attention to the biases of the judge's rating set.
\end{itemize}
\end{takeaways}

\chapter{Data, Provenance, and the Legal Record}
\label{sec:data}

Music generation should serve the musicians, songwriters, performers, and communities whose work makes it possible. We support research and tools that respect their creative agency, obtain meaningful consent for the use of their work and voices, recognize their contributions, and provide fair compensation. These are commitments of this monograph. Technical quality, commercial adoption, or a licensing announcement does not by itself demonstrate that they have been met. Our discussion of Suno, Udio, and other systems does not endorse their training practices or imply that creators' objections have been resolved.

Artist and industry organizations have raised concerns about authorization for training, the use of performers' voices and identities, transparency, and the economic consequences for creative work. The Human Artistry Campaign calls for permission, appropriate compensation, and traceable uses of protected material~\parencite{hac2026principles}. A joint letter of June 2026 from artist, songwriter, and manager organizations challenged default participation and pressure to accept AI uses without clear terms, and it called for meaningful consent, remuneration, and transparency~\parencite{emma2026consent}. These statements document the positions of their signatories. They should not be treated as evidence that all artists or all music companies have the same interests.

The evidence answers several different questions. Training-data provenance records where material came from, how it was obtained and processed, and what permissions or licenses are documented. Attribution methods estimate or record relationships between an output and particular inputs or training items. Memorization tests examine whether a model can reproduce training material. Watermarks and detectors provide evidence about an output's origin under specified conditions. Each kind of evidence answers its own question: a data statement describes the corpus, a memorization test measures what the model reproduces, an attribution estimate apportions influence, and a detector identifies generated audio. Each leaves the other questions open. Table~\ref{tab:evidence} lists what each kind of evidence can support and the inference to avoid.

\begin{table}[H]
  \caption{Kinds of evidence about training data and generated outputs, what each can support, and the inference to avoid.}
  \label{tab:evidence}
  \footnotesize
  \begin{tabular}{@{}>{\raggedright\arraybackslash}p{\dimexpr0.24\linewidth-1.34\tabcolsep\relax}>{\raggedright\arraybackslash}p{\dimexpr0.34\linewidth-1.33\tabcolsep\relax}>{\raggedright\arraybackslash}p{\dimexpr0.42\linewidth-1.33\tabcolsep\relax}@{}}
    \toprule
    Evidence & What it can support & Inference to avoid \\
    \midrule
    Source and processing records & An account of the material and its handling & A complete legal clearance, or proof that every creator participated willingly \\
    A stated license or agreement & Permissions described by its parties and terms & Authorization for all past uses, every right, or every model \\
    Input records or attribution methods & Recorded or estimated relationships in a specified workflow & A universal causal allocation of creative contribution or payment entitlement \\
    Memorization tests & Evidence of reproduction under a particular protocol & Proof that no protected material can ever be reproduced \\
    A watermark & A tested origin claim under defined conditions & A complete training history or proof of permission \\
    A binary generated-audio detector & Fallible evidence about generated versus recorded audio & Identification of the generator or its training recordings from that binary result alone \\
    Generator attribution & Evidence distinguishing outputs from specified generators & Identification of every recording used to train a model \\
    \bottomrule
  \end{tabular}
\end{table}

Of everything a system report describes, the training data is the component described least. In the data column of Tables~\ref{tab:t1} to~\ref{tab:t3}, 43 of the 105 rows carry only the code U: the source is undisclosed, in-house, or web-collected, with or without an hours figure, and the two commercial products are among them. Twenty rows carry the code L or CC, a license or a Creative Commons corpus, 11 are inference-time methods that add no training data of their own, and the remaining rows name a public or synthetic corpus, with one exception that states nothing. Many academic and most industrial models do not disclose their corpora, and dataset opacity has been an ethical concern since before the current generation of systems~\parencite{barnett2023ethical}. The Qwen-Music technical report states corpus scale, upward of five million hours, without provenance~\parencite{qwen2026music}. The datasets that are documented are skewed: a study of music generation datasets and papers found that only a small fraction of the total hours come from non-Western genres, with disparate model performance across genres as a consequence~\parencite{mehta2025music}.

These data codes describe what the sources disclose. U is not a finding of unlawful use, and L or CC is not an independent audit of every relevant right or of the terms on which creators participated.

The legal record, discussed in Section~\ref{sec:record}, is read here for its consequences for machine learning. First, the question the Munich judgments turned on is whether training items can be reproduced from the model, while the United States cases also concern how the recordings were acquired, the copies made during training, circumvention, market effect, and fair use. Memorization is therefore a measurement that a release should report, with the protocol stated (Section~\ref{sec:memorization}). Second, licensing agreements should be reported with their parties and stated scope. The systems with documented sources below are reference points for the systems with undisclosed data. The causal effect of licensed data would need a controlled comparison. Third, attribution and origin-verification tools (Section~\ref{sec:watermarking}) can support accountability, but they do not determine who authorized a use, who is entitled to payment, or whether payment occurred. The technical measurements and the legal record therefore need to be read together, with their different limits kept explicit.

Fourth, how creators should be paid when their catalogs train models is being formalized as a market design problem with data-attribution scores at its center~\parencite{zhang2026what}, and one proposed agent architecture makes attribution part of the generation workflow, with music organized into retrievable blocks whose every use triggers an attribution event~\parencite{kim2025generation}. The same group then tested what such a record can establish, in a study published after the eligibility cutoff and cited here for context. A stem generator conditioned only on audio of known provenance records every input, and the outputs are audited for reproductions of training tracks absent from the record with a version-identification model. In their experiments the recorded inputs shaped the outputs as intended (timbre followed the prompt audio and harmony followed the context), reproductions outside the record were rare, and on human-judged cases the audit's detector had higher precision and recall than the other memorization detectors tested~\parencite{kim2026tracing}. In this workflow, an input record and an output audit provide complementary evidence. The record identifies supplied material, while the audit tests for reproductions that the record alone would not show. Neither gives a complete account of training-data influence or creators' entitlements.

A minority of systems document the sources and licenses of their training data. Stable Audio Open generates from Creative Commons audio with open weights~\parencite{evans2025stableaudioopen}, and Diff-A-Riff is a production tool trained on a licensed catalog~\parencite{nistal2024diffariff}. Stable Audio 3 reports that its medium and large models are trained on licensed AudioSparx material and Creative Commons recordings from Freesound, and that the final training stages of its small music model use AudioSparx only. An audio tagger (PANNs) flagged music-like Freesound recordings, which an unnamed content-detection company then checked, and the authors removed the recordings that the service identified as copyrighted content~\parencite{evans2026stableaudio3}. This describes a source and filtering procedure. It does not establish that the corpus is free of copyright or that every relevant right has been cleared. A Creative Commons license is a copyright license and does not guarantee clearance of all other rights~\parencite{creativecommons2026faq}. Muse is an open song generator trained on a released licensed synthetic corpus~\parencite{jiang2026muse}, and the ICME challenge fixes a Creative Commons training set for every participant~\parencite{hsieh2026academic}. Open caption corpora over freely licensed music support the same goal~\parencite{roy2025jamendomaxcaps}. These systems are the reference points for comparisons with undisclosed-data models.

\paragraph{Running example.} MusicGen's data statement is the kind a reader should look for: about 20,000 hours of licensed music, named by source, with the weights released under a non-commercial license and the code under a permissive one~\parencite{copet2023musicgen}. The statement answers the first question of this section, what the model was trained on. On the second, whether the model reproduces any of it, the paper reports a prompted-continuation experiment: it prompts the model with excerpts of training items and counts exact and partial matches of the first-codebook tokens in the continuation. The scope of that test is narrower than a general test of whether generated music reproduces protected material, and a release that answered the question would state the protocol with the measurement.

\section{Litigation, Licensing, and Creators' Claims}
\label{sec:record}

The selected legal sources discussed here were checked on 23 October 2026. This account distinguishes court decisions from litigants' positions and company announcements. The coverage of the system comparison remains defined in Section~\ref{sec:method}.

In June 2024, major record labels sued the two leading song generation companies over training on copyrighted recordings~\parencite{riaa2024lawsuits,umg2024suno,umg2024udio}. Particular settlements have since coexisted with litigation. Universal Music Group and Udio announced a settlement and licensing agreements on 29 October 2025~\parencite{umg2025udio}. Warner Music Group announced an agreement with Udio on 19 November 2025~\parencite{wmg2025udio} and with Suno on 25 November 2025~\parencite{wmg2025suno,malik2025warner}. These announcements concern the identified parties and their proposed licensed products. They do not establish authorization for every prior use, resolve other parties' claims, or demonstrate consent and payment for all affected creators.

Creators' interests cannot be inferred from agreements between platforms and record companies. In June 2026, the American Federation of Musicians brought an action against Universal, Atlantic, and Warner entities, alleging breaches of obligations to notify the union and compensate musicians when recordings were licensed to Suno and Udio~\parencite{afm2026newuse}. The union's account states its allegations and its interpretation of the collective agreement, and no court has adjudicated them. The action illustrates why the existence of a label license and the treatment of the musicians whose performances were licensed are separate questions.

On 9 September 2026, Suno announced v6, named industry partners including Warner Music Group, BMG, and Believe, and said it would retire earlier models during the rollout~\parencite{suno2026v6}. The announcement does not provide an independent audit of the complete training corpus. On 18 September 2026, Universal and Sony entities filed a second United States action concerning 60,202 recordings~\parencite{umg2026sunosecond,ingham2026secondsuit}. The complaint alleges continuing unauthorized exploitation in developing v6, including through outputs and preference signals associated with earlier models. These allegations have not been adjudicated, and announcing a new model does not resolve them. In its September 2026 answer in the earlier action, Suno asserted fair use as a defense~\parencite{suno2026answer}. That pleading states the company's legal position and does not record a judicial approval of its practices.

The Munich proceedings produced findings on particular works and models. On 11 November 2025, the Munich Regional Court held that the lyrics at issue were reproduced in OpenAI's models and prompted outputs, and that the text-and-data-mining exception did not cover those reproductions~\parencite{lgmuenchen2025gema}. On 31 July 2026, the same court found infringement concerning six compositions in Suno's v3.5 and v4 models, addressing training copies, memorization, and outputs. It applied United States law to the training copies made in the United States and rejected the fair-use defense on those facts. The court substantially granted claims for injunctive relief, information, and damages~\parencite{lgmuenchen2026suno}. The court's announcements state that both judgments are not final, and their holdings should not be generalized to every model or training use.

The United States Copyright Office's 2025 training report provides legal and policy analysis and is not a judgment in these proceedings. It treats fair use as dependent on the works, their acquisition, the purpose of training, output controls, and market effects. The report regards large-scale commercial use of copyrighted material to generate competing expressive output, especially following illegal access, as exceeding established fair-use limits. The cited version is the Office's pre-publication report~\parencite{uscopyright2025training}.

\section{Memorization}
\label{sec:memorization}

Whether a model reproduces its training data is a measurable property, but few system authors measure it. Among major releases, YuE is an exception in reporting a memorization analysis alongside the model~\parencite{yuan2025yue}, and training-time mitigations such as beat-synchronous mixup were proposed early~\parencite{chen2024musicldm}. Independent studies supply the measurements. Memorization in audio latent diffusion models grows as training sets shrink, and mel-spectrogram similarity detects it more reliably than embedding retrieval~\parencite{bralios2023generation}. Anti-memorization guidance at sampling time reduces replication in an open text-to-audio model~\parencite{messina2025mitigating}. Phonetic prompting attacks show that lyric filters in commercial and open song generators can be bypassed by homophonic rewrites, because the models bind sub-lexical sound patterns to memorized songs~\parencite{roh2025bob}. Melody-aware similarity models built for plagiarism detection are the complementary tool on the output side~\parencite{lu2025melodysim,go2026music}.

\section{Watermarking and Detection}
\label{sec:watermarking}

Watermarking and detection provide different forms of evidence about an output's origin. A watermark inserted during generation can support an origin claim within the conditions in which it has been tested. A binary classifier trained to distinguish generated from recorded music provides fallible evidence for that distinction. By itself, such a classifier identifies neither the generating model nor its training recordings. Generator attribution is a separate task, studied with datasets that distinguish outputs from different models~\parencite{comanducci2025fakemusiccaps}. Neither task establishes training-data provenance or permission to use a recording. Watermarking is the first route. Localized watermarking with sample-level detection, developed for generated speech, is the reference design~\parencite{sanroman2024audioseal}, but a benchmark of watermarking methods under realistic attacks finds neural codec resynthesis the hardest case, even for methods trained against it~\parencite{ozer2025comprehensive}. Generative watermarking embedded during generation is one proposed approach~\parencite{yun2026musicmark}. Watermarking can also be applied to the training data, so that use of a catalog is detectable in a model's outputs~\parencite{epple2024watermarking}, and audio embeddings can attribute a generated piece to influential training recordings~\parencite{barnett2024exploring}.

Detection is the second route, and its literature grew quickly during the period reviewed. An early study showed that a music deepfake detector is easy to train to high accuracy on reconstructions and argued that the test score is not the end of the problem~\parencite{afchar2024detecting,afchar2025ai}. An overview organizes detection methods by the trace they exploit, at the signal, feature, watermark, or semantic-consistency level~\parencite{li2024audio}. Benchmarks and datasets followed: classifiers and benchmarks for fully synthetic songs~\parencite{rahman2025sonics}, datasets for attributing output to the generating model~\parencite{comanducci2025fakemusiccaps}, a multi-generator, multi-domain, multi-lingual database~\parencite{li2024m}, a semantically aligned dataset across ten generation services built to prevent shortcut learning~\parencite{pascu2026echoes}, and systematic evaluations with explanations~\parencite{li2024explainable}. The 2026 work addresses deployment conditions. Detectors that rely on codec artifacts collapse under speed and pitch manipulation, and a frequency-scaling-invariant design restores robustness~\parencite{dugelay2026improved}. Hybrid tracks that mix human and AI-generated stems defeat separation-then-detection pipelines~\parencite{rigaud2026detection}, and the task is reformulated as estimating the proportion of AI-generated content~\parencite{cruz2026much}. Broadcast monitoring, where music is short and masked by speech, degrades detectors trained on clean audio~\parencite{lopezayala2026ai,lopezayala2026assessing}. Zero-shot detection for generators unseen at training time is approached by one-class modeling of real music~\parencite{afchar2026finding,han2026musicdet}. Further work covers hard negatives, detector architectures, forensic residuals, and representation choices~\parencite{morosanu2026distinguishing,batra2025melody,oh2026artifactnet,kim2025segment,park2026probing}. Detection lags generation, as analyses of the interaction between the two describe~\parencite{crosvila2025armsrace}.

\begin{takeaways}
\begin{itemize}
  \item Training-source disclosure remains incomplete: 43 of the 105 table rows carry only the data code U (undisclosed, in-house, or web-collected data). The codes describe available documentation and are neither findings of legality nor assurances of creator consent and compensation.
  \item Licensing, provenance, memorization, and output identification answer different questions. A release should state its sources, the permissions it documents, and the protocols used to test reproduction of training material.
  \item Systems with documented sources provide useful reference points. Comparisons do not isolate the effect of data licensing when training data, scale, and evaluation also differ.
  \item Input records, attribution methods, and output audits can provide complementary evidence. Their conclusions depend on the workflow and test, and they do not by themselves determine who should be credited or paid.
  \item Watermarks and detectors provide fallible evidence whose robustness must be tested under relevant transformations and deployment conditions. They do not replace source documentation, creator authorization, or compensation arrangements.
\end{itemize}
\end{takeaways}

\chapter{Discussion: Trends and Open Challenges}
\label{sec:discussion}

\section{Trends Across Classes}
\label{sec:trends}

Read together, the tables and the sections above show the following patterns in the reviewed systems. First, song generation became more prominent in this collection: Table~\ref{tab:t2} holds no system from 2023 and 25 from 2025 and 2026, while Table~\ref{tab:t1} holds six from 2023 and five from 2025 and 2026. Second, planner-renderer designs became common, from MeLoDy and MusiCoT through SongBloom and DiffRhythm~2 to the 2026 industrial reports~\parencite{lam2023efficient,lam2025musicot,yang2025songbloom,jiang2025diffrhythm2,qwen2026music,dai2026pushing}, alongside single-stage flow models~\parencite{chen2026wansong} and single-stage token models~\parencite{liu2026shao}. Third, flow matching, consistency training, distillation, and adversarial post-training reduced sampling cost to real-time and streaming rates~\parencite{prajwal2024musicflow,liu2024audiolcm,novack2025presto,evans2026stableaudio3,caillon2025live,novack2026lmdm}. Fourth, preference alignment spread from a few systems in 2024~\parencite{cideron2024musicrl,majumder2024tango2,bai2024seedmusic} to 11 of the 25 song systems of 2025 and 2026 in Table~\ref{tab:t2}, with alignment targets widening from overall preference to lyric fidelity and multiple rewards~\parencite{zhang2025hallucination,ziv2025mr}. Fifth, open weights followed two prominent commercial capabilities: open song generation within a year of the commercial products~\parencite{yuan2025yue}, open streaming generation within a year of the first live models~\parencite{novack2026lmdm}. Openness is not permanent: the repository and weights that the LeVo papers name were withdrawn by July 2026. For this reason, the openness cells follow the dated artifact checks documented in Appendix~\ref{app:screening} and are not taken from the papers. Sixth, evaluation practice changed after the default metric was shown to correlate weakly with human preference, toward metrics validated against human preferences, learned judges, and arenas (Table~\ref{tab:eval}). Seventh, the legal context in the United States moved from lawsuits alone to settlements and licenses alongside continuing suits~\parencite{riaa2024lawsuits,malik2025warner}. In Germany, the Munich Regional Court issued judgments concerning song lyrics in OpenAI's language models and six compositions in Suno's models (Section~\ref{sec:record})~\parencite{lgmuenchen2025gema,lgmuenchen2026suno}. Section~\ref{sec:data} treats provenance, detection, and compensation as research topics in their own right.

\section{Open Challenges}

The following problems remain open. They are ordered by the question of Section~\ref{sec:intro} they belong to, and for each we state what a solution would need.

\emph{Live interaction (Q2, cost and latency).} Streaming models have made generation fast enough for live performance~\parencite{caillon2025live,novack2026lmdm,pasini2026liveband}. Latency suitable for the intended performance task, stability under reinforcement post-training~\parencite{wu2025generative}, and interfaces developed with musicians~\parencite{wu2025streaming,wu2026calliphony} remain design priorities. The instrument line of Section~\ref{sec:instruments} adds two criteria the text-to-music systems rarely report, embedded deployment and energy consumption~\parencite{esling2020diet,douwes2021energy}.

\emph{Controllability (Q3, conditioning).} Fine-grained control still trades against fidelity or requires per-control training~\parencite{wu2024musiccontrolnet,tal2024jasco}. Adapters, light conditioners, and inference-time optimization reduce the cost~\parencite{baker2025lilac,tsai2025musecontrollite,novack2024ditto}, and editing methods preserve structure with increasing reliability~\parencite{chang2026anchorsteer,fu2026echoedit}, but unifying the adapter, conditioner, inference-time, and plan-based approaches remains open.

\emph{Long-range structure (Q4, planning).} Models produce full-length songs, and explicit plans, layouts, and programs give users control over form~\parencite{lam2025musicot,li2026musiclayout,wang2025musicweaver,duan2026sketchsong}, with segment-level prompts extending that control along time~\parencite{wang2026segtune}. The existing measures score aspects of structure: a song judge's structure axis, which reports rank correlation with expert structure scores (Table~\ref{tab:eval})~\parencite{wu2026songbench}, structure preservation in editing~\parencite{vishe2025evaluating}, and a structural error rate against the target lyric structure reported with one system~\parencite{yang2025songbloom}. No widely adopted protocol evaluates long-range form, motif development, and tension over minutes and across genres, so the success of plan-based control can be scored only in part. A solution would need such a protocol validated against expert judgments and a demonstration that plan-conditioned generation improves it.

\emph{Evaluation (Q5, measurement).} Evaluation uses a growing range of preference-validated metrics, arenas, and learned judges in place of a default metric with weak human correlation~\parencite{huang2025aligning}. Full-song judges~\parencite{yao2025songeval,wu2026songbench,zhuang2026musecritic} address several musical aspects, but no shared protocol is widely adopted. Reusing an aesthetics predictor as both a training reward and an evaluation metric creates a risk of overfitting to that judge~\parencite{tjandra2025aesthetics,wu2025generative}. Judges inherit dataset biases~\parencite{zhang2026genre}, and controllability claims need counterfactual protocols~\parencite{wang2026text}. A solution would need a shared song-evaluation battery with documented listener recruitment and evaluators that were not used as training rewards for the systems compared.

\emph{Data and memorization (Q6, artifacts).} Reproducible research requires open, licensed corpora at scale~\parencite{evans2025stableaudioopen,jiang2026muse,hsieh2026academic}, and memorization reporting should become a release norm~\parencite{yuan2025yue,bralios2023generation}. Phonetic memorization attacks show that lexical filters are not sufficient~\parencite{roh2025bob}, and detection degrades under manipulation, mixing, and broadcast conditions~\parencite{dugelay2026improved,rigaud2026detection,lopezayala2026assessing}.

\emph{Coverage (Q6, artifacts and data).} Song models concentrate on English and Mandarin popular music, most of the world's musical traditions are under-represented in training data~\parencite{mehta2025music}, an evaluation of language models on South Asian music reports gaps in their performance~\parencite{kader2026exploring}, and the vocabulary people use to describe and prompt generated music differs across cultures~\parencite{park2026prompting}.

\emph{Interpretability (Q1 and Q3, representation and control).} Little is known about the internal structure of music models, leaving control and safety interventions empirical. Early work discovers concepts with sparse autoencoders and steers activations in audio generators~\parencite{singh2025discovering,facchiano2025activation,paek2025learning,mariotte2025sparse}, probes and patches symbolic generators~\parencite{pocwiardowski2026mi}, and uses the same findings for editing~\parencite{chang2026anchorsteer,yang2025melodia}. A solution would connect these findings to controllable and auditable generation.

\section{Recommendations}
\label{sec:recommendations}

The evidence of Sections~\ref{sec:instrumental} to~\ref{sec:data} supports a short list of recommendations, each addressed to the reader who can act on it.

\emph{To authors of system reports.} State the four slots of the generic problem (representation, factorization, conditioning route, sampler) and the attributes of the box in Section~\ref{sec:d5}, so that the system can be placed and compared without inference. Report the evaluation as the box in Section~\ref{sec:circularity} asks, with the encoder, the reference set, and the sample count of every distribution metric, and a listening test that follows Section~\ref{sec:listening}. If a learned judge supplied a training reward, say so and add a human comparison on the same items.

\emph{To evaluators and benchmark builders.} Validate measures used to support perceptual-quality claims against listeners, and report the scope of that validation. Report listener recruitment and characteristics, and use evaluators that were not used as training rewards for the systems being compared. Test controllability with counterfactual prompts rather than with agreement on requested attributes. A useful next step is a shared full-song battery with separate lyric, vocal, arrangement, and structure assessments, documented listener recruitment, and judges independent of the systems' training.

\emph{To those who release models and data.} State the training corpus by source, hours, and license, and keep the model card current when weights are withdrawn or replaced, because the tables of this monograph record withdrawals found only by checking cards after publication. Report a memorization measurement with its protocol at release. Report any watermark or other origin-verification mechanism, the conditions under which it was tested, and its limitations. Where inputs are recorded, keep the record, since input records and output audits are complementary evidence for attribution.

\emph{To readers choosing a system.} For an open backbone to adapt, MusicGen is the common choice among token models and ACE-Step and Stable Audio Open among latent models. For long-form stereo output, the Stable Audio reports provide examples with documented training-source statements. For training a song generator from scratch at academic scale, Muse supplies a released song-generation pipeline for training and evaluation~\parencite{jiang2026muse}. The ICME challenge provides a complementary setting for instrumental generation under fixed data and compute constraints~\parencite{hsieh2026academic}. For fine-tuning a song generator, ACE-Step, YuE, and HeartMuLa are open bases. The commercial song systems are closed, and the tables do not compare their quality. For a control added to a trained generator, an adapter keeps a token model's weights fixed, a control branch or inference-time optimization suits a diffusion model, and a plan is useful when structure over minutes must be specified or revised. For CPU or embedded deployment with audio or gesture control, Section~\ref{sec:instruments} provides starting points and deployment tools~\parencite{mitcheltree2025neutone}. Section~\ref{sec:streaming} covers text-conditioned streaming models and causal accompaniment to live input. The closing subsections of Sections~\ref{sec:instrumental} to~\ref{sec:realtime} give the evidence behind each line.

\chapter{Conclusion and Future Work}
\label{sec:conclusion}

This monograph has connected the musical requirements of generation, editing, and performance to the representations, objectives, and sampling procedures used by music audio models. The six questions introduced at the outset provide a way to read a system report: what it represents, how it generates, what controls it accepts, whether it plans before rendering, how its outputs are evaluated, and what is documented about its release.

Representation constrains sequence length, reconstruction, and the controls a model can support. Codec tokens make audio tractable for language modeling, while continuous latents support diffusion and flow-based generation. Intermediate plans and sketches separate decisions about musical structure from the production of audio detail. Runtime depends on the generation schedule as well as representation, model size, hardware, and implementation. Streaming systems must meet both a throughput requirement and a response-delay requirement.

Control enters through several routes, including additional training, adapters, inference-time optimization, inversion, and explicit plans. These routes differ in their computational cost and in how they preserve an existing recording (Section~\ref{sec:t3-lessons}). Several song generators use a planner-renderer design to make structure easier to specify and revise. Single-stage systems remain important alternatives. The choice of plan and the reliability with which a renderer follows it remain open design questions.

Distribution-level scores, prompt-adherence measures, learned judges, and listening studies provide different evidence (Table~\ref{tab:eval}). Depending on the method, results may depend on the representation, reference material, evaluation protocol, or population studied. A learned judge used during training cannot by itself provide independent evidence of improvement. Comparisons therefore need to state both what was measured and what conclusions the measurement supports.

Data and release conditions are equally consequential. The comparison tables expose substantial gaps in source documentation, and a stated license, an attribution estimate, a memorization test, and an output detector answer different questions. The legal record in Section~\ref{sec:data} shows why the scope and authority of each claim matter. Our position is that progress should support musicians' and artists' creative agency, meaningful consent and the ability to decline participation, recognition, and fair compensation. Commercial adoption and technical quality do not establish those outcomes.

The reviewed systems combine established generative methods with changes in representations, planning, and post-training. The tables cannot isolate the contribution of each change because systems also differ in data, scale, and evaluation, as Section~\ref{sec:t1-lessons} states for the one class in which the paradigms can be compared. Readers assessing new releases should therefore examine codecs, autoencoders, conditioning mechanisms, and training feedback as closely as the generator family.

Practical priorities follow. Full-song evaluation needs a shared protocol that reports lyrics, vocals, arrangement, and structure separately, with documented listener recruitment and evaluators independent of model training. Control and editing need comparisons that measure the requested change, preservation of the source, added training, and per-edit cost on the backbones to which each method applies. Releases need an auditable account of sources, permissions, memorization measurements, and the conditions under which any origin-verification mechanism works (Section~\ref{sec:d5}). For interactive systems, these requirements also include response latency, stable behavior, and evaluation with musicians in use.

Broader coverage of musical traditions and languages, together with a clearer connection between internal representations and controllable generation, remains part of the agenda of Section~\ref{sec:discussion}. The comparison framework can accommodate new systems under the rules of Section~\ref{sec:d4}.

\chapter*{Use of AI Tools}
\markboth{Use of AI Tools}{Use of AI Tools}
The authors used AI-based assistants while preparing this monograph: to search the literature and check bibliographic records, to gather system attributes from papers and model cards and to assemble the comparison tables from them, and to edit the language. The authors reviewed every claim, table cell, and reference against the cited sources, and against primary sources where available, and are responsible for the content and its correctness. The figures are drawn from the bibliography and from the comparison data described in Section~\ref{sec:background}.

\appendix
\chapter{Search Strings and Counts}
\label{app:search}
The structured searches of Section~\ref{sec:method} queried the arXiv API with the search expressions below, one request per expression, restricted to submissions from the group's start date onward and returning at most 100 results per expression sorted by submission date (newest first). The searches were repeated over the collection period and closed on 2026-09-11; an expression that returned the maximum of 100 results may have missed older eligible submissions, and the table gives the number of such expressions per group. Two expressions reached it. \texttt{ti:CLaMP}, in the targeted-titles group, was re-run with pagination on 2026-10-20 and returned 131 submissions since 2017, and none of the 31 beyond the first page concerns music. The \texttt{nn\textasciitilde{}} expression of the neural-synthesis-instruments group matches more than three thousand abstracts since 2023, because its \texttt{OR} binds more loosely than its \texttt{AND}; its results beyond the first page were not screened, and the term is covered by the narrower expression \texttt{abs:\textquotedbl{}nn\textasciitilde{}\textquotedbl{} AND abs:audio} of the instrument-lineage group. Candidates were de-duplicated against the bibliography by arXiv identifier and by normalized title. The counts are unique candidates per group after de-duplication within the group, the number already cited through citation chasing at the time of the search, and the number added after the abstract was read (a dedicated sentence or a grouped citation). A candidate can appear in several groups; the total of 1,115 unique candidates in Section~\ref{sec:method} counts each once. The counts are those at the close of the searches; a later run returns a different set, because each expression returns at most 100 results, newest first.
\begin{table}[H]
  \caption{Topic groups of the structured arXiv searches, with each group's start date.}
  \label{tab:lanes}
  \footnotesize
  \setlength{\tabcolsep}{3pt}
  \begin{tabular}{@{}p{3.6cm}p{1.2cm}p{1.5cm}p{1.6cm}p{1.5cm}p{1.1cm}@{}}
    \toprule
    Topic group & Start & At maximum & Candidates & Already cited & Added \\
    \midrule
    Recent months (all topics) & 2026-07 & 0 of 14 & 90 & 1 & 34 \\
    Surveys and taxonomies & 2017-01 & 0 of 9 & 21 & 2 & 12 \\
    Codecs and tokenizers & 2023-01 & 0 of 4 & 184 & 5 & 18 \\
    Instrumental text-to-music & 2023-01 & 0 of 4 & 117 & 23 & 31 \\
    Song generation & 2024-01 & 0 of 7 & 32 & 10 & 17 \\
    Control and editing & 2023-01 & 0 of 7 & 39 & 7 & 22 \\
    Real-time (title terms) & 2024-01 & 0 of 5 & 2 & 1 & 1 \\
    Cross-modal and unified & 2023-01 & 0 of 8 & 73 & 2 & 21 \\
    Evaluation & 2023-01 & 0 of 10 & 38 & 3 & 15 \\
    Provenance, detection, copyright & 2023-01 & 0 of 12 & 26 & 1 & 19 \\
    Real-time (abstract terms) & 2023-01 & 0 of 5 & 39 & 12 & 15 \\
    Alignment (abstract terms) & 2023-01 & 0 of 5 & 34 & 17 & 15 \\
    Targeted titles & 2017-01 & 1 of 44 & 168 & 6 & 32 \\
    Kernel Audio Distance & 2024-01 & 0 of 4 & 3 & 1 & 2 \\
    Neural synthesis instruments (IRCAM, NIME) & 2019-01 & 1 of 15 & 213 & 5 & 30 \\
    Motion, games, therapy, live coding & 2023-01 & 0 of 8 & 34 & 3 & 12 \\
    Multi-agent and swarm composition & 2023-01 & 0 of 8 & 84 & 5 & 17 \\
    Instrument lineage (DDSP, timbre spaces) & 2017-01 & 0 of 10 & 56 & 0 & 11 \\
    Alignment (title terms) & 2024-01 & 0 of 6 & 6 & 2 & 1 \\
    \bottomrule
  \end{tabular}
\end{table}

The search expressions per group, in arXiv API syntax (\texttt{ti} searches titles, \texttt{abs} abstracts), were:
\begin{itemize}\footnotesize\raggedright
  \item Recent months (all topics): \texttt{abs:\textquotedbl{}music generation\textquotedbl{}}; \texttt{abs:\textquotedbl{}text-to-music\textquotedbl{}}; \texttt{abs:\textquotedbl{}song generation\textquotedbl{}}; \texttt{abs:\textquotedbl{}music\textquotedbl{} AND abs:\textquotedbl{}diffusion\textquotedbl{}}; \texttt{abs:\textquotedbl{}music\textquotedbl{} AND abs:\textquotedbl{}language model\textquotedbl{} AND abs:\textquotedbl{}generat\textquotedbl{}}; \texttt{abs:\textquotedbl{}music\textquotedbl{} AND abs:\textquotedbl{}codec\textquotedbl{}}; \texttt{abs:\textquotedbl{}music\textquotedbl{} AND abs:\textquotedbl{}evaluation\textquotedbl{} AND abs:\textquotedbl{}generat\textquotedbl{}}; \texttt{abs:\textquotedbl{}AI-generated music\textquotedbl{}}; \texttt{abs:\textquotedbl{}music\textquotedbl{} AND abs:\textquotedbl{}watermark\textquotedbl{}}; \texttt{abs:\textquotedbl{}music\textquotedbl{} AND abs:\textquotedbl{}real-time\textquotedbl{} AND abs:\textquotedbl{}generat\textquotedbl{}}; \texttt{abs:\textquotedbl{}music editing\textquotedbl{}}; \texttt{abs:\textquotedbl{}accompaniment generation\textquotedbl{}}; \texttt{abs:\textquotedbl{}lyrics\textquotedbl{}}; \texttt{abs:\textquotedbl{}music\textquotedbl{} AND abs:\textquotedbl{}preference\textquotedbl{}}.
  \item Surveys and taxonomies: \texttt{ti:\textquotedbl{}music generation\textquotedbl{} AND ti:survey}; \texttt{ti:\textquotedbl{}music generation\textquotedbl{} AND ti:review}; \texttt{ti:music AND ti:taxonomy AND ti:generat}; \texttt{ti:\textquotedbl{}text-to-music\textquotedbl{} AND ti:survey}; \texttt{ti:\textquotedbl{}audio generation\textquotedbl{} AND ti:survey}; \texttt{ti:\textquotedbl{}music\textquotedbl{} AND ti:\textquotedbl{}foundation model\textquotedbl{} AND ti:survey}; \texttt{ti:\textquotedbl{}song generation\textquotedbl{} AND ti:survey}; \texttt{abs:\textquotedbl{}music generation\textquotedbl{} AND ti:survey}; \texttt{abs:\textquotedbl{}music\textquotedbl{} AND ti:\textquotedbl{}comprehensive survey\textquotedbl{}}.
  \item Codecs and tokenizers: \texttt{ti:codec AND (ti:audio OR ti:music) AND (ti:neural OR ti:tokenizer)}; \texttt{ti:tokenizer AND (ti:audio OR ti:music)}; \texttt{abs:\textquotedbl{}neural audio codec\textquotedbl{} AND abs:music}; \texttt{ti:autoencoder AND ti:audio AND abs:diffusion}.
  \item Instrumental text-to-music: \texttt{ti:\textquotedbl{}text-to-music\textquotedbl{}}; \texttt{ti:\textquotedbl{}music generation\textquotedbl{} AND (ti:diffusion OR ti:flow OR ti:transformer OR ti:latent)}; \texttt{ti:\textquotedbl{}long-form\textquotedbl{} AND ti:music}; \texttt{ti:\textquotedbl{}text-to-audio\textquotedbl{} AND abs:music}.
  \item Song generation: \texttt{ti:\textquotedbl{}song generation\textquotedbl{}}; \texttt{ti:\textquotedbl{}lyrics-to-song\textquotedbl{} OR ti:\textquotedbl{}lyrics to song\textquotedbl{} OR ti:\textquotedbl{}text-to-song\textquotedbl{}}; \texttt{ti:song AND ti:vocal AND ti:generat}; \texttt{ti:\textquotedbl{}cover song\textquotedbl{} AND ti:generat}; \texttt{ti:accompaniment AND ti:generat}; \texttt{ti:rap AND ti:generat}; \texttt{ti:singing AND ti:accompaniment}.
  \item Control and editing: \texttt{ti:\textquotedbl{}music editing\textquotedbl{}}; \texttt{ti:\textquotedbl{}audio editing\textquotedbl{} AND abs:music}; \texttt{ti:stem AND ti:generat AND abs:music}; \texttt{ti:inpainting AND ti:music}; \texttt{ti:controllable AND ti:music AND ti:generat}; \texttt{ti:control AND ti:\textquotedbl{}text-to-music\textquotedbl{}}; \texttt{ti:adapter AND abs:\textquotedbl{}music generation\textquotedbl{}}.
  \item Real-time (title terms): \texttt{ti:\textquotedbl{}real-time\textquotedbl{} AND ti:music AND ti:generat}; \texttt{ti:streaming AND ti:music AND ti:generat}; \texttt{ti:live AND ti:music AND ti:generat}; \texttt{ti:interactive AND ti:music AND ti:generat AND abs:latency}; \texttt{abs:\textquotedbl{}jam\textquotedbl{} AND abs:\textquotedbl{}music generation\textquotedbl{}}.
  \item Cross-modal and unified: \texttt{ti:\textquotedbl{}video-to-music\textquotedbl{} OR ti:\textquotedbl{}video to music\textquotedbl{}}; \texttt{ti:\textquotedbl{}background music\textquotedbl{} AND ti:generat}; \texttt{ti:\textquotedbl{}image-to-music\textquotedbl{} OR ti:\textquotedbl{}image to music\textquotedbl{}}; \texttt{ti:\textquotedbl{}music understanding\textquotedbl{} AND ti:generation}; \texttt{ti:\textquotedbl{}any-to-any\textquotedbl{} AND abs:music}; \texttt{abs:\textquotedbl{}music\textquotedbl{} AND ti:\textquotedbl{}omni\textquotedbl{} AND abs:generat}; \texttt{ti:\textquotedbl{}large language model\textquotedbl{} AND ti:music AND abs:generat}; \texttt{ti:\textquotedbl{}audio language model\textquotedbl{} AND abs:music}.
  \item Evaluation: \texttt{ti:\textquotedbl{}Fréchet audio distance\textquotedbl{} OR ti:\textquotedbl{}Frechet audio distance\textquotedbl{}}; \texttt{ti:\textquotedbl{}audio distance\textquotedbl{} AND abs:generat}; \texttt{ti:evaluat AND ti:\textquotedbl{}music generation\textquotedbl{}}; \texttt{ti:evaluat AND ti:\textquotedbl{}generated music\textquotedbl{}}; \texttt{ti:benchmark AND ti:\textquotedbl{}text-to-music\textquotedbl{}}; \texttt{ti:aesthetic AND (ti:audio OR ti:music)}; \texttt{ti:\textquotedbl{}reward model\textquotedbl{} AND abs:music}; \texttt{abs:\textquotedbl{}listening test\textquotedbl{} AND abs:\textquotedbl{}music generation\textquotedbl{}}; \texttt{ti:metric AND ti:music AND abs:generat}; \texttt{ti:arena AND abs:music}.
  \item Provenance, detection, copyright: \texttt{ti:memorization AND (ti:audio OR ti:music)}; \texttt{ti:replication AND ti:audio AND abs:diffusion}; \texttt{ti:watermark AND (ti:audio OR ti:music) AND abs:generat}; \texttt{ti:detect AND ti:\textquotedbl{}AI-generated music\textquotedbl{}}; \texttt{ti:detect AND ti:\textquotedbl{}generated music\textquotedbl{}}; \texttt{ti:\textquotedbl{}music\textquotedbl{} AND ti:deepfake}; \texttt{ti:copyright AND ti:music AND abs:generat}; \texttt{ti:\textquotedbl{}training data\textquotedbl{} AND ti:music AND abs:generat}; \texttt{ti:attribution AND ti:music AND abs:generat}; \texttt{ti:plagiarism AND ti:music}; \texttt{ti:dataset AND ti:music AND ti:caption}; \texttt{ti:licens AND ti:music AND abs:generat}.
  \item Real-time (abstract terms): \texttt{abs:\textquotedbl{}real-time\textquotedbl{} AND abs:\textquotedbl{}music generation\textquotedbl{}}; \texttt{abs:streaming AND abs:\textquotedbl{}music generation\textquotedbl{}}; \texttt{abs:live AND abs:accompaniment AND abs:generat}; \texttt{abs:latency AND abs:\textquotedbl{}music generation\textquotedbl{}}; \texttt{abs:\textquotedbl{}interactive\textquotedbl{} AND abs:\textquotedbl{}music generation\textquotedbl{} AND abs:\textquotedbl{}real time\textquotedbl{}}.
  \item Alignment (abstract terms): \texttt{abs:\textquotedbl{}preference optimization\textquotedbl{} AND abs:music}; \texttt{abs:\textquotedbl{}reinforcement learning\textquotedbl{} AND abs:\textquotedbl{}music generation\textquotedbl{}}; \texttt{abs:\textquotedbl{}human feedback\textquotedbl{} AND abs:music AND abs:generat}; \texttt{abs:reward AND abs:\textquotedbl{}music generation\textquotedbl{}}; \texttt{abs:\textquotedbl{}reward model\textquotedbl{} AND abs:music}.
  \item Targeted titles: \texttt{ti:\textquotedbl{}kernel audio distance\textquotedbl{}}; \texttt{ti:DRAGON AND abs:music}; \texttt{ti:\textquotedbl{}distributional reward\textquotedbl{}}; \texttt{ti:Riffusion}; \texttt{ti:TangoFlux}; \texttt{ti:\textquotedbl{}Make-An-Audio 2\textquotedbl{}}; \texttt{ti:AudioX}; \texttt{ti:\textquotedbl{}Stable Codec\textquotedbl{}}; \texttt{ti:Moshi AND abs:codec}; \texttt{ti:\textquotedbl{}X-Codec\textquotedbl{} OR ti:XCodec}; \texttt{ti:SemantiCodec}; \texttt{ti:SongComposer}; \texttt{ti:Melodist}; \texttt{ti:FreeStyler}; \texttt{ti:\textquotedbl{}NExT-GPT\textquotedbl{}}; \texttt{ti:AnyGPT}; \texttt{ti:AudioChat}; \texttt{ti:ChatMusician}; \texttt{ti:MuPT}; \texttt{abs:\textquotedbl{}sparse autoencoder\textquotedbl{} AND abs:music}; \texttt{ti:\textquotedbl{}Diffusion Model Alignment Using Direct Preference\textquotedbl{}}; \texttt{ti:\textquotedbl{}Aligning Generative Music AI\textquotedbl{}}; \texttt{ti:\textquotedbl{}musical roots\textquotedbl{}}; \texttt{ti:watermark AND ti:\textquotedbl{}training data\textquotedbl{} AND abs:music}; \texttt{ti:MidiCaps}; \texttt{ti:Text2midi}; \texttt{ti:Lyria}; \texttt{ti:AudioGen AND abs:\textquotedbl{}textually guided\textquotedbl{}}; \texttt{ti:\textquotedbl{}Music Arena\textquotedbl{}}; \texttt{ti:\textquotedbl{}Frechet Music Distance\textquotedbl{} OR ti:\textquotedbl{}Fréchet Music Distance\textquotedbl{}}; \texttt{ti:CLaMP}; \texttt{ti:consistency AND abs:\textquotedbl{}music generation\textquotedbl{}}; \texttt{ti:\textquotedbl{}Audio Flamingo\textquotedbl{}}; \texttt{ti:\textquotedbl{}MuseControlLite\textquotedbl{}}; \texttt{ti:\textquotedbl{}Symbotunes\textquotedbl{}}; \texttt{ti:\textquotedbl{}Foundation Models for Music\textquotedbl{}}; \texttt{ti:\textquotedbl{}AudioLDM 2\textquotedbl{}}; \texttt{ti:\textquotedbl{}Music ControlNet\textquotedbl{}}; \texttt{ti:\textquotedbl{}MusicHiFi\textquotedbl{}}; \texttt{ti:\textquotedbl{}Prevailing Research Areas\textquotedbl{}}; \texttt{ti:\textquotedbl{}MelodyFlow\textquotedbl{} OR ti:\textquotedbl{}Melody Flow\textquotedbl{}}; \texttt{ti:\textquotedbl{}Loop Copilot\textquotedbl{}}; \texttt{ti:\textquotedbl{}Stemphonic\textquotedbl{}}; \texttt{ti:StreamMUSE}.
  \item Kernel Audio Distance: \texttt{ti:KAD AND abs:audio}; \texttt{abs:\textquotedbl{}kernel audio distance\textquotedbl{}}; \texttt{ti:\textquotedbl{}No More FAD\textquotedbl{}}; \texttt{abs:\textquotedbl{}maximum mean discrepancy\textquotedbl{} AND abs:\textquotedbl{}audio generation\textquotedbl{} AND abs:metric}.
  \item Neural synthesis instruments (IRCAM, NIME): \texttt{au:Esling}; \texttt{au:Demerle OR au:Demerl}; \texttt{au:Caillon}; \texttt{au:Chemla}; \texttt{ti:RAVE AND (abs:audio OR abs:synthesis)}; \texttt{abs:\textquotedbl{}real-time\textquotedbl{} AND abs:\textquotedbl{}neural audio synthesis\textquotedbl{}}; \texttt{abs:\textquotedbl{}timbre transfer\textquotedbl{} AND abs:generat}; \texttt{abs:\textquotedbl{}neural audio\textquotedbl{} AND abs:instrument AND abs:\textquotedbl{}real-time\textquotedbl{}}; \texttt{abs:NIME AND abs:generat}; \texttt{abs:\textquotedbl{}new interfaces for musical expression\textquotedbl{} AND abs:neural}; \texttt{abs:\textquotedbl{}latent space\textquotedbl{} AND abs:\textquotedbl{}musical instrument\textquotedbl{} AND abs:neural}; \texttt{abs:\textquotedbl{}nn\textasciitilde{}\textquotedbl{} OR abs:\textquotedbl{}Max/MSP\textquotedbl{} AND abs:neural}; \texttt{abs:\textquotedbl{}Neutone\textquotedbl{} OR abs:\textquotedbl{}RAVE\textquotedbl{} AND abs:plugin}; \texttt{au:Lattner AND abs:generat}; \texttt{au:Pasini AND abs:music}.
  \item Motion, games, therapy, live coding: \texttt{abs:\textquotedbl{}dance-to-music\textquotedbl{} OR abs:\textquotedbl{}dance to music\textquotedbl{}}; \texttt{abs:motion AND abs:\textquotedbl{}music generation\textquotedbl{} AND abs:dance}; \texttt{abs:\textquotedbl{}video game\textquotedbl{} AND abs:\textquotedbl{}music generation\textquotedbl{}}; \texttt{abs:procedural AND abs:\textquotedbl{}music generation\textquotedbl{}}; \texttt{abs:\textquotedbl{}music therapy\textquotedbl{} AND abs:generat AND abs:model}; \texttt{abs:therapeutic AND abs:\textquotedbl{}music generation\textquotedbl{}}; \texttt{abs:\textquotedbl{}live coding\textquotedbl{} AND abs:music AND abs:\textquotedbl{}language model\textquotedbl{}}; \texttt{abs:TidalCycles OR abs:SuperCollider AND abs:agent}.
  \item Multi-agent and swarm composition: \texttt{abs:\textquotedbl{}multi-agent\textquotedbl{} AND abs:music AND abs:generat}; \texttt{abs:swarm AND abs:music}; \texttt{ti:agent AND ti:music AND abs:generat}; \texttt{abs:decentralized AND abs:music AND abs:generat}; \texttt{abs:collective AND abs:\textquotedbl{}music generation\textquotedbl{}}; \texttt{ti:agentic AND abs:music}; \texttt{abs:\textquotedbl{}LLM agents\textquotedbl{} AND abs:music}; \texttt{ti:\textquotedbl{}multi-agent\textquotedbl{} AND abs:music}.
  \item Instrument lineage (DDSP, timbre spaces): \texttt{ti:AFTER AND abs:audio AND abs:timbre}; \texttt{au:Demerl AND abs:audio}; \texttt{ti:\textquotedbl{}differentiable digital signal processing\textquotedbl{}}; \texttt{ti:Neurorack}; \texttt{ti:\textquotedbl{}generative timbre spaces\textquotedbl{}}; \texttt{abs:\textquotedbl{}nn\textasciitilde{}\textquotedbl{} AND abs:audio}; \texttt{au:Esling AND abs:timbre}; \texttt{ti:\textquotedbl{}neural audio synthesis\textquotedbl{} AND abs:\textquotedbl{}real-time\textquotedbl{}}; \texttt{ti:\textquotedbl{}DDSP-Piano\textquotedbl{} OR ti:\textquotedbl{}MIDI-DDSP\textquotedbl{}}; \texttt{abs:\textquotedbl{}neural audio effects\textquotedbl{} AND abs:\textquotedbl{}real-time\textquotedbl{}}.
  \item Alignment (title terms): \texttt{ti:preference AND ti:music AND abs:generat}; \texttt{ti:\textquotedbl{}reinforcement learning\textquotedbl{} AND ti:music AND abs:generat}; \texttt{ti:alignment AND ti:music AND abs:generat}; \texttt{ti:DPO AND abs:music}; \texttt{ti:\textquotedbl{}reward\textquotedbl{} AND ti:music AND abs:generat}; \texttt{ti:\textquotedbl{}post-training\textquotedbl{} AND abs:music}.
\end{itemize}

\begin{figure}[H]
  \centering
  \includegraphics[width=\linewidth,alt={Flow diagram from the unique search candidates through screening to the works added to the bibliography.}]{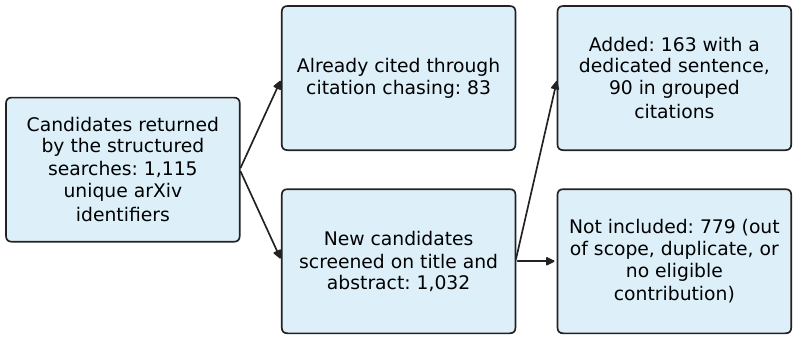}
  \caption{Flow from the candidates of the structured searches to the works included, counted over the collection period. Citation chasing and venue monitoring (Section~\ref{sec:method}) ran in parallel and are not counted here. Venue monitoring covered the conferences ISMIR, ICLR, NeurIPS, ICML, ICASSP, Interspeech, AAAI, and IJCAI and the journals TASLP, TISMIR, and TMLR.}
  \label{fig:flow}
\end{figure}

\section{Screening and Sources}
\label{app:screening}

Every candidate returned by a search expression was recorded with its identifier, title, authors, submission and update dates, categories, journal reference, and digital object identifier, the expressions that matched it, and whether it was already in the bibliography. Each candidate was screened on its title and abstract against the inclusion criteria of Section~\ref{sec:method} and received one of three decisions with a written reason: a dedicated sentence in the text, a grouped citation, or exclusion. The candidates not added during the collection period were read again in October 2026, after the classification was closed, with the same three decisions recorded. The additions that this second reading proposes are held for a later revision, including the five that predate the eligibility cutoff, so that the counts of this appendix describe one closed classification. The five are a time-compressed spectrogram generator for long accompaniments (SqueezeComposer, arXiv 2609.01046), two cross-modal systems (JenBridge, arXiv 2609.07217; VMAs, WACV 2025), a continuous-latent language-model method (CALM, ICLR 2026), and an evaluation method (PAM, Interspeech 2024). Table~\ref{tab:dates} lists the coverage, search, and source-observation dates.
\begin{table}[H]
  \caption{Coverage, search, and source-observation dates.}
  \label{tab:dates}
  \footnotesize
  \begin{tabular}{@{}p{6.2cm}p{4.6cm}@{}}
    \toprule
    Event & Date \\
    \midrule
    Eligibility cutoff for systems and publications & end of \windowend{} \\
    Closing date of the structured searches & 11 September 2026 \\
    Model cards, repositories, and product pages observed & 14 September 2026 \\
    Second reading of the remaining abstracts & October 2026 \\
    Date of this version & 26 October 2026 \\
    \bottomrule
  \end{tabular}
\end{table}

The comparison tables are generated from one set of system records with one row per system and one typed column per attribute of Section~\ref{sec:d5}, together with the source of each value (the paper, the model card, the repository, or product documentation, in that order of preference). The openness cells and the other artifact attributes follow the dated artifact checks of the model card or repository listed in Table~\ref{tab:dates}, because releases and withdrawals happen after the papers. The withdrawal of two systems was found this way. Systems released later can be added to the records and the tables under the same rules.

\backmatter
\begingroup\setlength{\emergencystretch}{2em}
\printbibliography[heading=bibintoc]
\endgroup
\end{document}